\documentclass[aps,twocolumn,groupedaddress]{revtex4}
\usepackage{graphicx}
\usepackage{amsmath}

\begin{document}

\title{Quantum optical implementation of quantum information processing
\footnote{Proceedings of the International School of Physics
Enrico Fermi {\bf 148}, {\em Experimental Quantum Computation and
Information}, Ed. F. De Martini and C. Monroe, IOS Press,
Amsterdam (2002)} }
\author{J. I. Cirac, Luming Duan, and P. Zoller}
\affiliation{Institute for Theoretical Physics, University of Innsbruck \\
Technikerstrasse 25-2, A-6020 Innsbruck, Austria}
\date{2001 10 17}

\begin{abstract}
We review theoretical proposals for implementation of quantum computing and
quantum communication with quantum optical methods.
\end{abstract}

\maketitle

\section{Introduction}

It is generally recognized that all the microscopic phenomena that we
observed can be described and explained by the principles of Quantum
Mechanics. These principles have been extensively tested, and some of them
are commonly used in several technological applications. Other principles,
like the ones related to the superposition principle and the measurement
process, and which are in the realm of most of the paradoxes and strange
phenomena related to Quantum Mechanics, have only recently become important
in some applications. In particular, they form the basis of a new theory of
information which may revolutionize the fields of communication and
computation \cite{Ni01}.

The basic ideas behind quantum communication and computation are very simple
\cite{Ni01}. Quantum communication deals with sending quantum states from
one place to another one in such a way that they arrive intact. The most
important application so far in this field is the one in which a sender
(traditionally called Alice) tries to convey a secret message to a receiver
(traditionally called Bob). The message is encoded in the state $|\Psi
\rangle $ of a quantum system. Due to the fact that the quantum state of a
system is distorted if somebody performs a measurement, Bob will receive a
wrong state if a third (malevolent) party (Eve) tries to read the message.
This way of secret communication is usually called quantum cryptography, and
it is the only provably secure way in which two partners can share secret
messages. In the context of quantum computation, the existence of entangled
states of several particles offers the possibility of performing certain
computational tasks in times much shorter than the ones taken by common
(classical) computers. By acting on a system entangled to other systems, one
modifies the state of the whole system at the same time, which leads to an
important speed up in several computations. In particular, if one could
build a quantum computer one would be able to decompose very large numbers
(of $n\gg 1$ digits) into prime factors in a time that scales polynomially
with $n$ ($t\simeq an^{b},$ with $a$ and $b$ constants) \cite{Sh94}, in
contrast to the exponential dependence of classical computers \cite{Ek96}. A
quantum computer would therefore allow to break all the classical
cryptographic protocols which are based on the impossibility of factorizing
large numbers in relatively short times. There are also other algorithms
which make use of the superposition principle in Quantum Mechanics and they
are more efficient that the classical counterparts.

The experimental situation in quantum computing and quantum communication is
very different. Whereas in the second case it is already a mature field,
close to reaching the commercial level, the first one is still in its
infancy. For the moment, it is possible to construct very small prototypes
of quantum computers which, of course, do not offer any advantage over
nowadays classical computers. In fact, it seems almost impossible that a
\textit{useful} quantum computer can be created in the next twenty or thirty
years. Nevertheless, pursuing research in this field does not have the only
goal of being useful in the near future, but there are several other goals
which can be attained in the way. In particular, creating a quantum computer
means that we can manipulate the quantum state of an enormous system at will
which apart from being capable of bringing some surprises to our present
knowledge of Quantum Mechanics, paves the way for some other applications
based on this theory which may be discovered in the future and do not
require a large system. On the other side, the first experiments on quantum
cryptography took place between location separated by few centimeters. At
present, quantum cryptography over distances of the order of 50 km is
possible. Other experiments on quantum communication achieve basically the
same distances. Their extension to longer distances does not seem to be
straightforward though, since the systems carrying the quantum states (i.e.
photons) are eventually absorbed and therefore the quantum states are
distorted before they arrive at their destination. A way to overcome this
problem is to use quantum repeaters \cite{Briegel98}, in which small quantum
computers amplify in a sense the quantum states so that they arrive safely
to their destination.

There are very few systems in which one can implement a small quantum
computer. Many of the ideas come from the field of Quantum Optics. The
reason is the spectacular experimental development of this field during the
last years, which has allowed, so to say, to dominate the quantum world. In
particular, the internal quantum levels of atoms and ions can be manipulated
very efficiently using lasers. One can basically stop them (i.e. cool them)
with laser cooling techniques. It is also possible to manipulate their
quantum state of motion by pushing them with laser light. These methods,
when combined appropriately, allow, at least in principle, to perform
quantum computations and to build quantum repeaters. On the other hand, very
recently it has been recognized that these goals can also be achieved using
atomic ensembles, instead of single atoms. The idea is to manipulate some
collective degrees of freedoms of atomic ensembles using lasers, the main
advantage being that the atoms do not need to be manipulated one by one, and
they can be at room temperature.

The aim of this paper is to review some of the quantum optical systems that
have been proposed to perform quantum computations, both using single atoms
and using atomic ensembles. In the next chapter we will give a brief
introduction to some of the main topics in quantum information theory, with
particular emphasis on the ones that are needed in the next chapters. In the
third chapter we will show how to use single atoms (ions) and photons in
order to perform several tasks related to quantum information processing,
whereas in the fourth chapter we will introduce several methods to deal with
atomic ensembles.

\section{Basic concepts in Quantum Information Theory}

\subsection{Introduction}

Most of the counter-intuitive predictions of Quantum Mechanics are related
to the superposition principle. For example, according to Quantum Mechanics
the properties of one object are generally not defined when we do not
observe it. This principle, when applied to more than one system, may lead
to very intriguing phenomena related to non-locality (actions in some system
may affect in a special way some other systems) which have called the
attention of philosophers and physicists since the advent of Quantum
Mechanics. The basic ingredient of such phenomena is \emph{entanglement},
i.e. the possibility of having two or more systems in a state which displays
(quantum) correlations. Apart from its fundamental interest, entanglement
plays an important role in most of the applications in the field of Quantum
Information \cite{Ni01}. In particular, entangled states are crucial for
quantum communication and computation. In this Section we will review the
concept of entanglement, as well as some of the basic concepts in quantum
communication and computation.

The characterization of this intriguing property of Quantum Mechanics,
entanglement, is one of the central theoretical issues in Quantum
Information Theory. In fact, there are still many open questions regarding
the entanglement properties of two or more quantum systems. Although for
pure states of two systems, entanglement is well understood, for more
systems we do not yet know how to quantify this property. The situation
becomes much more complicated if the state of the system is mixed. In that
case, we do not even know in general how to determine whether two systems
are entangled or not. This problem has important consequences in current
experiments in this field, since after preparing a quantum state of a system
one would like to determine whether it is entangled or not (as we will
explain, if it is not entangled this would mean that we could have created
the same state in a much cheaper and straightforward way). In the first
section of this chapter we will review this property, entanglement, both for
pure and mixed states and will give some of the known criteria to determine
whether a mixed state is entangled or not.

Mixed entangled states are not directly useful for quantum information
purposes. However, there exist some methods, known under the name
purification protocols, that allow us to make those states useful. We will
review some of the basic purification protocols in the second subsection.

Quantum computation and communication are the most visible applications in
the field of quantum information. In the last subsections of this section we
will review these two topics paying special attention to the physical
properties that a quantum system must possess in order to be useful for
these applications. In particular, we will give a set of requirements to
build a quantum computer which will serve us in the next chapters to show
that several quantum optical systems serve for this purpose. We will also
review one of the main tools in the field of quantum communication,
teleportation, which combined with certain purification protocols allow to
extend quantum communication over arbitrarily long distances.

\subsection{Entanglement}

\subsubsection{Entanglement of pure states}

The superposition principle is one of the basic concepts in Quantum
Mechanics: If a system can be in two different states (associated to the
vectors $|0\rangle ,|1\rangle \in H$) then it can also be in the state
described by a linear superposition $\alpha |0\rangle +\beta |1\rangle $.
This implies the existence of states in which properties are not well
defined. The situation is even more intriguing when we have a composite
system. For example, let us consider two subsystems A and B whose states are
associated to the elements of two Hilbert spaces, $H_{A}$ and $H_{B}$,
respectively. We will assume that these systems are located at different
places, although for most of our treatment this condition is not required.
Let us consider states of the whole system in which one subsystem is in
certain state $|i\rangle _{A}$ and the other in $|j\rangle _{B}$. One
denotes those states as $|i\rangle _{A}\otimes |j\rangle _{B}\in
H=H_{A}\otimes H_{B},$ or simply $|i,j\rangle \in H$. According to the
superposition principle, any superposition of these states must also be
possible, i.e. the state represented by $|\Psi \rangle \equiv \alpha
|0,0\rangle +\beta |1,1\rangle \in H$. A state of this form cannot be
described as a certain state for system A and some other state for system B;
that is, there exist no pair of vectors $|\phi _{1,2}\rangle _{A,B}\in
H_{A,B}$ such that $|\Psi \rangle =|\phi _{1},\phi _{2}\rangle $. States of
this form are called entangled states and play a fundamental role in Quantum
Information. Note that their existence arises from the fact that the states
of the whole system must be described as elements of a Hilbert space
themselves. One says that $H$ is the tensor product of $H_{A}$ and $H_{B}$,
i.e. $H=H_{A}\otimes H_{B}$. Thus entangled states are a direct consequence
of the tensor product structure of the Hilbert space describing composite
systems.

In the following, we will \ denote by \{$|k\rangle \}_{k=1}^{d_{A,B}}$ an
orthonormal basis in $H_{A,B},$ respectively. Although the definitions and
results apply for general dimensions, for most of the examples we will
consider qubits, i.e. systems where $d_{A}=d_{B}=2$. In that case we will
take as a basis \{$|0\rangle ,|1\rangle \}$. We will use the Pauli operators
\begin{eqnarray*}
\sigma _{x} &=&|1\rangle \langle 0|+|0\rangle \langle 1|, \\
\sigma _{y} &=&-i(|1\rangle \langle 0|-|0\rangle \langle 1|), \\
\sigma _{z} &=&|1\rangle \langle 1|-|0\rangle \langle 0|.
\end{eqnarray*}
For qubits, there are some entangled states which play a very important role
in quantum information, the so-called Bell states. They are
\begin{eqnarray*}
|\Psi ^{\pm }\rangle &=&\frac{1}{\sqrt{2}}(|0,1\rangle \pm |1,0\rangle ) \\
|\Phi ^{\pm }\rangle &=&\frac{1}{\sqrt{2}}(|0,0\rangle \pm |1,1\rangle )
\end{eqnarray*}

The most important properties of entangled states is that they carry
correlations. That is, if we measure an observable in A and another in B,
the outcomes will be, in general, correlated. For example, if we have the
state $|\Psi ^{-}\rangle $ and measure the observable $\sigma _{z}$ in both
systems we will obtain the opposite result. Actually, if we measure any
observable $\vec{\sigma}\cdot \vec{n}$ we will always obtain opposite
results in A and B, the reason being that $|\Psi ^{-}\rangle $ is invariant
under global rotations (i.e. $U\otimes U|\Psi ^{-}\rangle \propto |\Psi
^{-}\rangle $ for all unitary operators $U\in SU(2)$). Note that for all
entangled states there always exist some correlations. For product vectors,
however, the outcomes in A are independent of the outcomes in B. This can be
also viewed by noting that if $A$ and $B$ are two observables, then $\langle
A\otimes B\rangle =\langle A\rangle \langle B\rangle $ for product vectors,
but not (in general) for entangled states. The existence of correlations, by
itself, is not a property of entangled states. For example, if somebody
provides with two boxes A and B in which there are either two black or two
white balls, when we open the box we will see correlations. However, the
correlations carried by entangled states are, in some sense, different to
those, since they occur for any pair of observables. In fact, classical
correlations like the ones displayed by the balls in the boxes are
restricted by Bell's inequalities \cite{Be64}, whereas the ones
corresponding to entangled states may violate them. This is why with the
correlations contained in entangled states we can perform things that are
not possible using classical correlations.

In order to create entangled states out of product states we need
interactions. This can be easily understood as follows. If we do not have
interactions, the Hamiltonian describing the evolution of systems A and B
will be written as $H=H_{A}\otimes 1_{B}+1_{B}\otimes H_{B}$, where $1$ is
the identity operator. Since $H_{A}\otimes 1_{B}$ and $1_{B}\otimes H_{B}$ \
commute with each other, we have that the evolution operator can be always
written as $U(t)=U_{A}(t)\otimes U_{B}(T)$, and the product state $|\Psi
(0)\rangle =|\phi _{1}(0)\rangle _{A}\otimes |\phi _{2}(0)\rangle _{B}$ will
evolve into $|\Psi (t)\rangle =|\phi _{1}(t)\rangle _{A}\otimes |\phi
_{2}(t)\rangle _{B}$ which is a product state. Operators of the form $%
U=U_{A}\otimes U_{B}$ are called local operators. Similarly, we cannot get
entangled states by measuring observables in A and B, independently since
the state after the measurement will be changed by local operators. One says
that entanglement cannot be created by local operations (operations meaning
any action on the systems). Note, however, that product states can be
obtained by local operations (in particular, by measurements).

Let us show now how we can tell whether a state is entangled or not, and how
much. We consider a state of the form
\begin{equation}
|\Psi \rangle =\sum_{i=1}^{d_{A}}\sum_{j=1}^{d_{B}}c_{i,j}|i,j\rangle .
\end{equation}
All the information about the state is in the coefficients $c_{i,j}$ which
form a $d_{A}\times d_{B}$ matrix that we will call $C$. Note that we could
have chosen another orthonormal bases in $H_{A,B}$ to express this state. In
fact, there is a particular basis in which the matrix of the coefficients is
diagonal and positive. If we choose such a basis to write the state, it will
have the simple form
\begin{equation}
|\Psi \rangle =\sum_{k=1}^{d}d_{k}|u_{k},v_{k}\rangle ,
\end{equation}
where $d=$min$(d_{A},d_{B})$, and $\sum_{k=1}^{d}d_{k}^{2}=1,$with $%
d_{k}\geq 0.$ This form is called Schmidt decomposition. Its existence
directly follows from the singular value decomposition of the matrix $C$,
i.e., the existence of two unitaries $U$ and $V$ and a diagonal one $D$
whose diagonal elements are $d_{k}$ such that $C=UDV$. The $d_{k}$ are
called Schmidt coefficients and the bases $\{|u_{k}\rangle \}\in H_{A}$ and $%
\{|v_{k}\rangle \}\in H_{B}$ are called Schmidt bases. Once we have
expressed the state in the Schmidt decomposition, it is very simple to
obtain some other information. For example, if we are interested in
predicting expectation values or probabilities of outcomes if we only
measure system A (or B), all the information about them is in the reduced
density operator $\rho _{A}=$tr$_{B}$($|\Psi \rangle \langle \Psi |)$
(analogously for $\rho _{B}).$ We obtain
\begin{equation}
\rho _{A}=\sum_{k=1}^{d}d_{k}^{2}|u_{k}\rangle \langle u_{k}|,\quad \rho
_{B}=\sum_{k=1}^{d}d_{k}^{2}|v_{k}\rangle \langle v_{k}|
\end{equation}
Conversely, the Schmidt coefficients and the corresponding bases can be
easily found by simply diagonalizing both reduced density operators.

For a product state $|\phi _{1},\phi _{2}\rangle $, the reduced density
operators are rank one projectors, i.e. $\rho _{A,B}=|\phi _{1,2}\rangle
\langle \phi _{1,2}|.$ This means that there is only one Schmidt coefficient
which is different than zero. Conversely, if we have a state with only one
Schmidt coefficient then it must be a product state. Equivalently, $|\Psi
\rangle $ is a product state if and only if the corresponding reduced
density operators correspond to pure states. This means that if we have an
entangled state, the corresponding reduced density operators must correspond
to mixed states, or, equivalently, that there must be more than one nonzero
Schmidt coefficients.

Thus, we see that the entanglement of a state is directly related to the
mixedness of the reduced density operators. This is intuitively clear since,
as we mentioned above, entangled states give rise to correlations and if we
only observe one of the systems we loose information about these
correlations which results in the fact that we will effectively have a mixed
state. This suggests that we can measure the degree of entanglement by the
degree of mixedness of the reduced density operators. There are several
measures of mixedness of density operators; perhaps the most popular one is
the von Neumann entropy $S(\rho )=-$tr$(\rho \ln (\rho )).$ For a pure state
this entropy is zero, whereas for a maximally mixed state (described by the
identity operator, properly normalized) it gives $\log _{2}d$, where $d$ is
the dimension of the Hilbert space. The entropy is convex, i.e. for $p\in
\lbrack 0,1],$, $S[p\rho _{1}+(1-p)\rho _{2}]\geq $ $pS(\rho
_{1})+(1-p)S(\rho _{2})$, which means that it always increases by mixing
(i.e. by loosing information). This motivates the following definition:
Given a state $|\Psi \rangle $, we define the \emph{entropy of entanglement}%
, $E(\Psi )$ as the von Neumann entropy of the reduced density operator \cite%
{Be96}. Thus, we have
\begin{equation}
E(\Psi )=S(\rho _{A})=S(\rho _{B})=-\sum_{k=1}^{d}d_{k}^{2}\log
_{2}(d_{k}^{2}).
\end{equation}

The entropy of entanglement only depends on the Schmidt coefficients, but
not on the corresponding basis. This means that it is invariant under local
unitary operations. That is, if $|\Psi ^{\prime }\rangle =(U_{A}\otimes
U_{B})|\Psi \rangle $, then $E(\Psi ^{\prime })=E(\Psi )$. On the other
hand, one can show that it cannot increase in average by local operations
\cite{Po97}. That is, if we perform (independent) measurements in A and B
and obtain the state $|\Psi _{k}\rangle $ after the measurement with
probability $p_{k}$, we have that
\begin{equation}
E(\Psi )\geq \sum_{k}p_{k}E(\Psi _{k}).
\end{equation}
Note, however, that the previous inequality does not imply that none of the $%
E(\Psi _{k})$ can be larger than $E(\Psi )$, or even the maximum allowed $%
\log _{2}d$. States $|\Psi \rangle \in H_{AB}=C^{d}\otimes C^{d}$ for which $%
E(\Psi )=\log _{2}(d)$ are called \emph{maximally entangled states} in $d$
dimensions.

Let us now consider more systems $A_{1},A_{2},\ldots ,A_{N}$. Now, we can
have entangled states and product states of the different systems \cite{Li98}%
. For example, we can have a state of the form $|\Psi \rangle =|\phi
_{1}\rangle _{A_{1}A_{3}}\otimes |\phi _{2}\rangle _{A_{2}A_{5}A_{6}}\otimes
|\phi _{3}\rangle _{A_{4}}$, where $|\phi _{1,2,3}\rangle $cannot be written
as product states. It is clear that in a state like that, the parties $A_{1}$%
and $A_{3}$ are entangled with each other, but not to the rest; similarly,
the parties $A_{2},A_{5},$and $A_{6}$ are entangled among them, and the
party $A_{4}$ is completely disentangled.

In general we can consider all possible partitions of those systems in which
we group certain of them. For example, we can consider the partition $%
(A_{1}A_{3}),(A_{2}A_{5}A_{6})$, $(A_{4})$. We can classify the entangled
states according to the different partitions. That is, a state is entangled
according to some partition if it can be written as a product state of the
corresponding disjoint elements of the groups, but not within each of the
groups. In order to determine the partition corresponding to a particular
state we can calculate all possible reduced density operators and look
whether they correspond to mixed states or not.

The quantification of the multipartite entanglement is a more complicated
question which can be illustrated by the following example. Let us consider
three parties and the states \cite{Gr89},
\begin{eqnarray}
|\text{GHZ}\rangle &=&\frac{1}{\sqrt{2}}(|0,0,0\rangle -|1,1,1\rangle ) \\
|\text{W}\rangle &=&\frac{1}{\sqrt{3}}(|0,0,1\rangle +|0,1,0\rangle
+|1,0,0\rangle ).
\end{eqnarray}
Those are entangled states according to the partition $(A_{1}A_{2}A_{3})$.
However it is hard to say which one is more entangled. Certainly, the first
one possesses a sticking non-local behavior, in the sense that it can be
used to prove Bell's theorem without using inequalities \cite{Gr89}.
However, it is very weak in the sense that if one party does not participate
in the measurement (or is lost), then all the entanglement disappears.
However, the second one retains some entanglement even if one particle is
lost (in fact it is the most robust against particle losses) \cite{Du00W}.

\subsubsection{Entanglement of mixed states}

The states that we have considered in the previous Section are idealized. In
reality, all systems interact with some sort of environment. Thus, we should
include the state of the environment in our description in order to be
consistent. In fact, due to the interaction between system and environment
they will become entangled even if initially they were in a product state: $%
|\Psi _{S}(0)\rangle _{A}\otimes |\Psi _{E}(0)\rangle _{E}\rightarrow |\Psi
(t)\rangle _{SE}$. Since we are only interested in our system, all the
information that we can acquire (without performing measurements in the
environment) are contained in the reduced density operator $\rho _{S}(t)=$tr$%
_{E}[|\Psi (t)\rangle _{SE}\langle \Psi (t)|]$, which will correspond to a
mixed state. Actually, this process in which a pure state is converted into
a mixed state via its interaction \ with the environment is some times
called decoherence. The term decoherence comes from a process which makes
the coherence (non-diagonal elements of the density operator in a given
basis) to vanish. However, since this definition depends on the basis some
authors prefer to call decoherence any process which is not describable by a
unitary operator, i.e. which comes from the interaction of the system with
some other system.

Note that density operators can be always be written in the form
\begin{equation}
\rho =\sum_{k}p_{k}|\phi _{k}\rangle \langle \phi _{k}|  \label{decrho}
\end{equation}
where the $p_{k}$ are positive and add up to one. One particular
decomposition of this form is the spectral decomposition, in which,
additionally, the vectors $|\phi _{k}\rangle $ form an orthonormal basis. In
general, except for pure states (rank one density operators) there are
infinitely many decompositions. Note that a decomposition like (\ref{decrho}%
) tells us one way of creating a state described by $\rho $. We simply have
to prepare the system in state $|\phi _{k}\rangle $ with probability $p_{k}$%
. The fact that there exist infinitely many decompositions of a state means
that it can be prepared in infinitely many different forms. For example, the
state $\rho =I/2$ of one qubit can be prepared by choosing randomly one
among the states \{$|0\rangle ,|1\rangle \}$ or one among the states $%
\{|+\rangle ,|-\rangle \},$ where $|\pm \rangle =(|0\rangle \pm |1\rangle )/%
\sqrt{2}$. It is also worth stressing that even though the systems are
prepared in different forms, they are completely indistinguishable. The
reason is that the probability of any outcome after a measurement is
completely determined by the density operator, so that if two systems have
the same density operator they cannot be distinguished by performing any
measurement (and therefore by any means). Density operators are linear and
self-adjoint ($\rho =\rho ^{\dagger }$), have trace one (tr($\rho )=1)$),
and are positive ($\rho \geq 0$).

We thus have to define entanglement for mixed states \cite{We89}. Following
what happens with pure states, it makes sense to define entangled states as
those that require interactions between the systems in order to be prepared,
and non-entangled (or separable) as the ones that can be created without
interactions. More specifically, in Quantum Information a state is called
entangled if it cannot be prepared by local operations (and classical
communication) out of a product state. As we will see, this definition is
equivalent to imposing that mixtures of product states are not entangled.
Let us give some examples with two qubits: The state described by $\rho
=|0,0\rangle \langle 0,0|$ is not entangled since it is already a product
state. Any density operator of the form $\rho =\rho _{A}\otimes \rho _{B}$
is not entangled since the states $\rho _{A,B}$ can be prepared locally out
of the state $|0\rangle $. To see his, if we write $\rho $ as in (\ref%
{decrho}) we can simply transform the state $|0\rangle $ into $|\phi
_{k}\rangle $ with probability $p_{k}$. The state
\begin{equation*}
\rho =\frac{1}{2}(|0,0\rangle \langle 0,0|+|1,1\rangle \langle 1,1|)
\end{equation*}
is not entangled since it can be locally prepared as follows. We choose
randomly 0 or 1. If we have 0, we prepare both A and B in state $|0\rangle $
and otherwise in $|1\rangle $. Obviously, in this way we do not need any
interaction between the systems. We just need classical communication
between the location of A and B so that the corresponding preparers can
agree on the state they prepare. With these examples we see that the
definition of entanglement is equivalent to the following mathematical
characterization: $\rho $ is separable if and only if there exist $p_{k}\geq
0$ and $\{|a_{k}\rangle \}\in H_{A}$ and $\{|b_{k}\rangle \in H_{B}$ such
that
\begin{equation*}
\rho =\sum_{k}p_{k}|a_{k},b_{k}\rangle \langle a_{k},b_{k}|.
\end{equation*}
Otherwise it is entangled. It turns out that it is very hard in practice to
determine whether a given state is entangled or not. There exists, however,
an important sufficient criterion that may be useful in some occasions. It
states \cite{Pe96} that if $\rho ^{T_{A}}$ has a negative eigenvalue (i.e.
it is not positive) then $\rho $ is entangled. Here $\rho ^{T_{A}}$ stands
for the partial transpose of $\rho $ with respect to the first system in the
basis $\{|k\rangle \}_{k=1}^{d_{A}}\in H_{A},$ i.e. $_{A}\langle k|\rho
^{T_{A}}|k^{\prime }\rangle _{A}=_{A}\langle k^{\prime }|\rho |k\rangle .$
In general, the converse of this criterion is not true \cite{Ho97}. That is,
there exist entangled states fulfilling $\rho ^{T_{A}}\geq 0.$ However, in
low dimensions (if $d_{A}\times d_{B}\leq 6)$ this criterion (called
Peres-Horodecki criterion \cite{Pe96,Ho96}) gives a necessary and sufficient
condition:\ $\rho $ is separable if and only if $\rho ^{T_{A}}\geq 0.$ For
other separability criteria see \cite{Le00}.

Given some state, sometimes we would like to know 'how close' it is to some
pure state, like for example a Bell \ state. In order to measure this
quantity we define the fidelity of $\rho $ with respect to some state $|\Phi
\rangle $ as

\begin{equation*}
F=\langle \Phi |\rho |\Phi \rangle .
\end{equation*}
A fidelity $F\simeq 1$ means that our state is very close to the desired
one. Note that a completely random state has a fidelity $F=1/d,$where $d$ is
the dimension of the total Hilbert space.

A particular useful family of mixed states are the so-called Werner-like
states. Let us consider two systems A and B with corresponding Hilbert
spaces of dimension $d$ both. Given a state $\rho $ we depolarize it locally
by applying the same random unitary operator to system A and system B. One
can show that the state after this process has the form
\begin{equation}
\rho _{F}=F\frac{P_{a}}{d_{a}}+(1-F)\frac{P_{s}}{d_{s}}
\end{equation}
where $P_{s}=(1+\pi _{AB})/2\ $and $P_{a}=(1-\pi _{AB})/2$ are the projector
onto the symmetric and antisymmetric subspaces ($\pi _{AB}$ is the
permutation operator), and $d_{s}=d(d+1)/2$ and $d_{a}=d(d-1)/2$ the
corresponding dimensions. Here, $F=$ tr$(P_{a}\rho )$. $\rho _{F}$ is called
Werner-like state, since Werner \cite{We89} was the first one who introduced
them for the case of qubits. One can easily show that $\rho _{F}$ is
entangled if and only if $\rho _{F}^{T_{A}}$is not positive.

\subsection{Purification}

Most of the applications in the field of Quantum Information are based on
the use of superpositions of pure states. However, in practice, the state
that one has at disposal are mixed. For example, if one would like to
perform quantum cryptography over long distances using entangled photons,
when they arrive at the final location their state will also be entangled to
the environment and therefore mixed. The longer the distance the photons
have to travel, the more mixed they will become. Unfortunately, if they are
significantly mixed, the security of the corresponding cryptographic
protocol will no longer be ensured. This fact considerably limits the
distances over which one can perform secure quantum cryptography.
Fortunately, there is a method that allows to make the states more pure, and
even more entangled. The idea is to use several copies of a state which is
not useful for the applications of Quantum Information, but that it is still
entangled \cite{Bennett96,Gi96}. Using local operations and classical
communication it is sometimes possible to obtain fewer copies of particles
in a state which is closer to a maximally entangled states, for example the
state $|\Phi ^{+}\rangle $. This process is called entanglement purification
(or distillation), and will be the subject of the present section.

Let us consider the two-qubit Werner state,
\begin{equation*}
\rho _{F}=F|\Psi ^{-}\rangle \langle \Psi ^{-}|+\frac{(1-F)}{3}(|\Psi
^{+}\rangle \langle \Psi ^{+}|+|\Phi ^{+}\rangle \langle \Phi ^{+}|+|\Phi
^{-}\rangle \langle \Phi ^{-}|),
\end{equation*}
where all the vectors appearing here are Bell states. In the
present scenario, Alice and Bob share two pairs of qubits in that
state. Let us denote by $A_{1}$ and $A_{2}$ Alice's particles and
by $B_{1}$ and $B_{2}$ Bob's, so that their state is $\rho
_{F}\otimes \rho _{F}$. The distillation procedure presented in
Ref.~\cite{Bennett96} is as follows. First, Alice applies the
unitary transformation $\sigma _{y}$ to her two qubits. This
transforms in each pair $|\Psi ^{\pm }\rangle \rightarrow |\Phi
^{\mp }\rangle .$ Thus, the resulting state will have now the
maximum contribution coming from $|\Phi ^{+}\rangle .$ Then, both
apply locally a controlled-NOT operation to their two particles,
where $A_{1}$ and $B_{1}$ act like sources, and the other two
($A_{2}$ and $B_{2}$) as targets. The control-NOT operation acts
as follows:
\begin{eqnarray}
|0\rangle _{A_{1}}|0\rangle _{A_{2}} &\rightarrow &|0\rangle
_{A_{1}}|0\rangle _{A_{2}} \\
|0\rangle _{A_{1}}|1\rangle _{A_{2}} &\rightarrow &|0\rangle
_{A_{1}}|1\rangle _{A_{2}} \\
|1\rangle _{A_{1}}|0\rangle _{A_{2}} &\rightarrow &|1\rangle
_{A_{1}}|1\rangle _{A_{2}} \\
|1\rangle _{A_{1}}|1\rangle _{A_{2}} &\rightarrow &|1\rangle
_{A_{1}}|0\rangle _{A_{2}}
\end{eqnarray}
and similarly with $B$. Alice and Bob then measure the state of their target
particle in the basis $\{|0\rangle ,|1\rangle \}$ (i.e., they measure $%
\sigma _{z}^{A_{2}}$ and $\sigma _{z}^{B_{2}})$ and broadcast their results.
If the results are the same, they keep the source particles, and otherwise
they discard them. One can easily see that this is equivalent to projecting
the initial states onto the subspace in which either both the sources and
the targets are $\Phi $ states or both are $\Psi $ states. The probability
of having at the end of the process the state $|\Phi ^{+}\rangle $ is
\begin{equation}
F^{\prime }=\frac{F^{2}+(1-F)^{2}/9}{F^{2}+2F(1-F)/3+5(1-F)^{2}/9}
\end{equation}

For $1>F>1/2$, we have that $F^{\prime }>F$. Therefore, the fidelity after
this operation increases. To finish the process, the output states should be
left in a Werner state, so that this process can be continued. Alice applies
the operation $\sigma _{y}$ to his source particle, which transforms $|\Phi
^{+}\rangle \rightarrow |\Psi ^{-}\rangle $ and then they depolarize. In
summary, if the process is successful, Alice and Bob are left with a single
pair in a Werner state but with fidelity $F^{\prime }$. Then, they can take
two successful pairs and repeat the same procedure to obtain a higher
fidelity. By proceeding in this way they can reach a fidelity as close to
one as they wish, but at the expenses of wasting many pairs. In Fig. \ref%
{figupuri} we have plotted $F^{\prime }$ as a function of $F$ and show how
the fidelity increases as one repeats it with the successful pairs.

\begin{figure}[tbp]
\label{figupuri}
\par
\begin{center}
\includegraphics[width=8.0cm]{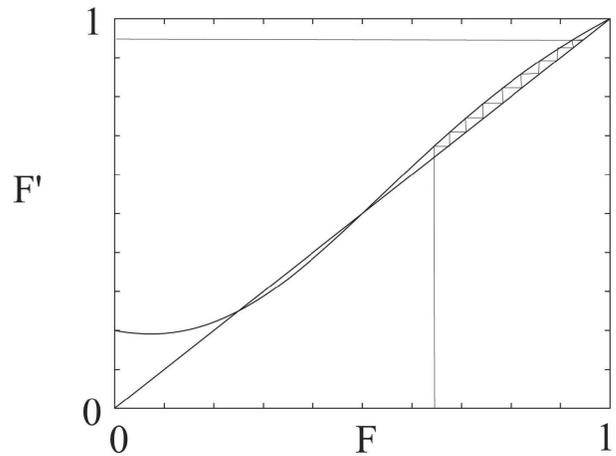}
\end{center}
\caption{New fidelity in terms of the old fidelity for the purification
protocol. Successive applications lead to a fidelity as close to one as one
wishes.}
\end{figure}

\begin{figure}[tbp]
\label{Figpuriimperf}
\par
\begin{center}
\includegraphics[width=8.0cm]{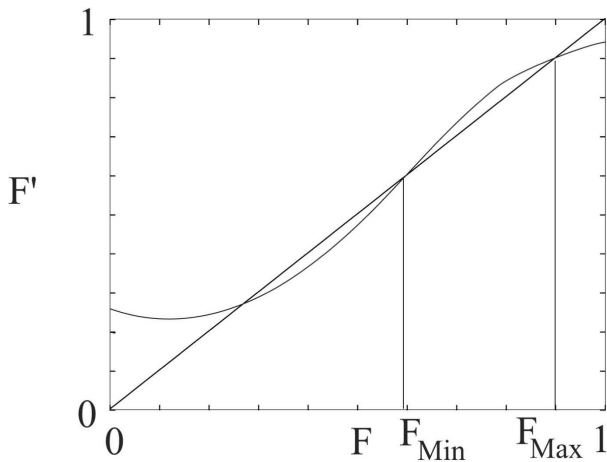}
\end{center}
\caption{Same as in the previous figure, but with imperfections.}
\end{figure}

So far we have assumed that the operations that take place during the
purification protocol (Controlled-NOT, measurements, etc.) are perfect. In
reality there will be imperfections in all these operations. One can take
them into account by using some explicit models \cite{Briegel98} or by
studying the worst case scenario \cite{Gi98}. The result is schematized in
Fig. \ref{Figpuriimperf}. Now, there is a minimum value of the original
fidelity of the state $F_{\text{min}}$ for which purification is possible.
Apart from that, there is a maximum achievable fidelity $F_{\text{max}}$due
to the imperfections.

\bigskip

\subsection{Quantum computing}

\subsubsection{What is a quantum computer?}

A computation can be considered as a physical process that transforms an
input into an output. A classical computation is that in which the physical
process is based on classical laws (without coherent quantum phenomena). A
quantum computation is that based on quantum laws (and in particular on the
superposition principle). In quantum computation, inputs and outputs are
represented by states of the system. For example, enumerating the state of a
given basis as \{$|1\rangle ,|2\rangle ,\ldots \}$, the number $N$ would be
represented by the $N$th state of this basis. A quantum computation consists
of evolving the system with a designed Hamiltonian interaction, such that
the states are transformed as we want. Note that the operation that
transforms input into outputs has to be unitary. For example, the operation
that gives $1$ if a number is odd and $2$ if it is even could not be
implemented: $|2n+1\rangle \rightarrow |1\rangle \ $and $|2n\rangle
\rightarrow |2\rangle $ (where $n$ is an integer). This operation cannot be
unitary since it does not conserve the scalar product (i.e. $\langle
1|3\rangle =0$ but the corresponded mapped states are not orthogonal). One
can, however, use an auxiliary system so that the output is written in that
system while keeping the unitarity of the operation: $|2n+1\rangle \otimes
|0\rangle \rightarrow |2n+1\rangle \otimes |1\rangle $ and $|2n\rangle
\otimes |0\rangle \rightarrow |2n\rangle \otimes |2\rangle $ (where $n$ is
an integer). In general, if our algorithm consists of evaluating a given
function $f$, we can design an interaction Hamiltonian such that the
evolution operator transforms the input states according to the following
table:
\begin{eqnarray*}
|1\rangle \otimes |0\rangle &\rightarrow &|1\rangle \otimes |f(1)\rangle \\
|2\rangle \otimes |0\rangle &\rightarrow &|2\rangle \otimes |f(2)\rangle \\
&&\ldots \\
|N\rangle \otimes |0\rangle &\rightarrow &|N\rangle \otimes |f(N)\rangle
\end{eqnarray*}

Note that using this transformation we can, at least, do the same
computations with quantum computers as with classical computers. However,
with quantum computers we can do even more. We can prepare the input state
that in a superposition
\begin{equation*}
|\Psi \rangle =\frac{1}{\sqrt{N}}\sum_{k=1}^{N}|k\rangle \otimes |0\rangle
\rightarrow \frac{1}{\sqrt{N}}\sum_{k=1}^{N}|k\rangle \otimes |f(k)\rangle
\end{equation*}
after a single run. In principle, all the values of $f$ are present in this
superposition. Note, however, that we do not have access to this information
since if we perform a measurement we will only obtain a result (with certain
probability). Nevertheless, we see that with a quantum computer we can do at
least the same as with a classical computer, ... and even more. This
property of using quantum superpositions to run only once the computer was
termed by Feynman quantum parallelism.

\subsubsection{\protect\bigskip Requirements}

A quantum computer consists of a quantum register (a quantum system) that
can be manipulated and measured in a controlled way. In order to build a
quantum computer, one needs the following elements (see also Ref. \cite{Di00}%
):

\begin{enumerate}
\item \textbf{A set of qubits: }These are two-level systems perfectly
identified and which form the quantum register. We denote by \{$|0\rangle
_{k},|1\rangle _{k}$\} two orthogonal states of the k-th qubit, so that the
state of all the qubits (the quantum register) can be written as
\begin{equation*}
|\Psi \rangle =\sum_{k_{1},k_{2},\ldots k_{N}=0}^{1}c_{k_{1}k_{2}\ldots
k_{N}}|k_{1}\rangle _{1}\otimes |k_{2}\rangle _{2}\otimes \ldots \otimes
|k_{N}\rangle _{N}.
\end{equation*}
In the following, we will simplify the notation and write $%
|k_{1},k_{2},\ldots k_{N}\rangle $ instead of the cumbersome notation that
uses tensor products. Note that these qubits can be in superposition and
entangled states, which gives the extraordinary power to the quantum
computer. Note that the state of the qubits must be kept almost pure since
otherwise the power of the superpositions would not be effective. This means
that the qubits must be well isolated from the environment in such a way
that process of decoherence is sufficiently slow.

\item \textbf{Universal set of quantum gates:} The controlled manipulation
of the qubits means that we can perform any unitary operation $U$ on the
qubits so that $|\Psi \rangle \rightarrow U|\Psi \rangle .$ In principle, if
we want to perform general operations we should be able to engineer
arbitrary interactions between the qubits. Fortunately, this task is
enormously simplified given the fact that any $U$ can be decomposed as a
product of gates belonging to a small set, the so-called universal set of
gates. This means that if we are able to perform the gates of this set we
will be able to perform any computation (unitary operations on the register)
by simply applying a sequence of them. There are many sets of universal

gates, and of course, they are all equivalent. The most convenient set is
the one that contains one two-qubit gate (an operation acting on two qubits
only) plus a set of single-qubit gates. Let us recall here some of the gates
of this sort ($\sigma _{x,y,z}$ are the Pauli operators acting on a qubit):

\begin{enumerate}
\item Single-qubit gates: act on a single qubit.

\begin{enumerate}
\item Phase gate: $U_{z}^{(1)}(\varphi )=e^{-i\varphi \sigma _{Z}},$%
\begin{eqnarray*}
|0\rangle &\rightarrow &e^{i\varphi /2}|0\rangle \\
|1\rangle &\rightarrow &e^{-i\varphi /2}|1\rangle
\end{eqnarray*}

\item Excitations: $U_{z}^{(1)}(\theta )=e^{-i\theta \sigma _{x}},$%
\begin{eqnarray*}
|0\rangle &\rightarrow &\cos \theta |0\rangle -i\sin \theta |1\rangle \\
|1\rangle &\rightarrow &-i\sin \theta |0\rangle +\cos \theta |1\rangle
\end{eqnarray*}
\end{enumerate}

\item Two-qubit gates: act on two qubits.

\begin{enumerate}
\item Controlled-not: $U_{\text{CNOT}}^{(2)}=|0\rangle \langle 0|\otimes
1+|1\rangle \langle 1|\otimes \sigma _{x}$%
\begin{eqnarray*}
|0,0\rangle &\rightarrow &|0,0\rangle \\
|0,1\rangle &\rightarrow &|0,1\rangle \\
|1,0\rangle &\rightarrow &|1,1\rangle \\
|1,1\rangle &\rightarrow &|1,0\rangle
\end{eqnarray*}
This gate changes the state of the second qubit conditioned to the state of
the first one. The first qubit is therefore called control qubit, whereas
the second one is called target.

\item Controlled-phase: $U_{\pi }^{(2)}=|0\rangle \langle 0|\otimes
1-|1\rangle \langle 1|\otimes \sigma _{z}$%
\begin{eqnarray*}
|0,0\rangle &\rightarrow &|0,0\rangle \\
|0,1\rangle &\rightarrow &|0,1\rangle \\
|1,0\rangle &\rightarrow &|1,0\rangle \\
|1,1\rangle &\rightarrow &-|1,1\rangle
\end{eqnarray*}
Two different universal sets of gates are \cite{Ni01}:
\begin{eqnarray*}
S_{1} &=&\{U_{\text{CNOT}}^{(2)},U_{x}^{(1)}(\pi /4),U_{z}^{(1)}(\varphi
),\varphi \in \lbrack 0,2\pi )\}, \\
S_{2} &=&\{U_{\pi }^{(2)},U_{z}^{(1)}(\pi /4),U_{x}^{(1)}(\varphi ),\varphi
\in \lbrack 0,2\pi )\}.
\end{eqnarray*}
Note that the two-qubit gates require interactions between the qubits, and
therefore are the more difficult ones in practice. The fact that the
operations are unitary (and therefore reversible) means that the
uncontrolled interaction with any other part of the quantum computer must be
avoided.
\end{enumerate}
\end{enumerate}

\item \textbf{Detection: }One should be able to measure $\sigma _{z}$ on
each of the qubits (or, equivalently, to detect whether they are in state $%
|0\rangle $ or $|1\rangle ).$ Note that this process requires the
interaction with a measurement apparatus in an irreversible way.

\item \textbf{Erase:} We must be able to prepare the initial state of the
system, for example the state $|0,0,\ldots ,0\rangle $. Actually, this is
not an extra requirement since if one is able to detect and to apply the
single-qubit gate $U_{x}^{(1)}(\pi )$ this is enough.

\item \textbf{Scalability:}The difficulty of performing gates, measurements,
etc., should not grow (exponentially) with the number of qubits. Otherwise,
the gain in the quantum algorithms would be lost.
\end{enumerate}

For the moment, we know very few systems which fulfill the requirements to
implement a quantum computer with them. Perhaps, the most important problem
is related to the necessity of finding a quantum system which is
sufficiently isolated, and for which the required controlled interactions
can be produced. For the moment, there exist three kind of physical systems
that fulfill, at least, most of the requirements (see Ref. \cite{fort}):

\begin{enumerate}
\item \textbf{Quantum optical systems }: Qubits are atoms (ions), and the
manipulation takes place with the help of a laser. This systems are very
clean in the sense that with them it is possible to observe quantum
phenomena very clearly. In fact, with them several groups have managed to
prepare certain states which lead to phenomena that present certain
analogies with the Schr\"{o}dinger cat paradox, Zeno effect, etc. Moreover,
those systems are currently used to create atomic clocks, and with them one
can perform the most precise measurements that exist nowadays. For the
moment, experimentalist have been able to perform certain quantum gates, and
to entangle 3 or 4 atoms. The most important difficulty with those systems
is to scale up the models so that one can perform computations with many
atoms.

\item \textbf{Solid state systems }: There have been several important
proposals to construct quantum computers using Cooper pairs or quantum dots
as qubits. The highest difficulty in these proposals is to find the proper
isolation of the system, since in a solid it seems hard to avoid
interactions with other atoms, impurities, phonons, etc. For the moment,
only single quantum gates have been experimentally reported. However, these
systems posses the advantage that they are easily scalable.

\item \textbf{Nuclear magnetic resonance systems}: In this case the qubits
are represented by atoms within the same molecule, and the manipulation
takes place using the NMR technique. Initially, these systems seemed to be
very promising for quantum computation, since it was thought that the
cooling of the molecules was not required, which otherwise would make the
experimental realization very difficult. However, it seems that without
cooling, these systems loose all the advantages of quantum computation.
\end{enumerate}

\subsubsection{Error correction}

In any computation (classical and quantum) or during storing of information
there will be errors. One way to fight against these errors is to improve
the hardware and make it better. However, this is expensive, and not always
possible. Shannon realized that instead of trying to avoid the errors it is
much better to correct them. This is done by giving redundant information,
and using this extra information to find out if an error occur. One can
distinguish two kind of errors:

\emph{Memory errors:} Those that occur to the information that is stored,
regardless of whether an operation takes place or not.

\emph{Operation errors:} Those that occur during an operation.

Here we will concentrate on memory errors, since the corresponding
correction procedures are easier to understand. On the other hand, they play
an important role not only in quantum computing, but also in quantum
communication and information. Once one knows how memory errors can be
corrected, (with some modifications) one can understand how to correct
operation errors. We will first revise the most straightforward way of
correcting errors in a classical computer, and then we will show how to do
it in a quantum computer.

\textbf{Classical error correction:}

Imagine that one wants to store a single bit for a time $t$ (we will call
this bit a \emph{logical bit}). Let us denote by $P_{\tau }$ the probability
that one error occurs in a time interval $\tau $; that is, the probability
that the bit flips (if it was 0 then it changes to 1 and vice versa). If $%
P_{\tau }\simeq 1$ there will be problems in achieving the goal. One way to
correct the errors is based on what is called \emph{redundant coding}. This
consists of using three bits to store the logical bit. That is, we encode
the information such that if the logical bit is 0 the three bits are 0, and
if it is 1, the three bits are 1: $0_{L}\equiv $ $000$, and $1_{L}\equiv
111. $. These logical qubits are called \emph{code words}. After at time $%
\tau $, we will have

\begin{itemize}
\item Probability of no errors: $(1-P_{\tau })^{3}$ (for example, if we had
initially $000$, after the time $\tau $ it is $000$).

\item Probability of error in one bit: $3P_{\tau }(1-P_{\tau })^{2}$ (for
example, if we had initially $000$, after the time $\tau $ it is $100$, $010$
or $001$).

\item Probability of two or more errors:\ $3P_{\tau }^{2}(1-P_{\tau
})+P_{\tau }^{3}.$
\end{itemize}

The error correction consists of measuring if the three bits are in the same
state or not. If they are in the same state, then we do nothing. If they are
in a different state, we use majority vote to change the bit that is
different. For example, if we have that the first and the third bit are
equal and the second is different ($010$ or $101$), we flip the second bit ($%
000$ and $111$, respectively) . After the correction we will have the
correct state with a probability $P_{\tau }^{c}=1-3P_{u}^{2}+2P_{\tau }^{3}$%
. Thus, one gains if $P_{\tau }^{c} < 1-P_{\tau }$, that is, if (roughly) $%
P_{\tau }<1/3.$ If one wants to keep the state for very long times $t$, one
has to perform many measurements. More precisely, assume that $P_{\tau
}=1-e^{-\gamma \tau }\simeq \gamma \tau $ for times $\tau $ sufficiently
short. Let us divide $t$ in $N$ intervals of duration $\tau =t/N $. For $N$
sufficiently large, the probability of having the correct state after
performing the correction after the time $t$ will be
\begin{equation}
P_{t}^{c}\geq \left[ 1-3\left( \frac{\gamma t}{N}\right) ^{2}\right] ^{N}
\end{equation}
For $N$ large, this probability can be made as close to one as desired. One
can generalize this method to the case in which one wants to store $k$
logical bits and allow for errors in $l$ bits. For example, encoding $%
0_{L}\equiv 00000,$ $1_{L}\equiv 11111$, one can allow for two errors.

\textbf{Quantum error correction:}

Imagine that one wants to store a single quantum bit in an unknown state $%
|\Psi \rangle =c_{0}|0\rangle +c_{1}|1\rangle $ for a time $t$ (we will call
this qubit a \emph{logical qubit}). Let us assume that after a time $\tau $
with a probability $1-P_{\tau }$ the qubit remains intact and that with a
probability $P_{\tau }$ it changes to $|\Psi ^{\prime }\rangle
=c_{0}|1\rangle +c_{1}|0\rangle .$This error is called spin flip, and it can
be represented by the action of $\sigma _{x}$ onto the state of the qubit.
One can correct the above error by using \emph{redundant coding} \cite%
{Sh95,St96}. For example, one can encode the state of the logical qubit in $%
3 $ qubits as $|0\rangle _{L}\equiv |000\rangle $ and $|1\rangle _{L}\equiv
|111\rangle $ (code words). The subspace spanned by these states is called
subspace of code words. After time $\tau $, we will have:

\begin{itemize}
\item Probability of no errors: $(1-P_{\tau })^{3}$ (the state will be $%
|\Psi \rangle _{L}$).

\item Probability of error in one bit: $3P_{\tau }(1-P_{\tau })^{2}$ (the
state will be $\sigma _{x}^{1}|\Psi \rangle _{L},$or $\sigma _{x}^{2}|\Psi
\rangle _{L},$ or $\sigma _{x}^{3}|\Psi \rangle _{L}$).

\item Probability of two or more errors:\ $3P_{\tau }^{2}(1-P_{\tau
})+P_{\tau }^{3}.$
\end{itemize}

Note that in order to correct the errors, we cannot do the same as in the
classical case, since measuring the state of the qubit will collapse it in a
different state (for example $|000\rangle $), and therefore the
superposition will be destroyed. What we can do is to detect whether the
three bits are in the same state or not, without disturbing the state. If
the qubits are in the same state, then we do nothing. If they are a
different state, we use majority vote to change the bit that is different.
All these measurements have to be performed without destroying the
superposition. This can be done as follows: first we measure the projector $%
P=|000\rangle \langle 000|+|111\rangle \langle 111|$ (which corresponds to
an incomplete measurement). If we obtain $1$, then we leave the qubits as
they are. If we obtain $0$ then we measure the projector $P_{1}=|100\rangle
\langle 100|+|011\rangle \langle 011|$: if we obtain $1$ we apply the local
unitary operator $\sigma _{x}^{1}$ and if not we proceed. We measure $%
P_{2}=|010\rangle \langle 010|+|101\rangle \langle 101|$; if we obtain $1$
we apply the local unitary operator $\sigma _{x}^{2}$ and if not we apply
the operator $\sigma _{x}^{3}$ (note that if we measure the operator $P_{3}$
we would obtain $1$ with probability $1$). As a result, if there was either
no error or one error, it will be corrected. If there were two or more
errors, they will not be corrected. Using this method, we achieve the same
results as in the classical correction method, namely, by correcting very
often we can keep the unknown state of a qubit for as long as we want. The
idea of the method for quantum error correction is based on designing the
code words in such a way that every possible error (in the first, second, or
third qubit) transforms the subspace of code words onto another subspace
which is orthogonal to it, but without modifying its internal structure.
Then, by performing an incomplete measurement, we can detect in which
subspace our state is, and therefore we know how to correct the error. This
method can be generalized to the case in which other kinds of errors can
occur. For example, imagine that with a small probability we can have errors
consisting of applying the operator $\sigma _{\alpha }$ ($\alpha =x,y,z$) to
a qubit. We want to preserve the state of $k$ qubits against arbitrary
errors in $t$ different qubits. We will denote by $E$ the possible operators
corresponding to the errors that we want to correct. For example, $\sigma
_{x}^{1}\otimes \sigma _{y}^{4}$. We encode the $k$ logical qubits in $n$
qubits. The subspace of code words $H_{L}$ has dimension $2^{k}$, whereas
the Hilbert space $H$ of all the qubits has dimensions $2^{n}$. Each of the
possible error operators (consisting of up to $t$ tensor products of Pauli
operators) transform $H_{L}$ into a subspace of dimension $2^{k}$ (Note that
the $E$'s are unitary and therefore they conserve the dimension of the
subspace on which they are applied). The subspace of code words has to be
such that all these subspaces are mutually orthogonal. This condition
imposes a minimum bound (the quantum Hamming bound) to the number of qubits
needed, since all these orthogonal subspaces have to fit in $H$. One can
easily show that this bound implies that

\begin{equation}
2^{k}\sum_{l=0}^{t}3^{l}\left(
\begin{array}{c}
n \\
l%
\end{array}
\right) <2^{n}.
\end{equation}
For $k=1$ the minimum $n$ is $5$. Methods have been devised to construct
codewords for each of these cases \cite{La96,Go96}. On the other hand, one
can take into account the errors that are produced while errors are being
corrected, as well as the ones produced during operations. There is a whole
theory dealing with the so-called fault tolerant error correction \cite{Sh96}%
, which basically shows that this is always possible provided the error per
gate is smaller than some error threshold, which lies between $%
10^{-4}-10^{-6}$ depending on the error model. This result implies that if
the error per gate is smaller than this threshold, then quantum computation
is possible by using fault tolerant error correction.

The above error correction schemes work in the presence of (undesired)
coupling to the environment which leads to decoherence. In order to show
that, one can expand the operator that describes the evolution of the $i$th
qubit with its local environment as
\begin{equation}
U^{i}=\alpha ^{i}1^{i}\otimes E_{0}^{i}+\epsilon _{1}^{i}\sigma
_{x}^{i}\otimes E_{1}^{i}+\epsilon _{2}^{i}\sigma _{y}^{i}\otimes
E_{2}^{i}+\epsilon _{3}^{i}\sigma _{z}^{i}\otimes E_{3}^{i}
\end{equation}
where the $E$'s are operators acting on the environment, and $\alpha ^{i}$
and $\epsilon _{1,2,3}^{i}$ are constant numbers. Note that we can always
use this expansion given the fact that the Pauli operators (plus the
identity) form a basis in the space of operators acting on a qubit. We will
consider that the time is sufficiently short so that all $\alpha ^{i}\simeq
1 $ and $\epsilon _{1,2,3}^{i}\ll 1$. The state of all the qubits after some
interaction time $U|\psi \rangle |E\rangle =\otimes _{i=1}^{n}U^{i}|\psi
\rangle |E\rangle $ can be expanded in terms of the epsilon keeping only the
lowest orders. The error correction explained above will project the state
onto only one of the terms of the resulting expression. The state of the
environment will therefore factorize, and all the analysis made before
remains valid.

\subsection{Quantum communication}

The situation one has in mind in quantum communication is the following:
Alice wants to send Bob an unknown state $|\Psi \rangle .$ One way of doing
this is to send the particle carrying the state directly. However, the
particle will very likely interact with the environment which may result in
a different state, generally mixed. There are some ways of avoiding this. In
the following we will describe a basic tool in quantum communication which
allows to send one quantum state from one place to another provided one has
a maximally entangled state shared between the two places, and is able to
communicate classically without errors.

\subsubsection{Teleportation}

By teleportation we define to transfer an intact quantum state from one
place to another, by a sender who knows neither the state to be teleported
nor the location of the intended receiver \cite{Bennett93}. The term
teleportation comes from Science Fiction meaning to make a person or object
disappear while an exact replica appears somewhere else. The first
teleportation experiments have recently taken place. Consider two partners,
Alice and Bob, located at different places. Alice has a qubit in an unknown
state $|\phi \rangle $, and she wants to teleport it to Bob, whose location
is not known. Prior to the teleportation process, Alice and Bob share two
qubits in a Bell state $|\Psi ^{-}\rangle =\sqrt{\frac{1}{2}}(|0,1\rangle
-|1,0\rangle )$. The idea is that Alice performs a joint measurement of the
two-level system to be teleported and her particle. Due to the nonlocal
correlations contained in the Bell state, the effect of the measurement is
that the unknown state appears instantaneously in Bob's hands, except for a
unitary operation which depends on the outcome of the measurement. If Alice
communicates to Bob the result of her measurement, then Bob can perform that
operation and therefore recover the unknown state (for experiments, see \cite%
{Boumeester97,Boschi98,Furusawa98}).

Let us call particle 1 that which has the unknown state $|\phi \rangle _{1}$%
, particle 2 the member of the EPR that Alice possesses and particle 3 that
of Bob. We write the state of particle 1 as $|\phi \rangle _{1}=a|0\rangle
_{1}+b|1\rangle _{1}$ where $a$ and $b$ are (unknown) complex coefficients.
The state of particles 2 and 3 is the Bell state $|\Psi ^{\text{-}}\rangle $%
. The complete state of particles 1, 2 and 3 is therefore
\begin{equation*}
|\Psi \rangle _{123}=\frac{a}{\sqrt{2}}(|0,0,1\rangle -|0,1,0\rangle )+\frac{%
b}{\sqrt{2}}(|1,0,1\rangle -|1,1,0\rangle ).
\end{equation*}

In order to teleport the state, Alice and Bob follow this procedure:

(1) \emph{Alice measurement}: Alice makes a joint measurement of her
particles (1 and 2) in the Bell basis
\begin{eqnarray*}
|\Psi ^{\pm }\rangle &=&\frac{1}{\sqrt{2}}(|0,1\rangle \pm |1,0\rangle ) \\
|\Phi ^{\pm }\rangle &=&\frac{1}{\sqrt{2}}(|0,0\rangle \pm |1,1\rangle )
\end{eqnarray*}

(2) \emph{Alice broadcasting}: Then she broadcasts (classically) the outcome
of her measurement. That is, she has to send Bob two bits of classical
information which indicate the outcome of the measurement.

(3) \emph{Bob restoration}: Bob then applies a unitary operation to his
particle to obtain $|\phi \rangle _{3}$. According to the state of the
particles the possible outcomes are:

\begin{description}
\item With probability $1/4$, Alice finds ${|}\Psi ^{\text{-}}\rangle _{12}$%
. The state of the third particle is automatically projected onto $%
a|0\rangle _{3}+b|1\rangle _{3}$. Thus, in this case Bob does not have to
perform any operation.

\item With probability $1/4$, Alice finds $|\Psi ^{\text{+}}\rangle _{12}$.
The state of the third particle is automatically projected onto $-a|0\rangle
_{3}+b|1\rangle _{3}$. Teleportation occurs if Bob applies $\sigma _{z}$ to
his particle.

\item With probability $1/4$, Alice finds $|\Phi ^{\text{-}}\rangle _{12}$.
The state of the third particle is automatically projected onto $a|1\rangle
_{3}+b|0\rangle _{3}$. Teleportation occurs if Bob applies $\sigma _{x}$ to
his particle.

\item With probability $1/4$, Alice finds $|\Phi ^{\text{+}}\rangle _{12}$.
The state of the third particle is automatically projected onto $a|1\rangle
_{3}-b|0\rangle _{3}$. Teleportation occurs if Bob applies $\sigma _{z}$ to
his particle.
\end{description}

Note that Alice ends up with no information about her original state so that
no violation of the no-cloning theorem occurs. In \ this sense, the state of
particle 1 has been transferred to particle 3. On the other hand, there is
no instantaneous propagation of information. Bob has to wait until he
receives the (classical) message from Alice with her outcome. Before he
receives the message, his lack of knowledge prevents him from having the
state. Note that no measurement can tell him whether Alice has performed her
measurement or not. Since teleportation is a linear operation applied to a
state, it will also work for statistical mixtures, or in the case in which
particle 1 is entangled with other particles.

\subsection{Quantum repeaters}

We have now all necessary tools available to introduce the concept of the
quantum repeater. Our goal is to create an EPR pair of high fidelity between
two distant locations. Since nonlocal entanglement between distant particles
cannot be created using only local operations, this involves the usage of a
quantum channel, which is noisy in general. The bottleneck for communication
over large distances is the scaling of the error probability with the length
of the channel. When using, for example, optical fibers and single photons
as a quantum channel, both the absorption losses and the depolarization
errors scale exponentially with the length of the channel. The state of the
photon or the photon itself will therefore be destroyed with almost
certainty if the channel is longer than a few half-lengths of the fiber.

To overcome this problem, one can use quantum repeaters. The idea of such a
repeater is to divide a long quantum channel into shorter segments, which
are purified separately, before they are connected. Connecting two segments
of a channel means here to build up quantum correlations across the compound
channel from correlations that exist across the individual segments. This
can be done by teleportation of entanglement. A quantum repeater must
therefore combine the methods of entanglement purification and
teleportation. Although the combination of these methods should, in
principle, allow to create entanglement over arbitrary distances, it is
another question how much this ''costs'' in terms of resources needed for
purification. Resources means here the number of low fidelity entangled
pairs that have to be provided for purification of each channel segment.
This quantity is related to the number of particles that have to be
manipulated locally (at the connection points between the segments) in a
coherent fashion. If the resources grow too fast with the length of the
channel, not much will be gained by the whole procedure. A further important
quantity is the error tolerance for the local operations. In every real
situation, the local operations applied to one or more particles will bear
some imperfections. Since such operations are the building blocks for any
entanglement purification protocol, their imperfections will limit the
maximum attainable fidelity for an EPR pair and the efficiency of the
protocol. In the context of the quantum repeater, a maximum fidelity $F<1$
corresponds to a residual amount of noise for each segment. When the
segments are connected, this noise accumulates.

To overcome this limitation \cite{Briegel98}, we can divide the long channel
into $N$ smaller segments and create less distant entangled pairs across
each segment. The number of segments $N$ is thereby chosen in such a way
that it is possible to create entangled pairs with sufficiently high initial
fidelity $F>F_{\text{min}}$ over the distance of such a segment, such that
they can be purified, according to our previous discussion. In a next step,
we connect these ``elementary'' pairs by using teleportation. For example if
we have an entangled purified pair between the nodes $A_{1}$and $A_{2}$, and
another one between $A_{2}$ and $A_{3},$we teleport the state of the first
particle in $A_{2}$ to $A_{3}$ by using this second pair. The result of this
teleportation will be that the nodes $A_{1}$and $A_{3}$ will now share an
entangled state. Of course, due to imperfections during the teleportation
procedure, as well as the fact that the pairs used were not perfectly pure,
the new entangled state will not be pure. But as long as its fidelity is
larger than $F_{\text{min}}$, it will be possible to purify it to a value
close to $F_{\text{max}}.$ Thus, the crucial point is that, on the one hand,
the operations that are performed cannot be too noisy since otherwise they
could decrease the fidelity below $F_{\text{min}}$, which would make the
process impossible; on the other hand, the distance between nodes has to be
such that purification be possible. This limits the number of pairs one can
connect before purification becomes impossible. We therefore connect a
smaller number $L\ll N$ of pairs so that the resulting fidelity $F_{L}$
stays above the threshold value for purification ($F_{L}\geq F_{\text{min}}$%
) and purification is possible.

The general strategy will be to design an alternating sequence connection
and (re-)purification procedures in such a way that the number of resources
needed remains as small as possible, and in particular does not grow
exponentially with $N$ and thus with $l$. This is possible, in principle,
using a nested purification protocol \cite{Briegel98}.

\section{Quantum information processing with single atoms and photons}

\subsection{Introduction}

This chapter discusses various schemes of quantum information processing
with single trapped atoms and photons, i.e., manipulation of atoms and
photons on the level of the single quantum level. Experimental realization
includes laser cooled trapped ions, either in a linear trap or in arrays of
micro traps, and neutral atoms stored in far-off-resonance traps or optical
lattices. Single atoms can be stored in high-Q cavities, providing an
interface between atoms and photons. The models discussed below share the
feature that long lived internal atomic states, such as atomic hyperfine
ground states or metastable states, serve as quantum memory to store the
qubits. Furthermore, we assume that single qubit rotations can be performed
by coupling the qubit states to laser light for an appropriate time period.
In general, this requires that single atoms can be addressed by laser light.
The discussion during the last few years has focused on developing various
schemes for two-qubit gates. The models discussed in the literature can be
classified in two categories. The first version relies on the concept of a
quantum data bus: in this case the qubits are coupled to a collective
auxiliary quantum mode, and entanglement of qubits is achieved by swapping
qubits to excitations of the collective mode. Examples for such systems are
the collective phonon modes in ion traps \cite{Cirac95}, and photons in
cavity QED \cite{Pellizzari95,Cirac97}. Requirements often include the
initialization of the quantum data bus in a pure initial state, e.g. laser
cooling to the motional ground state in ion traps. However, recently
specific protocols for ``hot gates'' have been developed which loosen these
requirements \cite{Molmer99,Poyatos98}. The second concept for performing
the two-qubit gate is controllable internal-state dependent two-body
interactions between atoms. Examples for this latter scheme are coherent
cold collisions of atoms in optical traps and optical dipole-dipole
interactions \cite{Jaksch99,Brennen99}. A third example is the ``fast''
two-qubit gate based on large permanent dipole interactions between laser
excited Rydberg atoms in static electric fields \cite{Jaksch00}. We note
that the unitary operations, which can be decomposed in a series of single
and two-qubit operations on the qubits, can either be performed \emph{%
dynamically}, i.e. based on the time evolution generated by a specific
Hamiltonian, or \emph{geometrically} as in holonomic quantum computing \cite%
{Duan01,Zanardi00}. Finally, a common feature of the quantum optical models
is that read out of the atomic qubit is performed using the method of
quantum jumps.

This chapter is arranged as follows: we start with a detailed description of
trapped ions as a physical system to implement quantum computing in Sec. 2.
This is followed by a Section on cavity QED which discusses optical
interconnects between atoms as quantum memory and photons for transmission
of quantum information. Finally, Sec. 3 discusses examples of two-qubit
gates with neutral atoms based on cold coherent collisions and interaction
between laser excited Rydberg atoms in electric fields.

\subsection{Trapped Ions}

In this section we give a theoretical description of quantum state
engineering \cite{Cirac96} and entanglement engineering \cite%
{Cirac95,Steane97} in a system of trapped and laser cooled ions. The
development of the theory starts with the description of Hamiltonians, state
preparation, laser cooling and state measurements first for single ions,
which is then generalized to the case of many ions. This serves as the basis
of our discussion of quantum computer models \cite%
{{Cirac95,Poyatos98,Molmer99,Cirac00,Calarco01,Duan01}}.

\subsubsection{The Model}

\subsubsection{Single Trapped Ion}

\begin{figure}[tbp]
\label{pzfigion1}
\par
\begin{center}
\includegraphics[width=8.0cm]{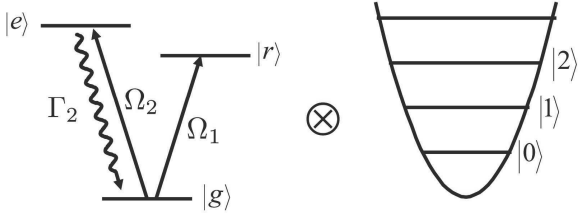}
\end{center}
\caption{Energy levels of an ion trap. Left: internal level structure with $%
|g\rangle \rightarrow |r\rangle$ a metastable transition, and $|g\rangle
\rightarrow |e\rangle$ a strong dissipative transition coupled by Rabi
frequencies $\Omega_1$ and $\Omega_2$, respectively. Right: quantized energy
levels in the harmonic trapping potential}
\end{figure}

\emph{Motional degrees of freedom:} We consider a single ion confined in a
harmonic trap and interacting with laser light \cite{Cirac96}. We assume
that the lasers are directed along one of the principal axes of the harmonic
potential, which allows us to consider ion motion in only one dimension.
Hence, the Hamiltonian describing the free motion of the ion in the trap is
\begin{equation}
H_{0T}=\frac{\hat{p}^{2}}{2M}+\frac{1}{2}M\nu ^{2}\hat{x}^{2}.  \label{Htp1}
\end{equation}
Here $\hat{x}$ and $\hat{p}$ are the position and momentum operators
respectively, $M$ is the ion mass, $\nu $ is the oscillation frequency (Fig.~%
\ref{pzfigion1}. We can rewrite this Hamiltonian in the familiar form $%
H_{0T}=\nu (a^{\dagger }a+1/2)$ with raising and lowering operators $a$ and $%
a^{\dagger }$, defined according to $\hat{x}=\sqrt{1/2M\nu }(a+a^{\dagger })$
and $\hat{p}=i\sqrt{M\nu /2}(a^{\dagger }-a)$ (we set $\hbar =1$).

\emph{Internal degrees of freedom:} We assume that the internal electronic
structure of the ion is modelled by a three-level system, with levels $%
|g\rangle $, $|e\rangle $ and $|r\rangle $, where the first transition $%
|g\rangle \rightarrow |r\rangle $ is a dipole-forbidden, and $|g\rangle
\rightarrow |e\rangle $ is dipole-allowed (Fig.~\ref{pzfigion1}). In our
model system we will employ the transition to the metastable state $%
|r\rangle $ for \emph{quantum state engineering}, while the strongly
dissipative transition coupling $|e\rangle $ to the ground state will be
used for \emph{laser cooling} and \emph{state measurement}. These
transitions can be excited by laser beams of frequencies close to the
corresponding resonance frequencies. Obviously, emission or absorption of
laser photons will modify the atomic motion. We confine our discussion to
the Lamb-Dicke limit (LDL), i.e., to the limit where the ion motion is
restricted to a region much smaller than the wavelength of the laser light
exciting a given transition \cite{Stenholm86}. This allows us to expand the
Hamiltonian describing the interaction of the ion with the laser light in
terms of the Lamb-Dicke parameter $\eta _{i}=2\pi a_{0}/\lambda _{i}$, where
$a_{0}=1/(2M\nu )^{1/2}$ is the size of the ground state of the harmonic
potential, and $\lambda _{i}$ is the wavelength of the laser light exciting
transition $i$. We will now write out in details the Hamiltonians describing
the coupling of the ion to laser light in the LDL.

\emph{Dipole forbidden transition $|g\rangle \rightarrow |r\rangle $
transition:} We first consider the situation in which only the laser driving
the dipole-forbidden transition $|g\rangle \rightarrow |r\rangle $ is on. We
assume for the moment that the interaction time with the laser beam is much
shorter than the lifetime of level $|r\rangle $, so that we can neglect
dissipation. The corresponding Hamiltonian is
\begin{equation}
H_{1}=H_{0T}+H_{0A_{1}}+H_{A_{1}L},  \label{H1}
\end{equation}
where the first two terms are the bare trap and atomic Hamiltonian, and $%
H_{A_{1}L}$ describes the interaction with the laser. In a frame rotating
with the laser frequency, we have $H_{0A_{1}}=\delta _{1}|g\rangle \langle
g| $ and
\begin{eqnarray}
H_{A_{1}L} &=&\frac{1}{2}\Omega _{1}\sin \left[ \eta _{1}(a+a^{\dagger
})+\phi _{1}\right] (|r\rangle \langle g|+|g\rangle \langle r|),  \label{H_-}
\\
H_{A_{1}L}^{\pm } &=&\frac{1}{2}\Omega _{1}\{|r\rangle \langle g|\exp [\pm
i\eta _{1}(a+a^{\dagger })]+\text{h.c.}\},
\end{eqnarray}
for standing-wave and travelling-wave configurations, respectively. Here, $%
\delta _{1}=\omega _{L_{1}}-\omega _{rg}$ is the laser detuning from the
internal transition, $\Omega _{1}$ is the Rabi frequency, and $\eta _{1}$ is
the Lamb-Dicke parameter for this particular transition. The index $+$ ($-$)
denotes that the laser plane wave propagates in the positive (negative) $x$
direction, while $\phi _{1}$ defines the position of the center of the trap
in the laser standing wave.

\begin{figure}[hbp]
\label{pzfigion2}
\par
\begin{center}
\includegraphics[width=8.0cm]{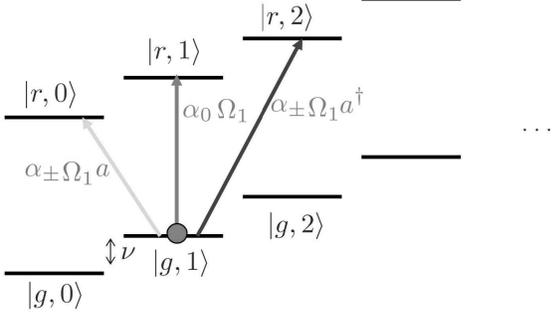}
\end{center}
\caption{Coupling to the atom + trap levels according to the Hamiltonians (%
\protect\ref{h0}), (\protect\ref{kone} and (\protect\ref{kmone},
respectively, in lowest order Lamb-Dicke expansion.}
\end{figure}

In lowest order in the Lamb-Dicke expansion we have
\begin{equation}
H_{A_{1}L}=\frac{1}{2}\Omega _{1}\{|r\rangle \langle g|[\alpha _{0}+\alpha
_{\pm }(a+a^{\dagger })+\mathcal{O}(\eta _{1}^{2})]+\text{h.c.}\}.
\label{LDL1}
\end{equation}
where $\alpha _{0}=1$, $\alpha _{\pm }=\pm i\eta _{1}$ and $\alpha _{0}=\sin
(\phi _{1})$, $\alpha _{\pm }=\eta _{1}\cos (\phi _{1})$ for a travelling
and standing-wave configuration, respectively. The Hamiltonian (\ref{LDL1})
can be further simplified if the laser field is sufficiently weak so that
only pairs of bare atom + trap levels are coupled resonantly. We denote by $%
|n,g\rangle $ and $|n,r\rangle $ the eigenstates of the bare Hamiltonian $%
H_{0T}+H_{0A_{1}}$, where the internal two-level system is in the ground
(excited) state and $n$ is the excitation number of the harmonic oscillator.
These states are degenerate for $\omega _{L_{1}}-\omega _{rg}=k\nu $ ($%
k=0,\pm 1,\ldots $), i.e., whenever the laser is tuned to one of the
``motional sidebands'', corresponding to a degeneracy between $|n,g\rangle $
and $|n+k,r\rangle $. In the presence of the laser these degeneracies become
avoided crossings, and for sufficiently weak laser excitation these avoided
crossings will be isolated (non-overlapping). For example, for $|\omega
_{L_{1}}-\omega _{rg}|\ll \nu $, i.e. $k=0$, transitions changing the
harmonic oscillator quantum number $n$ are off-resonance and can be
neglected. In this case the Hamiltonian (\ref{H1}) can be approximated by
\begin{equation}
H_{0}=\nu a^{\dagger }a-\frac{1}{2}\delta _{1}\sigma _{z}+\frac{1}{2}\Omega
_{1}(\alpha _{0}\sigma _{+}+\text{h.c.}),  \label{h0}
\end{equation}
where we have used the spin-$\frac{1}{2}$ notation $\sigma _{+}=(\sigma
_{-})^{\dagger }=|r\rangle \langle g|$, $\sigma _{z}=|r\rangle \langle
r|-|g\rangle \langle g|$. For laser frequencies close to the lower motional
sideband resonance $|\omega _{L_{1}}-(\omega _{rg}-\nu )|\ll \nu $ ($k=-1$),
only transitions decreasing the quantum number $n$ by one are important, and
$H_{1}$ can be approximated by a Hamiltonian of the Jaynes-Cummings type:
\begin{equation}
H_{\text{JC}}{}_{\pm }=\nu a^{\dag }a-\frac{1}{2}\delta _{1}\sigma _{z}+%
\frac{1}{2}\Omega _{1}(\alpha _{\pm }\sigma _{+}a+\text{\textrm{h.c.}}).
\label{kone}
\end{equation}
Similarly, for $|\omega _{L_{1}}-(\omega _{rg}+\nu )|\ll \nu $, only
transitions increasing the quantum number $n$ by one ($k=+1$) contribute, so
that $H_{1}$ can be approximated by the \emph{anti}-Jaynes-Cummings
Hamiltonian
\begin{equation}
H_{\text{AJC}}{}_{\pm }=\nu a^{\dagger }a-\frac{1}{2}\delta _{1}\sigma _{z}+%
\frac{1}{2}\Omega _{1}(\alpha _{\pm }\sigma _{+}a^{\dagger }+\text{h.c.}).
\label{kmone}
\end{equation}
(see Fig.~\ref{pzfigion2}) For the above approximations to be valid we
require that the effective Rabi frequencies to the non-resonant states have
to be much smaller than the trap frequency $(\alpha _{i}\Omega _{1}/\nu
)^{2}\ll 1$ ($i=0,\pm $). Note in particular that for an ion at the node of
a standing light wave corrections to the JC Hamiltonian (\ref{kone}) are of
the order $(\eta _{1}\Omega _{1}/\nu )^{2}\ll 1$, i.e. the conditions of
validity are greatly relaxed.

Eigenstates of the Hamiltonians $H_{0}$, $H_{\text{JC}}{}_{\pm }$ and $H_{%
\text{AJC}}{}_{\pm }$ are the dressed states familiar from cavity QED, which
are obtained by diagonalizing the $2x2$ matrices of nearly degenerate
states. Applying a laser pulse on resonance, $\omega _{L_{1}}=\omega _{rg}$,
will according to (\ref{h0}) induce Rabi flopping between the states $%
|n,g\rangle $ and $|n,r\rangle $, while a laser tuned for example to the
lower motional sideband $\omega _{L_{1}}=\omega _{rg}-\nu $ will lead to
Rabi oscillations coupling $|n,g\rangle $ and $|n-1,r\rangle $. The above
Hamiltonians are the basic building blocks to engineer quantum states. As an
example, Refs.~\cite{Law96,Gardiner97} give a protocol to build the general
motional superposition states $\sum_{n=0}^{M}c_{n}|g,n\rangle $ starting
from the (pure) ground state $|g,0\rangle $. The unitary operation effecting
this transformation can be decomposed into unitary operations generated by
the Hamiltonians (\ref{kone}) and (\ref{h0}), i.e. by applying a sequence of
laser pulses with proper detunings and duration.

\emph{The dipole-allowed transition $|g\rangle \rightarrow |e\rangle $
transition:} For a laser beam exciting the dipole-allowed transition $%
|g\rangle \rightarrow |e\rangle $ spontaneous emission is expected to play a
significant role, and thus the dynamics must be described in terms of a
master equation \cite{Gardiner99,Stenholm86,Cirac92},
\begin{equation}
\dot{\rho}=-i[H_{2},\rho ]+\mathcal{L}_{2}\rho .  \label{ME0}
\end{equation}
In a rotating frame the Hamiltonian is $H_{2}=\nu a^{\dag }a+\delta
_{2}|g\rangle \langle g|+H_{A_{2}L}$ with $\delta _{2}=\omega
_{L_{2}}-\omega _{eg}$ laser detuning from the internal transition. The
laser atom coupling $H_{A_{2}L}$ has a structure analogous to (\ref{H1})
with the replacements $|r\rangle \rightarrow |e\rangle $, a Rabi coupling $%
\Omega _{2}$, position of the center of the trap in the laser standing wave $%
\phi _{2}$ , and Lamb-Dicke parameter $\eta _{2}$. The dissipative part of
the master equation (\ref{ME0}) can be written as \cite{Cirac92}
\begin{equation}
\mathcal{L}_{2}\rho =\Gamma _{2}|g\rangle \langle g|\langle e|\tilde{\rho}%
|e\rangle -\frac{1}{2}\Gamma _{2}(|e\rangle \langle e|\rho +\rho |e\rangle
\langle e|),  \label{Liouvil}
\end{equation}
where $\Gamma _{2}$ is the spontaneous emission rate from level $|e\rangle $%
, and
\begin{equation}
\tilde{\rho}=\int_{-1}^{1}duN(u)e^{-i\eta _{2}u(a+a^{\dagger })}\rho
e^{i\eta _{2}u(a+a^{\dagger })},  \label{recycling}
\end{equation}
with $N(u)$ the dipole emission pattern for spontaneous emission from $%
|e\rangle $ to $|g\rangle $. For example, for a $\Delta m_{J}=\pm 1$
transition, $N(u)=3/8(1+u^{2})$. Master equations of this type have been
derived and studied in the context of laser cooling \cite%
{Gardiner99,Stenholm86,Cirac92}. Physically speaking, Eqs. (\ref{ME0},\ref%
{Liouvil}) describe the excitation of the atomic electron by the laser,
which can either return to the ground state by either a laser induced
process or by spontaneous emission. The emission of the spontaneous photon
according to the angular distribution $N(u)$ is accompanied by a momentum
transfer to the atom, as described by the recycling term (\ref{recycling}).

\emph{Spontaneous emission and laser cooling to the motional ground state:}
A prerequisite for many of the schemes for quantum engineering of
nonclassical states of motion, or entangled atomic states is that the
initial motional state of the ion is prepared in a well defined pure state,
e.g., the ground state $|0\rangle $ \cite{Cirac96,Cirac95}. The standard
approach to preparing such a pure state of the atomic motion is sideband
cooling \cite{thebible}.

The theoretical description is particularly simple in the Lamb-Dicke limit,
where the separation of the time scale for the internal and external
dynamics allows the adiabatic elimination of the internal degree of freedom
\cite{Stenholm86}. For details of the calculations we refer to \cite{Cirac92}
and references cited therein. The physical picture of laser cooling in the
Lamb-Dicke limit is as follows: in the rest frame of the ion, the ion
``sees'' a laser field consisting of a carrier at frequency $\omega _{L_{2}}$
and small (motional) sidebands at frequencies $\omega _{L_{2}}\pm \nu $.
Cooling occurs when absorption of laser photons from the upper laser
sideband (at $\omega _{L_{2}}+\nu $) is stronger than from the lower
sideband, since the former absorption reduces the external energy, whereas
the latter increases the external energy. Hence, laser cooling is
particularly efficient in the strong confinement limit when one is able to
tune the laser frequency in such a way that the upper laser sideband is on
resonance with the two-level transition, since in this case only the photons
of the upper sideband are absorbed, and consequently the ion ends up in the
ground state of the harmonic trapping potential. This is the basic mechanism
of \textit{sideband cooling}. For many of the ions currently in use in Paul
traps, the strong confinement condition $\nu \gg \Gamma _{2}$ is not
fulfilled for dipole-allowed transitions, and hence sideband cooling is not
possible. However, employing auxiliary internal atomic levels and additional
laser excitation allows one to effectively ``design'' two-level atoms for
sideband cooling \cite{thebible,Marzoli94}.

\emph{State measurement by the quantum jump technique:} Implementation of
quantum computing and communication protocols require measurement of the
internal state of the atom \cite{Cirac95}. In an ion trap this can be
achieved with essentially $100\%$ efficiency using the method of quantum
jumps \cite{thebible,Gardiner99}. The theoretical understanding of quantum
jumps is based on the continuous measurement theory, and we refer to \cite%
{Gardiner99} for a detailed mathematical description of the underlying
theory. For our purpose it suffices to summarize the results as follows.
Consider a \emph{single} ion prepared initially in a superposition state on
the metastable transition, $\alpha |g\rangle +\beta |r\rangle $. Switching
on the laser on the strongly dissipative transition will give with
probability $|\alpha |^{2}$ a burst of photon emissions $|e\rangle
\rightarrow |g\rangle $ on the time scale $1/\Gamma _{2}$, or with
probability $|\beta |^{2}$ the appearance on an emission window on the
strong line. Measuring an emission window, or no window thus corresponds to
a projective measurement of $|r\rangle $ or $|g\rangle $.

\subsubsection{Ions in a linear trap}

The above model is readily extended to describe a string of $N$ ions in a
linear trap \cite{thebible} which is the basis of the ion trap '95 quantum
computer proposal \cite{Cirac95,Steane97} described below. A linear trap
corresponds to a confinement of the motion along $x,y$ and $z$ directions in
an (anisotropic) harmonic potential of frequencies $\nu \equiv \nu _{x}\ll
\nu _{y},\nu _{z}$. The equilibrium position of the ions will be given by
the confining forces of the trapping potential balancing the Coulomb
repulsion between the ions. If the ions have been previously laser cooled in
all three dimensions they undergo small oscillations around these
equilibrium position. In this case, the motion of the ions is described in
terms of normal modes.

As an example, a 1D model for two ions in a linear trap with internal levels
$|g\rangle $ and $|r\rangle $ and coupled to laser light is given by the
following Hamiltonian
\begin{eqnarray}
H &=&\nu a_{\mathrm{cm}}^{\dagger }a_{\mathrm{cm}}+\sqrt{3}\nu a_{\mathrm{r}%
}^{\dagger }a_{\mathrm{r}}-\delta _{1}|r\rangle _{1}{}_{1}\!\langle
r|-\delta _{2}|r\rangle _{2}{}_{2}\!\langle r| \\
&&+\frac{1}{2}\Omega _{1}(t)[|r\rangle _{1}{}_{1}\!\langle g|e^{-i\eta _{%
\mathrm{cm}}(a_{\mathrm{cm}}+a_{\mathrm{cm}}^{\dagger })}e^{-i\eta _{\mathrm{%
r}}(a_{\mathrm{r}}+a_{\mathrm{r}}^{\dagger })}+\text{h\textrm{.c}}\mathrm{.}]
\\
&&+\frac{1}{2}\Omega _{1}(t)[|r\rangle _{2}{}_{2}\!\langle g|e^{-i\eta _{%
\mathrm{cm}}(a_{\mathrm{cm}}+a_{\mathrm{cm}}^{\dagger })}e^{+i\eta _{\mathrm{%
r}}(a_{\mathrm{r}}+a_{\mathrm{r}}^{\dagger })}+\text{h\textrm{.c.}}].
\end{eqnarray}
Here, the first line are the Hamiltonians for the harmonic collective
oscillations of center-of-mass and stretch mode with oscillation frequency $%
\nu $ and $\sqrt{3}\nu $, respectively, and the bare atomic Hamiltonian for
the first and second ion in the rotating frame with $\delta _{1,2}$ the
laser detunings. The second and third line are the laser couplings to the
ions with $\Omega _{1,2}$ Rabi frequencies of the laser acting on each ion,
respectively, and $\eta _{\text{cm,r}}$ are the corresponding Lamb-Dicke
parameters. In the general case of $N$ ions there will be a set of $N$
collective modes of ion motion, where again the minimum frequency of
collective mode oscillations is that of the center-of-mass (CM) mode in the $%
x$-direction $\nu _{x}\equiv \nu $, and the next frequency is the stretch
mode $\sqrt{3}\nu _{x}$, and all the others are larger. It is an important
feature for the quantum computer proposal below that the frequency spacing
of the low lying modes is essentially \emph{independent} of the number of
ions $N$ in the trap.

Our previous discussion of the Lamb-Dicke expansion of the Hamiltonian is
readily extended to the string of $N$ ions. In a similar way, we can model
spontaneous emission of the ions by a master equation. In a typical
experimental situation, the distance between the ions will be much larger
than the optical wave length, so that the spontaneous emission of the ions
is independent, and the master equation will contain a sum of the
independent spontaneous emission terms of the ions of the form (\ref{Liouvil}%
).

\subsubsection{Ion trap quantum computer '95}

As first proposed in Ref.~\cite{Cirac95}, $N$ cold ions interacting with
laser light and moving in a linear trap provide a realistic physical system
to implement a quantum computer. The distinctive features of this system
are: (i) it allows the implementation of a complete set of quantum gates
between any set of (not necessarily neighboring) ions; (ii) decoherence is
comparatively small, and (iii) the final readout can be performed with
essentially unit efficiency \cite%
{Steane97,Monroe95,King98,Roos00,Nagerl99,Turchette98,Sackett00,Kielpinski01,Rowe01}%
.

\begin{figure}[hbp]
\label{pzfigion3}
\par
\begin{center}
\includegraphics[width=8.0cm]{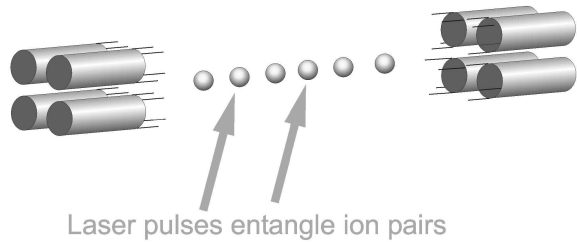}
\end{center}
\caption{Ion trap quantum computer (schematic).}
\end{figure}

Fig.~\ref{pzfigion3} illustrates the basic setup. The qubits are represented
by the long-lived internal states of the ions, with $|g\rangle _{j}\equiv
|0\rangle _{j}$ representing the ground state, and $|r_{0}\rangle _{j}\equiv
|1\rangle _{j}$ a metastable excited state ($j=1,,N$). (In addition, we
assume that there is a second metastable excited state $|r_{1}\rangle $
which serves below the role of an auxiliary state.) In this system
independent manipulation of each individual qubit is accomplished by
addressing the ions with individual laser beams and inducing a Rabi
rotation. The heart of the proposal is the implementation of a two-qubit
gate between two (or more) arbitrary ions in the trap by exciting the
collective quantized motion of the ions with lasers, i.e. the collective
phonon mode plays the role of a quantum data bus. For this we assume that
the collective phonon modes have been cooled to the ground state \cite%
{King98,Roos00}.

Single qubit rotations can be performed tuning a laser on resonance with the
internal transition ($\delta _{j}=0$) with polarization $q=0$. In an
interaction picture the corresponding Hamiltonian is
\begin{equation}
\hat{H}_{j}=(\Omega /2)\left[ |r_{0}\rangle _{j}\langle g|e^{-i\phi
}+|g\rangle _{j}\langle r_{0}|e^{i\phi }\right] .  \label{Ha}
\end{equation}
For an interaction time $t=k\pi /\Omega $ (i.e., using a $k\pi $ pulse),
this process is described by the following unitary evolution operator
\begin{equation}
\hat{V}_{j}^{k}(\phi )=\exp \left[ -ik\frac{\pi }{2}(|e_{0}\rangle
_{j}\langle g|e^{-i\phi }+h.c.)\right] \;,  \label{Vn}
\end{equation}
so that we achieve a Rabi rotation
\begin{eqnarray}
|g\rangle _{j} &\longrightarrow &\cos (k\pi /2)|g\rangle _{j}-ie^{i\phi
}\sin (k\pi /2)|r_{0}\rangle _{j},  \notag \\
|r_{0}\rangle _{j} &\longrightarrow &\cos (k\pi /2)|r_{0}\rangle
_{j}-ie^{-i\phi }\sin (k\pi /2)|g\rangle _{j}.  \notag
\end{eqnarray}

If the laser addressing the $j$-th ion is tuned to the lower motional
sideband of, for example, the center-of-mass mode, we have in the
interaction picture the Hamiltonian
\begin{equation}
{H}_{j,q}=\frac{\eta }{\sqrt{N}}\frac{\Omega }{2}\left[ |r_{q}\rangle
_{j}\langle g|ae^{-i\phi }+|g\rangle _{j}\langle r_{q}|a^{\dagger }e^{i\phi }%
\right] .  \label{Hb}
\end{equation}
Here $a^{\dagger }$ and $a$ are the creation and annihilation operator of CM
phonons, respectively, $\Omega $ is the Rabi frequency, $\phi $ the laser
phase, and $\eta $ is the Lamb-Dicke parameter. The subscript $q=0,1$ refers
to the transition excited by the laser, which depends on the laser
polarization. Equation (\ref{Hb}) follows from the Hamiltonian for the case
of a linear trap, similar to the derivation of (\ref{kone}). The factor $%
\sqrt{N}$ appears since the effective mass of the CM motion is $NM$, and the
amplitude of the mode scales like $1/\sqrt{NM}$ (M{\"{o}}ssbauer effect).

If this laser beam is on for a certain time $t=k\pi /(\Omega \eta /\sqrt{N})$
(i.e., using a $k\pi $ pulse), the evolution of the system will be described
by the unitary operator:
\begin{equation}
\hat{U}_{j}^{k,q}(\phi )=\exp \left[ -ik\frac{\pi }{2}(|r_{q}\rangle
_{j}\langle g|ae^{-i\phi }+h.c.)\right] .  \label{Un}
\end{equation}
It is easy to prove that this transformation keeps the state $|g\rangle
_{j}|0\rangle $ unaltered, whereas
\begin{eqnarray}
|g\rangle _{j}|1\rangle &\longrightarrow &\cos (k\pi /2)|g\rangle
_{j}|1\rangle -ie^{i\phi }\sin (k\pi /2)|r_{q}\rangle _{j}|0\rangle ,  \notag
\\
|r\rangle _{j}|0\rangle &\longrightarrow &\cos (k\pi /2)|r_{q}\rangle
_{j}|0\rangle -ie^{-i\phi }\sin (k\pi /2)|g\rangle _{j}|1\rangle ,  \notag
\end{eqnarray}
where $|0\rangle $ ($|1\rangle $) denotes a state of the CM mode with no
(one) phonon.

\begin{figure}[hbp]
\label{pzfigion4}
\par
\begin{center}
\includegraphics[width=8.0cm]{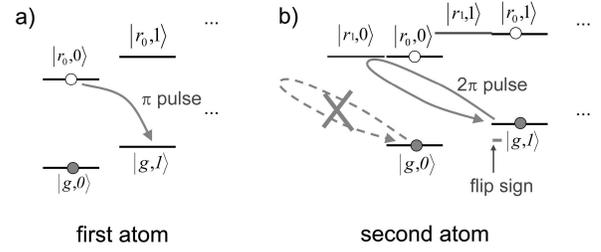}
\end{center}
\caption{The two-qubit quantum gate. a) First step according to (\protect\ref%
{bigone}): the qubit of the first atom is swapped to the photonic data bus
with a $\protect\pi$-pulse on the lower motional sideband, b) Second step:
the state $|g,1\rangle$ acquires a minus sign due to a $2\protect\pi$%
-rotation via the auxiliary atomic level $|r_1\rangle$ on the lower motional
sideband.}
\end{figure}

Let us now show how a two-bit gate can be performed using this interaction.
We consider the following three--step process (see Fig.~\ref{pzfigion4}): (%
\textbf{i}) A $\pi $ laser pulse with polarization $q=0$ and $\phi =0$
excites the $m$th ion. The evolution corresponding to this step is given by $%
\hat{U}_{m}^{1,0}\equiv \hat{U}_{m}^{1,0}(0)$ (Fig.~\ref{pzfigion4}a). (%
\textbf{ii}) The laser directed on the $n$--th ion is then turned on for a
time of a $2\pi $-pulse with polarization $q=1$ and $\phi =0$. The
corresponding evolution operator $\hat{U}_{n}^{2,1}$ changes the sign of the
state $|g\rangle _{n}|1\rangle $ (without affecting the others) via a
rotation through the auxiliary state $|e_{1}\rangle _{n}|0\rangle $ (Fig.~%
\ref{pzfigion4}b). (\textbf{iii}) Same as (i). Thus, the unitary operation
for the whole process is $\hat{U}_{m,n}\equiv \hat{U}_{m}^{1,0}\hat{U}%
_{n}^{2,1}\hat{U}_{m}^{1,0}$ which is represented diagrammatically as
follows:
\begin{equation}
\begin{array}[b]{rrrrrrr}
& \hat{U}_{m}^{1,0} &  & \hat{U}_{n}^{2,1} &  & \hat{U}_{m}^{1,0} &  \\
|g\rangle _{m}|g\rangle _{n}|0\rangle & \longrightarrow & |g\rangle
_{m}|g\rangle _{n}|0\rangle & \longrightarrow & |g\rangle _{m}|g\rangle
_{n}|0\rangle & \longrightarrow & |g\rangle _{m}|g\rangle _{n}|0\rangle , \\
|g\rangle _{m}|r_{0}\rangle _{n}|0\rangle & \longrightarrow & |g\rangle
_{m}|r_{0}\rangle _{n}|0\rangle & \longrightarrow & |g\rangle
_{m}|r_{0}\rangle _{n}|0\rangle & \longrightarrow & |g\rangle
_{m}|r_{0}\rangle _{n}|0\rangle , \\
|r_{0}\rangle _{m}|g\rangle _{n}|0\rangle & \longrightarrow & -i|g\rangle
_{m}|g\rangle _{n}|1\rangle & \longrightarrow & i|g\rangle _{m}|g\rangle
_{n}|1\rangle & \longrightarrow & |r_{0}\rangle _{m}|g\rangle _{n}|0\rangle ,
\\
|r_{0}\rangle _{m}|r_{0}\rangle _{n}|0\rangle & \longrightarrow &
-i|g\rangle _{m}|r_{0}\rangle _{n}|1\rangle & \longrightarrow & -i|g\rangle
_{m}|r_{0}\rangle _{n}|1\rangle & \longrightarrow & -|r_{0}\rangle
_{m}|r_{0}\rangle _{n}|0\rangle .%
\end{array}
\label{bigone}
\end{equation}
The effect of this interaction is to change the sign of the state only when
both ions are initially excited. Note that the state of the CM mode is
restored to the vacuum state $|0\rangle $ after the process. Equation (\ref%
{bigone}) is phase gate $|\epsilon _{1}\rangle |\epsilon _{2}\rangle
\rightarrow (-1)^{\epsilon _{1}\epsilon _{2}}|\epsilon _{1}\rangle |\epsilon
_{2}\rangle $ ($\epsilon _{1,2}=0,1$) which together with single qubit
rotations becomes equivalent to a controlled-NOT.

Final readout of the quantum register (state measurement of the individual
qubits) at the end of the computation can be accomplished using the quantum
jumps technique with unit efficiency \cite{thebible,Gardiner99}.

The above proposal for the implementation of quantum computing with trapped
ions has been the basis and has provided the stimulus for a significant body
of experimental and theoretical work over the last few years \cite%
{Steane97,Monroe95,King98,Roos00,Nagerl99,Turchette98,Sackett00,Kielpinski01,Rowe01}%
. Highlights of the series of experimental work in particular by Chris
Monroe and Dave Wineland at NIST Boulder \cite%
{Monroe95,King98,Turchette98,Sackett00,Kielpinski01,Rowe01}, and Rainer
Blatt at the University of Innsbruck \cite{Roos00,Nagerl99}, are the
implementation of a 2 qubit quantum gate with a single ion \cite{Monroe95},
ground state cooling of a string of ions \cite{King98,Roos00}, addressing of
single ions to perform single qubit operations \cite{Nagerl99}, the
generation of Bell states \cite{Rowe01}, and - as the most remarkable
achievement - preparation of a maximally entangled state of four ions by the
NIST Boulder group \cite{Sackett00}. For a detailed discussion of the
experimental situation we refer to the lecture notes by D. Wineland and R.
Blatt in this volume. On theory side various extensions to finite
temperature gates have been given, as well as schemes which might allow
faster gate operation times \cite{Poyatos98,Molmer99,Knight00}. One of the
most interesting contributions is a scheme by M{\o }lmer and S{\o }rensen
\cite{Molmer99} which allows preparation of a maximally entangled state of $%
N $ ions with a single pulse without addressing the individual ions, as
employed in the four-ion NIST experiment \cite{Sackett00}. A quantum
computer model based on geometric variants of the gate \cite%
{Zanardi99,Zanardi00,Pachos01} was recently proposed by Duan \emph{et al.}
\cite{Duan01}. Error correction of the phonon data bus was studied in \cite%
{Cirac96b}.

\subsubsection{Ion trap quantum computer 2000}

\begin{figure}[hbp]
\label{pzfigion5}
\par
\begin{center}
\includegraphics[width=8.0cm]{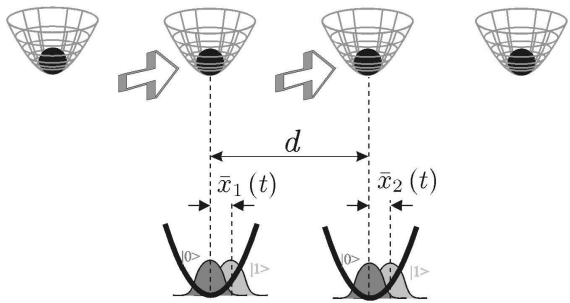}
\end{center}
\caption{Ions stored in an array of microtraps. By addressing two adjacent
ions with an external field the ion wave packet is displaced conditional to
its internal state.}
\end{figure}

While in the ion trap '95 scheme a two-qubit gate was realized using the
collective phonon mode as an auxiliary quantum degree of freedom, we
describe now briefly a version on an ion trap computer where entanglement is
achieved by designing an \emph{internal-state dependent two-body interaction}
between the ions \cite{Cirac00,Calarco01}. This proposal has the advantage
being conceptually simpler (e.g. there is no zero temperature requirement),
and obviously scalable. The model assumes that ions are stored in an \emph{%
array of microtraps} (Fig.~\ref{pzfigion5}). Similar to the ion trap '95
proposal, it is assumed that long lived internal states of the ions serve as
carriers of the qubits, and that single qubit operations can be performed by
addressing ions with a laser.

\begin{figure}[hbp]
\label{pzfigion6}
\par
\begin{center}
\includegraphics[width=8.0cm]{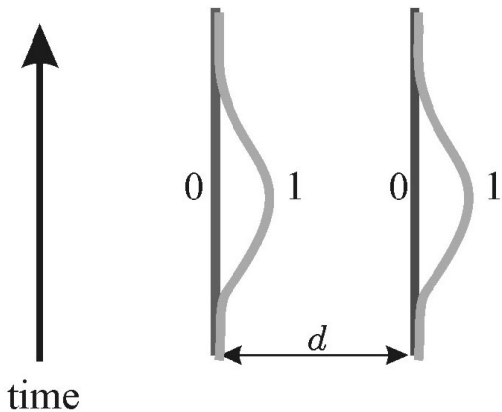}
\end{center}
\caption{Trajectories of the qubits as a function of time. Depending on the
internal state different phases are accumulated.}
\end{figure}

The model assumes a set of $N$ ions confined in independent harmonic
potential wells separated by some constant distance $d$, where $d$ is large
enough so that: (i) the Coulomb repulsion is not able to excite the
vibrational state of the ions; (ii) the ions can be individually addressed.
Two--qubit gates between two neighboring ions can be performed by slightly
displacing them for a short time $T$ if they are in a particular internal
state, say $|1\rangle $. In that case, and provided the ions come back to
their original motional state after being pushed, the Coulomb interaction
will provide the internal wave function (quantum register) with different
phases depending on the internal states of the ions. Choosing the time
appropriately, the complete process will give rise to the two-qubit gate $%
|\epsilon _{1}\rangle |\epsilon _{2}\rangle \rightarrow e^{i\epsilon
_{1}\epsilon _{2}\phi }|\epsilon _{1}\rangle |\epsilon _{2}\rangle $. In
order to analyze this in a more quantitative way, we consider two ions $1$
and $2$ of mass $m$ confined by two harmonic traps of frequency $\omega $ in
one dimension (Fig.~\ref{pzfigion5}). We denote by $\hat{x}_{1,2}$ the
position operators of the two ions. The potential see by the ions is
\begin{equation}
V=\sum_{i=1,2}\frac{1}{2}m\omega ^{2}\left( \hat{x}_{i}-\bar{x}%
_{i}(t)|1\rangle _{i}\langle 1|\right) ^{2}+\frac{e^{2}}{4\pi \epsilon _{0}}%
\frac{1}{|d+\hat{x}_{2}-\hat{x}_{1}|}  \label{Hamiltonian}
\end{equation}
where $\bar{x}_{i}(t)$ is the state dependent displacement induced by the
force. We are interested in the limit, where the displacements are much
smaller than the distance between the traps, $|\hat{x}_{1,2}|\ll d$, and
where the Coulomb energy is small compared with the trapping potentials, $%
\epsilon |\hat{x}_{1}\hat{x}_{2}|/a_{0}^{2}\ll 1$. Furthermore, we assume
that the motional state of the pushed ions will change adiabatically with
the potential. Expansion of the Coulomb term in powers of $\hat{x}_{1,2}/d$
gives rise to a term $-m\omega ^{2}\epsilon \hat{x}_{1}\hat{x}_{2}$ in the
potential (\ref{Hamiltonian}). It is this term which is responsible for
entangling the atoms, giving rise to a conditional phase shift, which can be
simply interpreted as arising from the energy shifts due to the Coulomb
interactions of atoms accumulated on different trajectories according to
their internal states (Fig.~\ref{pzfigion6}),
\begin{equation*}
\phi =-\frac{e^{2}}{4\pi \epsilon _{0}}\int_{0}^{T}dt\left[ \frac{1}{d+\bar{x%
}_{2}-\bar{x}_{1}}-\frac{1}{d+\bar{x}_{2}}-\frac{1}{d-\bar{x}_{1}}+\frac{1}{d%
}\right] ,  \label{phi}
\end{equation*}
where the four terms are due to atoms in $|1\rangle _{1}|1\rangle _{2}$, $%
|1\rangle _{1}|0\rangle _{2}$, $|0\rangle _{1}|1\rangle _{2}$ and $|0\rangle
_{1}|0\rangle _{2}$, respectively. The expression (\ref{phi}) depends only
on mean displacement of the atomic wavepacket and thus is insensitive to the
temperature (the width of the wave packet) which will appear only in the
problem in higher orders in $x_{1,2}/d$ of our expansion of the potential (%
\ref{Hamiltonian}), or in cases of non--adiabaticity. A detailed theory of
this proposal including an analysis of imperfections has been given by
Calarco \emph{et al.} \cite{Calarco01}.

\subsection{Cavity QED}

Cavity QED (CQED) realizes a situation where one or a few atoms interact
strongly with a single quantized high-Q cavity mode, where the light field
can be either in the optical or in the microwave domain (see the lecture
notes by G. Rempe, S. Haroche and H. Walther). The coupling of atoms via the
cavity mode can be used to engineer entanglement between the atoms \cite%
{Pellizzari95,Cirac97,Pellizzari97,Turchette95,Maitre97,Rauschenbeutel99,Rauschenbeutel00,Osnaghi01,Walther01}%
) . The underlying physics is described by the Jaynes-Cummings Hamiltonian
\cite{Gardiner99} which for a single two-level atom has the form
\begin{equation}  \label{eq:H_JCM}
H=\frac{1}{2}\hbar \omega _{\mathrm{0}}\sigma _{z}+\hbar \omega _{\mathrm{c}%
}a^{\dagger }a+\hbar \frac{1}{2}\left[ \Omega _{eg}(\mathbf{r})a^{\dagger
}\sigma _{-}+\Omega _{eg}(\mathbf{r})^{\ast }\sigma _{+}a\right] ,
\end{equation}
where the $\sigma$'s are the Pauli spin operators describing the two-level
atom, and $a^{\dagger }$ and $a$ are the boson creation and annihilation
operators of the field mode, respectively. The term proportional to $\Omega
_{0}(\mathbf{r})$ in (\ref{eq:H_JCM}) describes the coherent oscillatory
exchange of energy between the atom and the field mode. In optical cavity
QED dissipation in this system due to spontaneous emission of the atom in
the excited state, and cavity decay, which can be modeled by a master
equation.

From a formal point of view there is a close analogy between the ion trap
models discussed in the previous section and the Jaynes-Cummings type models
of CQED, where the role of the collective phonon modes of trapped ions is
now taken by photons in the high-Q cavity mode. Thus, in many cases schemes
for quantum state and entanglement engineering proposed in the ion traps are
readily translated to their CQED counterparts. A first example of a CQED
model of a quantum computer is the scheme by Pellizzari \emph{et al.} \cite%
{Pellizzari95}. In this scheme atoms representing the qubits are stored in a
high-Q cavity, and the photon mode in the cavity plays the role of an
auxiliary quantum degree of freedom. This allows entanglement of pairs of
atoms, where the specific feature of Ref.~\cite{Pellizzari95} is that photon
exchange between atoms is performed using adiabatic passage along (the
decoherence free subspace of) dark states, so that spontaneous emission as a
decoherence channel is completely eliminated. On the other hand, CQED
provides a natural interface between atoms as carriers of qubits and
photons, which - once they leave the cavity - can be guided by optical
fibers to transport qubits to distant locations. CQED thus serves as the
paradigm for an \emph{optical interconnect} between atoms as quantum memory
and photons as carriers of qubits for quantum communication \cite{Cirac97,
Pellizzari97}.

\subsubsection{Optical interconnects}

\begin{figure}[hbp]
\label{pzfigcqed1}
\par
\begin{center}
\includegraphics[width=8.0cm]{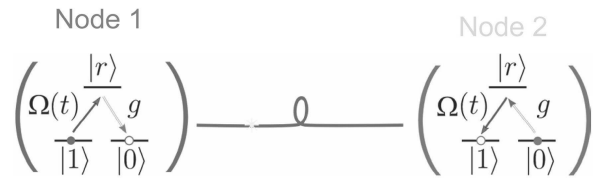}
\end{center}
\caption{Transmission of a qubit from an atom at the first node to an atom
at the second node according to (\protect\ref{transfer}) and (\protect\ref%
{transfer1}).}
\end{figure}

The goal of quantum communications is to transmit an unknown quantum state
from a first to a second node in a quantum network We consider a situation
where the state of a qubit is stored in internal state of atoms. The task is
to transmit the qubit according to
\begin{equation}  \label{transfer}
(\alpha |0\rangle _{1}+\beta |1\rangle _{1})\otimes |0\rangle
_{2}\rightarrow |0\rangle _{1}\otimes (\alpha |0\rangle _{2}+\beta |1\rangle
_{2})
\end{equation}
from the first to the second atom. Below we study a model of an optical
interconnect based on storing atoms in high--Q optical cavities (see Fig.~%
\ref{pzfigcqed1}). By applying laser beams, one first transfers the internal
state of an atom (qubit) at the first node to the optical state of the
cavity mode. The generated photons leak out of the cavity, propagate as a
wavepacket along the transmission line, and enter an optical cavity at the
second node. Finally, the optical state of the second cavity is transferred
to the internal state of an atom. Multiple-qubit transmissions can be
achieved by sequentially addressing pairs of atoms (one at each node), as
entanglements between arbitrarily located atoms are preserved by the
state-mapping process.

The distinguishing feature of the protocol described below \cite{Cirac97} is
that by controlling the atom-cavity interaction, one can absolutely avoid
the reflection of the wavepackets from the second cavity, effectively
switching off the dominant loss channel that would be responsible for
decoherence in the communication process. For a physical picture of how this
can be accomplished, let us consider that a photon leaks out of an optical
cavity and propagates away as a wavepacket. Imagine that we were able to
``time reverse'' this wavepacket and send it back into the cavity; then this
would restore the original (unknown) superposition state of the atom,
provided we would also reverse the timing of the laser pulses. If, on the
other hand, we are able to drive the atom in a transmitting cavity in such a
way that the outgoing pulse were already symmetric in time, the wavepacket
entering a receiving cavity would ``mimic'' this time reversed process, thus
``restoring'' the state of the first atom in the second one.

The simplest possible configuration of quantum transmission between two
nodes consists of two three-level atoms 1 and 2 which are strongly coupled
to their respective cavity modes (see Fig.~\ref{pzfigcqed1}). The qubit is
stored in a superposition of the two degenerate ground states $|g\rangle
\equiv |0\rangle $ and $|e\rangle \equiv |1\rangle $. The states $|e\rangle $
and $|g\rangle $ are coupled by a Raman transition, where a laser excites
the atom from $|e\rangle $ to $|r\rangle $ with to a time-dependent Rabi
frequency, followed by a transition $|r\rangle \rightarrow |e\rangle $ which
is accompanied by emission of a photon into the corresponding cavity mode.
In order to suppress spontaneous emission from the excited state during the
Raman process, we assume that the laser is strongly detuned from the atomic
transition. In such a case, one can eliminate adiabatically the excited
states $|r\rangle$. The Hamiltonian for the dynamics of the two ground
states becomes, in a rotating frame for the cavity modes at the laser
frequency, is of the Jaynes-Cummings form with time-dependent coupling $%
g_{i}(t)$
\begin{equation}
\hat{H}_{i}=-\delta \hat{a}_{i}^{\dagger }\hat{a}_{i}-ig_{i}(t)\left[
|e\rangle _{i}\;{}_{i}\langle g|a_{i}-\mathrm{h.c.}\right] ,\quad (i=1,2)
\end{equation}
where $\hat{a}_{i}$ are the destruction operators for the cavity modes $%
i=1,2 $, and $\delta $ denotes the Raman detuning between the ground state
levels. For simplicity we ignore here AC--Stark shifts of the ground states
due to the cavity mode and laser field, which can be easily include in a
complete model \cite{Cirac97}. The last term is a Jaynes--Cummings
interaction, with an effective time-dependent coupling constant $g_{i}(t)$.

The goal is now to select a laser pulse shape $g_{i}(t)$ to accomplish \emph{%
ideal quantum transmission}
\begin{equation}  \label{transfer1}
\big(c_{g}|g\rangle _{1}+c_{e}|e\rangle_ {1}\big)|g\rangle _{2}\otimes
|0\rangle _{1}|0\rangle _{2}|\text{vac}\rangle \rightarrow |g\rangle _{1}%
\big(c_{g}|g\rangle _{2}+c_{e}|e\rangle _{2}\big)\otimes |0\rangle
_{1}|0\rangle _{2}|\text{vac}\rangle ,
\end{equation}
where $c_{g,e}$ are complex numbers. In (\ref{transfer1}), $|0\rangle _{i}$
and $|$vac$\rangle $ represent the vacuum state of the cavity modes and the
free electromagnetic modes connecting the cavities. Transmission will occur
by photon exchange via these modes.

\begin{figure}[hbp]
\label{pzfigcqed2}
\par
\begin{center}
\includegraphics[width=8.0cm]{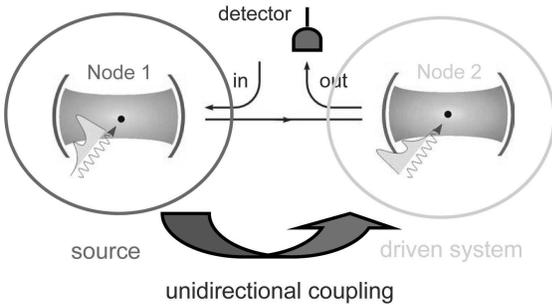}
\end{center}
\caption{Transmission of a qubit between two atoms as a cascaded quantum
system}
\end{figure}

It is useful to formulate this problem in the language of cascaded quantum
systems. A cascaded quantum system consists of a quantum \emph{source}
driving in a unidirectional coupling a quantum \emph{system}. In our case
the source is the first node emitting a photon while the system is the
second node (Fig.~\ref{pzfigcqed2}). In case of perfect transmission of the
qubit we require that the photon is not reflected from the second cavity,
and thus there is no back reaction on the first node, i.e. the coupling
becomes unidirectional. A theory of cascaded quantum systems has been
developed independently by Gardiner and Carmichael \cite%
{Gardiner99,Gardiner93,Carmichael93}. In the present context it is
convenient to use a quantum trajectory formulation \cite{Gardiner99} of
cascaded quantum system \cite{Carmichael93}. To this end , we consider a
fictitious experiment where the output field of the second cavity is
continuously monitored by a photodetector (Fig.~\ref{pzfigcqed2}). The
evolution of the quantum system under continuous observation, conditional to
observing a particular trajectory of counts, can be described by a pure
state wavefunction $|\Psi _{c}(t)\rangle $ in the system Hilbert space of
the two nodes (where the radiation modes outside the cavity have been
eliminated). During the time intervals when no count is detected, this
wavefunction evolves according to a Schr\"{o}dinger equation with \emph{%
non--hermitian} effective Hamiltonian
\begin{equation}
\hat{H}_{\mathrm{eff}}(t)=\hat{H}_{1}(t)+\hat{H}_{2}(t)-i\kappa \left( \hat{a%
}_{1}^{\dagger }\hat{a}_{1}+\hat{a}_{2}^{\dagger }\hat{a}_{2}+2\hat{a}%
_{2}^{\dagger }\hat{a}_{1}\right) .  \label{Heff}
\end{equation}
The detection of a count at time $t_{r}$ is associated with a quantum jump
according to $|\Psi _{c}(t_{r}+dt)\rangle \propto \hat{c}|\Psi
_{c}(t_{r})\rangle $, where $\hat{c}=\hat{a}_{1}+\hat{a}_{2}$, while he
probability density for a jump (detector click) to occur during the time
interval from $t$ to $t+dt$ is $\langle \Psi _{c}(t)|\hat{c}^{\dagger }\hat{c%
}|\Psi _{c}(t)\rangle dt$ \cite{Gardiner99}.

We wish to design the laser pulses in both cavities in such a way that ideal
quantum transmission condition (\ref{transfer1}) is satisfied. A necessary
condition for the time evolution is that a quantum jump (detector click, see
Fig.~\ref{pzfigcqed2}) never occurs, i.e.~ $\hat{c}|\Psi _{c}(t)\rangle =0$ $%
\forall t$, and thus the effective Hamiltonian will become a hermitian
operator. In other words, the system will remain in a \textit{dark} state of
the cascaded quantum system. Physically, this means that the wavepacket is
not reflected from the second cavity. We expand the state of the system as
\begin{eqnarray}
|\Psi _{c}(t)\rangle &=&|c_{g}|gg\rangle |00\rangle _{c}  \notag
\label{ansatz} \\
&+&|c_{e}\Big[\alpha _{1}(t)|eg\rangle |00\rangle _{c}+\alpha
_{2}(t)|ge\rangle |00\rangle _{c}  \notag \\
&&\hspace*{1cm}+\beta _{1}(t)|gg\rangle |10\rangle _{c}+\beta
_{2}(t)|gg\rangle _{c}|01\rangle _{c}\Big].
\end{eqnarray}
Ideal quantum transmission (\ref{transfer1}) will occur for $\alpha
_{1}(-\infty )=\alpha _{2}(+\infty )=1.$ We can now easily derive evolution
equations for the amplitudes $\alpha _{i}(t)$, $\beta _{i}(t)$ ($i=1,2$).
Starting from these equations Cirac \emph{et al.} derives a class of
solutions for pulses shapes in analytical form, guided by the physical
expectation that the time evolution in the second cavity should reverse the
time evolution in the first one, i.e. one looks for solutions satisfying the
\emph{symmetric pulse condition }$g_{2}(t)=g_{1}(-t)$. Numerical examples
and a discussion of imperfection due to photon loss and spontaneous emission
can be found in the quoted references.

Motivated by this CQED scheme various transmission protocols have been
discussed based on this \emph{photonic channel} \cite{Enk97,Enk98},
including a protocol for a \emph{quantum repeater} \cite{Briegel98}.

\subsection{Collisional interactions for neutral atoms}

In atomic physics with \emph{neutral atoms} recent advances in cooling and
trapping have led to an exciting new generation of experiments with Bose
condensates, experiments with optical lattices, and atom optics and
interferometry (see the lecture notes by E. Cornell and S. Rolston). The
question therefore arises, to what extent these new experimental
possibilities and the underlying physics can be adapted to the field of
experimental quantum computing. In the previous section we have outlined
possibilities of entangling neutral atoms in Cavity QED schemes. Here we
will discuss two examples of entangling atoms directly using controlled
two-body interactions. The specific examples to be discussed are cold
coherent collisions between ground state atoms \cite{Jaksch99}, and
interaction of atoms via large dipole-dipole coupling of Rydberg atoms \cite%
{Jaksch00}. A scheme for entangling atoms with laser induced dipole-dipole
interactions was proposed by G. K. Brennen \emph{et al.}\cite%
{Brennen99,Brennen00,Brennen01}.

\subsubsection{Entanglement via coherent ground state collisions}

Below we study coherent cold collisions \cite{Weiner99} as the basic
mechanism to entangle neutral atoms \cite{Jaksch99}. The picture of \emph{%
atomic collisions} as \emph{coherent interactions} has emerged in the field
of Bose Einstein condensation (BEC) of (ultracold) gases. In a field
theoretic language these interactions correspond to Hamiltonians which are
quartic in the atomic field operators, analogous to Kerr nonlinearities
between photons in quantum optics. By storing ultracold atoms in arrays of
microscopic potentials provided, for example, by optical lattices these
collisional interactions can be controlled via laser parameters \cite%
{Jaksch99}. Furthermore, these nonlinear atom-atom interactions can be large
\cite{Jaksch99}, even for interactions between individual pairs of atoms,
thus providing the necessary ingredients to implement quantum logic.

\begin{figure}[hbp]
\label{pzfigatom1}
\par
\begin{center}
\includegraphics[width=8.0cm]{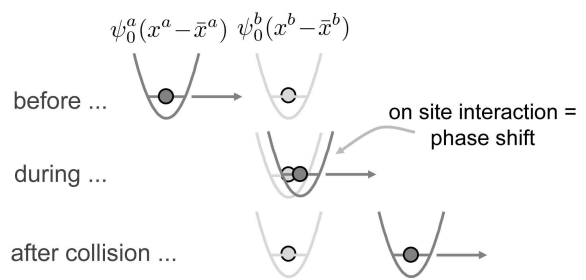}
\end{center}
\caption{We collide a first atom in the internal state $|a\rangle$ with a
second atom in state $|b\rangle$. In the collision the wave function
accumulates a phase according to (\protect\ref{transf}}
\end{figure}

Consider a situation where two atoms with electrons populating the internal
states $|a\rangle $ and $|b\rangle $, respectively, are trapped in the
ground states $\psi _{0}^{a,b}$ of two potential wells $V^{a,b}$ (Fig.~\ref%
{pzfigatom1}). Initially, these wells are centered at positions $\bar{x}^{a}$
and $\bar{x}^{b}$, sufficiently far apart (distance $d=\bar{x}_{b}-\bar{x}%
_{a}$) so that the particles do not interact. The positions of the
potentials are moved along trajectories $\bar{x}^{a}(t)$ and $\bar{x}^{b}(t)$
so that the wavepackets of the atoms overlap for certain time, until finally
they are restored to the initial position at the final time. This situation
is described by the Hamiltonian
\begin{equation}
H\!=\!\sum_{\beta =a,b}\left[ \frac{(\hat{p}^{\beta })^{2}}{2m}+V^{\beta
}\left( \hat{x}^{\beta }\!-\!\bar{x}^{\beta }(t)\right) \right] +u^{\mathrm{%
ab}}(\hat{x}^{a}\!-\!\hat{x}^{b}).  \label{Hamil}
\end{equation}
Here, $\hat{x}^{a,b}$ and $\hat{p}^{a,b}$ are position and momentum
operators, $V^{a,b}\left( \hat{x}^{a,b}-\bar{x}^{a,b}(t)\right) $ describe
the displaced trap potentials and $u^{\mathrm{ab}}$ is the atom--atom
interaction term. Ideally, we would like to implement the transformation
from before to after the collision,
\begin{equation}
\psi _{0}^{a}(x^{a}\!-\!\bar{x}^{a})\psi _{0}^{b}(x^{b}\!-\!\bar{x}%
^{b})\rightarrow e^{i\phi }\psi _{0}^{a}(x^{a}\!-\!\bar{x}^{a})\psi
_{0}^{b}(x^{b}\!-\!\bar{x}^{b}),  \label{transf}
\end{equation}
where each atom remains in the ground state of its trapping potential and
preserves its internal state. The phase $\phi $ will contain a contribution
from the interaction (collision). The transformation (\ref{transf}) can be
realized in the \emph{adiabatic limit}, whereby we move the potentials
slowly on the scale given by the trap frequency, so that the atoms remain in
the ground state. Moving non-interacting atoms will induce kinetic single
particle kinetic phases. In the presence of interactions ($u^{\mathrm{ab}%
}\neq 0$), we define the time--dependent energy shift due to the interaction
as
\begin{equation}
\Delta E(t)=\frac{4\pi a_{s}\hbar ^{2}}{m}\int dx|\psi _{0}^{a}\left( x-\bar{%
x}^{a}(t)\right) |^{2}|\psi _{0}^{b}\left( x-\bar{x}^{b}(t)\right) |^{2},
\label{deltaE}
\end{equation}
where $a_{s}$ is the $s$--wave scattering length. We assume that $|\Delta
E(t)|\ll \hbar \nu $ with $\nu $ the trap frequency so that no sloshing
motion is excited. In this case, (\ref{transf}) still holds with $\phi =\phi
^{a}+\phi ^{b}+\phi ^{\mathrm{ab}}$, where in addition to (trivial) single
particle kinetic phases $\phi ^{a}$ and $\phi ^{b}$ arising from moving the
potentials, we have a \emph{collisional phase shift}
\begin{equation}
\phi ^{\mathrm{ab}}=\int_{-\infty }^{\infty }dt\Delta E(t)/\hbar .
\label{phicol}
\end{equation}

\begin{figure}[hbp]
\label{pzfigatom2}
\par
\begin{center}
\includegraphics[width=8.0cm]{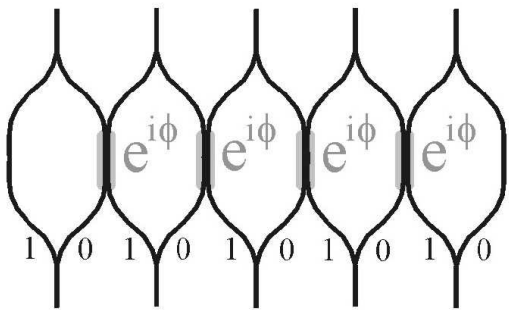}
\end{center}
\caption{By moving an optical lattice in a state-dependent way neighboring
atoms collide and acquire a phase shift.}
\end{figure}

The assumption behind the \emph{colliding atoms by hand}, as described
above, is that different internal states of the atom see a different
trapping non-dissipative potential. In practice, this can be achieved by
trapping atoms in a far-off-resonant trap or optical lattice, where the
lasers are tuned in such a way that the atomic states $|a\rangle $ and $%
|b\rangle $ couple differently to the excited atomic states, so that there
AC Stark shifts differ. A specific laser configuration achieving this state
dependent trapping has been analyzed in Ref. \cite{Jaksch99} for Alkali
atoms, based on tuning the laser between the fine structure excited states.
The trapping potentials can be moved by changing the laser parameters. Such
trapping potentials could also be realized with magnetic and electric
microtraps \cite{Calarco00}.

So far, we have argued that one can use cold collisions as a coherent
mechanism to induce phase shifts in two--atom interactions in a controlled
way. We can use these interactions to implement conditional dynamics. By the
above arguments, the atomic wave packet in a superposition of the two
internal levels can be ''split'' by moving the state dependent potentials,
very much like a beam splitter in atom interferometry. Thus we can move the
potentials of neighboring atoms such that only the $|a\rangle $ component of
the first atom ``collides'' with the state $|b\rangle $ of the second atom
\begin{eqnarray}
|a\rangle _{1}|a\rangle _{2} &\rightarrow &e^{i2\phi ^{a}}|a\rangle
_{1}|a\rangle _{2},  \notag  \label{gate} \\
|a\rangle _{1}|b\rangle _{2} &\rightarrow &e^{i(\phi ^{a}+\phi ^{b}+\phi ^{%
\mathrm{ab}})}|a\rangle _{1}|b\rangle _{2},  \notag \\
|b\rangle _{1}|a\rangle _{2} &\rightarrow &e^{i(\phi ^{a}+\phi
^{b})}|b\rangle _{1}|a\rangle _{2},  \notag \\
|b\rangle _{1}|b\rangle _{2} &\rightarrow &e^{i2\phi ^{b}}|b\rangle
_{1}|b\rangle _{2},
\end{eqnarray}
where the motional states remain unchanged in the adiabatic limit, and $\phi
^{a}$ and $\phi ^{b}$ are single particle kinetic phases. The transformation
(\ref{gate}) corresponds to a fundamental two--qubit gate. The fidelity of
this gate is limited by nonadiabatic effects, decoherence due to spontaneous
emission in the optical potentials and collisional loss to other unwanted
states, or collisional to unwanted states. According to Ref. \cite{Jaksch99}
the fidelity of this gate operation is remarkably close to one in a large
parameter range. We note that loading of single atoms in laser traps has
been achieved recently \cite{Grangier01}, and movable arrays of trapping
potential have been created with arrays of microlenses \cite{Birkl01}.
Finally, by filling the lattice from a Bose condensate, and using the ideas
related to Mott transitions in optical lattices \cite{Jaksch98} it is
possible to achieve uniform lattice occupation (``optical crystals'') or
even specific atomic patterns, as well as the low temperatures necessary for
performing the experiments proposed above.

It is interesting to generalize these ideas to the dynamics of a
two-component Bose gas in an optical lattice. For example, in the adiabatic
regime we denote by $a_{i}$ and $b_{i}$ the annihilation operators for a
particle in the ground state of the potential centered at the position $i$,
and corresponding to the internal levels $|a\rangle $ and $|b\rangle $,
respectively. The effective Hamiltonian in this regime is a time dependent
Bose-Hubbard model \cite{Jaksch98},
\begin{eqnarray}
H &=&\sum_{i}\left[ \omega ^{a}(t)a_{i}^{\dagger }a_{i}+\omega
^{b}(t)b_{i}^{\dagger }b_{i}+u^{\mathrm{aa}}(t)a_{i}^{\dagger
}a_{i}^{\dagger }a_{i}a_{i}+\right.  \notag \\
&&\left. u^{\mathrm{bb}}(t)b_{i}^{\dagger }b_{i}^{\dagger }b_{i}b_{i}\right]
+\sum_{i,j}u_{ij}^{\mathrm{ab}}(t)a_{i}^{\dagger }a_{i}b_{j}^{\dagger }b_{j},
\end{eqnarray}
where the $\omega $'s and $u$'s depend on the specific way the potentials
are moved. In quantum optics this Hamiltonian corresponds to a quantum
non-demolition situation \cite{Gardiner99}, whereby the particle number can
be measured non-destructively.

\subsubsection{Rydberg atoms}

Entanglement via cold coherent collisions, as described above, requires
moving atoms and accumulating a phase due to (comparatively small) onsite
interactions, which makes this a slow process. For neutral atoms there is a
difficulty in identifying \emph{strong and controllable long-range two-body
interactions}, which is required to design a gate. Furthermore, the strength
of two-body interactions does not necessarily translate into a useful fast
quantum gate: large interactions are usually associated with strong
mechanical forces on the trapped atoms. Thus, internal states of the trapped
atoms (the qubits) may become entangled with the motional degrees of freedom
during the gate, resulting effectively in an additional source of
decoherence. This leads to the typical requirement that the process is \emph{%
adiabatic} on the time scale of the oscillation period of the trapped atoms
in order to avoid entanglement with motional states. As a result, extremely
tight traps and low temperatures are required. A fast two-qubit phase gate
for neutral trapped atoms, which addresses these problems, was proposed in
Ref.~\cite{Jaksch99}. The scheme is based on (i) the very large interactions
of permanent dipole moments of laser excited Rydberg states in a constant
electric field to entangle atoms, and (ii) allows gate operation times set
by the time scale of the laser excitation or the two particle interaction
energy, which can be significantly shorter than the trap period. Among the
attractive features of the gate are the insensitivity to the temperature of
the atoms and to the variations in atom-atom separation.

Rydberg states \cite{Gallagher94} of a hydrogen atom within a given manifold
of a fixed principal quantum number $n$ are degenerate. This degeneracy is
removed by applying a constant electric field $\mathcal{E}$ along the $z$%
-axis (linear Stark effect). For electric fields below the Ingris-Teller
limit the mixing of adjacent $n$-manifolds can be neglected, and the energy
levels are split according to $\Delta E_{nqm}=3nqea_{0}\mathcal{E}/2$ with
parabolic and magnetic quantum numbers $q=n-1-|m|,n-3-|m|,\ldots ,-(n-1-|m|)$
and $m$, respectively, $e$ the electron charge, and $a_{0}$ the Bohr radius.
These Stark states have permanent dipole moments $\mathbf{\mu }\equiv \mu
_{z}\mathbf{e}_{z}=3/2\,nqea_{0}\mathbf{e}_{z}$. In alkali atoms the $s$ and
$p$-states are shifted relative to the higher angular momentum states due to
their quantum defects, and the Stark maps of the $m=0$ and $m=1$ manifolds
are correspondingly modified \cite{Gallagher94}.

\begin{figure}[hbp]
\label{pzfigatom3}
\par
\begin{center}
\includegraphics[width=8.0cm]{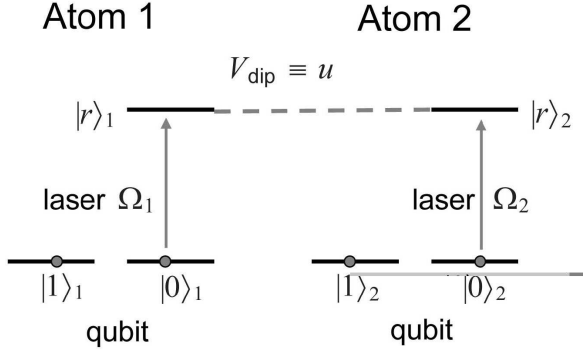}
\end{center}
\caption{Atomic level scheme of the two-qubit gate. The laser excites the
atoms in state $|1\rangle$ to Rydberg states in an electric field. The
Rydberg states interact via a dipole-dipole interaction.}
\end{figure}

Let us consider two atoms $1$ and $2$ initially prepared in Rydberg Stark
eigenstates, with a dipole moment along $z$ and a given $m$, as selected by
the polarization of the exciting laser. They interact and evolve according
to the dipole-dipole potential
\begin{equation}
V_{\mathrm{dip}}(\mathbf{r})=\frac{1}{4\pi \epsilon _{0}}\left[ \frac{%
\mathbf{\mu }_{1}\cdot \mathbf{\mu }_{2}}{|\mathbf{r}|^{3}}-3\frac{(\mathbf{%
\mu }_{1}\cdot \mathbf{r})(\mathbf{\mu }_{2}\cdot \mathbf{r})}{|\mathbf{r}%
|^{5}}\right] ,
\end{equation}
with $\mathbf{r}$ the distance between the atoms. We are interested in the
limit where the electric field is sufficiently large so that the energy
splitting between two adjacent Stark states is much larger than the
dipole-dipole interaction. For two atoms in the given initial Stark
eigenstate, the diagonal terms of $V_{\mathrm{dip}}$ provide an energy shift
whereas the non-diagonal terms couple $(m,m)\rightarrow (m\pm 1,m\mp 1)$
adjacent $m$ manifolds with each other. We will assume that these
transitions are suppressed by an appropriate choice of the initial Stark
eigenstate. For a hydrogen state $|r\rangle =|n,q=n-1,m=0\rangle ,$ for
example, and we find for a fixed distance $\mathbf{r}=R\mathbf{e}_{z}$ of
the two atoms a dipole-dipole interaction $u(R)=\langle r|\otimes \langle
r|V_{\mathrm{dip}}(R\mathbf{e}_{z})|r\rangle \otimes |r\rangle $ with $%
u(R)=-9[n(n-1)]^{2}(a_{0}/R)^{3}(e^{2}/8\pi \epsilon _{0}a_{0})\propto n^{4}$%
. In alkali atoms we have to replace $n$ by the effective quantum number $%
\nu $ \cite{Gallagher94}. We will use this large energy shift to entangle
atoms.

\begin{figure}[hbp]
\label{pzfigatom4}
\par
\begin{center}
\includegraphics[width=8.0cm]{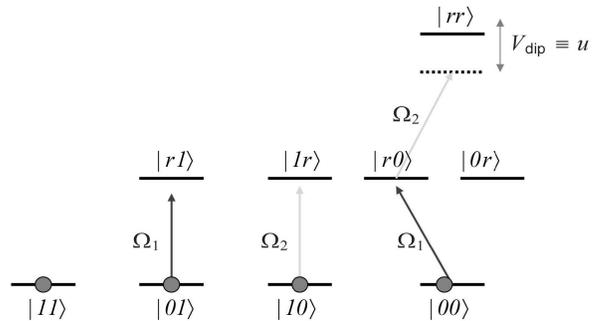}
\end{center}
\caption{Laser excitation sequence of the dipole-dipole gate with Rydberg
atoms. Qubits are stored in two internal atomic ground states denoted by $%
|0\rangle _{j}$ and $|1\rangle _{j}$.}
\end{figure}

We study a configuration where two atoms are for the moment assumed to be at
fixed positions $\mathbf{x}_{j}$ (with $j=1,2$ labelling the atoms), at a
distance at a distance $R=|\mathbf{x}_{1}-\mathbf{x}_{2}|$. We store qubits
in two internal atomic ground states denoted by $|g\rangle _{j}\equiv
|0\rangle _{j}$ and $|e\rangle _{j}\equiv |1\rangle _{j}$. The ground states
$|g\rangle _{j}$ are coupled by a laser to a given Stark eigenstate $%
|r\rangle _{j}$. The internal dynamics is described by a model Hamiltonian
\begin{equation}
H^{i}(t,\mathbf{x}_{1},\mathbf{x}_{2})=u|r\rangle _{1}\langle r|\otimes
|r\rangle _{2}\langle r|+\sum_{j=1,2}\left[ (\delta _{j}(t)-i\gamma
)|r\rangle _{j}\langle r|-\frac{1}{2}\Omega _{j}(t,\mathbf{x}_{j})\left(
|g\rangle _{j}\langle r|+\mathrm{h.c.}\right) \right] ,  \notag
\end{equation}
Here $\Omega _{j}(t,\mathbf{x}_{j})$ are Rabi frequencies, $\delta _{j}(t)$
are the detunings of the exciting lasers, and $\gamma $ accounts for loss
from the excited states $|r\rangle _{j}$.

When we include the atomic motion, the complete Hamiltonian has the
structure
\begin{eqnarray}
H(t,\mathbf{\hat{x}}_{1},\mathbf{\hat{x}}_{2}) &=&H^{T}(\mathbf{\hat{x}}_{1},%
\mathbf{\hat{x}}_{2})+H^{i}(t,\mathbf{\hat{x}}_{1},\mathbf{\hat{x}}_{2})
\label{Hbb} \\
&\equiv &H^{e}(t,\mathbf{\hat{x}}_{1},\mathbf{\hat{x}}_{2})+H^{i}(t,\mathbf{x%
}_{1},\mathbf{x}_{2}),
\end{eqnarray}
where $H^{T}$ describes the motion of the trapped atoms, and $\mathbf{\hat{x}%
}_{j}$ are the atomic position operators, and we define $\mathbf{\hat{r}}=%
\mathbf{\hat{x}_{1}}-\mathbf{\hat{x}_{2}}$. Our goal is to design a phase
gate for the internal states with a gate operation time $\Delta t$ with the
internal Hamiltonian $H^{i}(t,\mathbf{x}_{1},\mathbf{x}_{2})$ in Eq.~(\ref%
{Hbb}), where (the c--numbers) $\mathbf{x}_{j}$ now denote the centers of
the initial atomic wave functions as determined by the trap, while avoiding
motional effects arising from $H^{e}(t,\mathbf{\hat{x}}_{1},\mathbf{\hat{x}}%
_{2})$. This requires that the gate operation time $\Delta t$ is short
compared to the typical time of evolution of the external degrees of
freedom, $H^{e}\Delta t\ll 1$. Under this condition, the initial density
operator of the two atoms evolves as $\rho _{e}\otimes \rho _{i}\rightarrow
\rho _{e}\otimes \rho _{i}^{\prime }$ during the gate operation. Thus the
motion described by $\rho _{e}$ does not become entangled with the internal
degrees of freedom given by $\rho _{i}$. Typically, the Hamiltonian $H^{T}$
will be the sum of the kinetic energies of the atoms and the trapping
potentials for the various internal states. We will assume that the
potentials are harmonic with a frequency $\omega $ for the ground states,
and $\omega ^{\prime }$ for the excited state.

Physically, for the splitting of the Hamiltonian according to Eq.~(\ref{Hbb}%
) to be meaningful we require the initial width of the atomic wave function $%
a$, as determined by the trap, to be much smaller than the mean separation
between the atoms $R$. We expand the dipole-dipole interaction around $R$, $%
V_{\mathrm{dip}}(\hat {\mathbf{r}})=u(R)-F (\mathbf{\hat r}_z-R)+\ldots$,
with $F=3 u(R)/R$. Here the first term gives the energy shift if both atoms
are excited to state $|r\rangle$, while the second term contributes to $H^e$
and describes the mechanical force on the atoms due to $V_{\mathrm{dip}}$.
Other contributions to $H^e$ arise from the photon kick in the absorption $%
|g\rangle\rightarrow|r\rangle$, but these terms can be suppressed in a
Doppler-free two photon absorption, for example. We obtain $H^e(\mathbf{\hat
x}_1, \mathbf{\hat x}_2)=H^T-F(\mathbf{\hat r}_z-R) |r\rangle_1 \langle
r|\otimes |r\rangle_2 \langle r|. $

We can study our model for a phase gate according to dynamics induced by $%
H^{i}$ in various parameter regimes. First, in the limit $\Omega _{j}\gg u$,
it is possible to perform a gate by exciting the Rydberg atoms with $\pi $%
-pulses, so that the state $|r\rangle _{1}|r\rangle _{2}$ picks up the extra
phase $\varphi =u\Delta t$ from the dipole-dipole interaction. Thus, this
scheme realizes a fast phase gate operating on the time scale $\Delta
t\propto 1/u$. We note, however, that the accumulated phase depends on the
precise value of $u$, i.e.\ is sensitive to the atomic distance. In
addition, during the gate operation there are large mechanical effects due
to the force $F$.

These problems can be avoided by operating the gate in the parameter regime $%
u\gg \Omega _{j}$. In the simplest case we assume that the two atoms can be
addressed individually, i.e.~$\Omega _{1}(t)\neq \Omega _{2}(t)$. We set $%
\delta _{j}=0$ and perform the gate operation in three steps: (i) We apply a
$\pi $-pulse to the first atom, (ii) a $2\pi $-pulse (in terms of the
unperturbed states, i.e.~it has twice the pulse area of pulse applied in
(i)) to the second atom, and, finally, (iii) a $\pi $-pulse to the first
atom. As can be seen from Fig.~\ref{pzfigatom4}, the state $|ee\rangle $ is
not affected by the laser pulses. If the system is initially in one of the
states $|ge\rangle $ or $|eg\rangle $ the pulse sequence (i)-(iii) will
cause a sign change in the wave function. If the system is initially in the
state $|gg\rangle $ the first pulse will bring the system to the state $%
i|rg\rangle $, the second pulse will be \emph{detuned} from the state $%
|rr\rangle $ by the interaction strength $u$, and thus accumulate a \emph{%
small} phase $\tilde{\varphi}\approx \pi \Omega _{2}/2u\ll \pi $. The third
pulse returns the system to the state $e^{i(\pi -\tilde{\varphi})}|gg\rangle
$, which realizes a phase gate with $\varphi =\pi -\tilde{\varphi}\approx
\pi $ (up to trivial single qubit phases). The time needed to perform the
gate operation is of the order $\Delta t\approx 2\pi /\Omega _{1}+2\pi
/\Omega _{2}$. It is possible to formulate an \emph{adiabatic} version of
this gate with the advantage that individual addressing of the two atoms is
not required, $\Omega _{1,2}(t)\equiv \Omega (t)$ and $\delta
_{1,2}(t)\equiv \delta (t)$.

A remarkable feature of the second model is that, in the ideal limit, the
doubly excited state $|rr\rangle $ is never populated, because the double
excited state is shifted by the large dipole-dipole interaction, i.e. we
have a dipole-blockade mechanism (reminiscent of the Coulomb blockade in
quantum dots). Hence, the mechanical effects due to atom-atom interaction
are greatly suppressed. Furthermore, this version of the gate is only weakly
sensitive to the exact distance between the atoms, since the
distance-dependent part of the entanglement phase $\tilde{\varphi}\ll \pi $.

We now turn to a discussion of decoherence mechanisms, which include
spontaneous emission, transitions induced by black body radiation,
ionization of the Rydberg states due to the trapping or exciting laser
fields, and motional excitation of the trapped atoms. While dipole-dipole
interaction increases with $\nu ^{4}$, the spontaneous emission and
ionization of the Rydberg states by optical laser fields decreases
proportional to $\nu ^{-3}$. The Rabi frequency coupling the ground to the
Rydberg states scales as $n^{-3/2}$. For $\nu <20$ the black body radiation
is negligible in comparison with spontaneous emission, and similar
conclusions hold for typical ionization rates from the Rydberg states. For
typical numbers one expects errors on the percent level \cite{Jaksch99}.

\section{Quantum information processing with atomic ensembles}

\subsection{Introduction}

In the previous chapter, all the schemes for quantum information processing
are based on laser manipulation of single trapped particles. In this
chapter, we will show that many quantum information protocols can also be
implemented simply by laser manipulation of atomic ensembles containing a
large number of identical neutral atoms (see also the lecture notes by M.
Fleischhauer). The experimental candidate systems for the atomic ensembles
can be either laser-cooled atoms confined in a magnetic optical trap \cite%
{Hald99,Roch97,Hau99,Liu01}, or room-temperature atoms contained in a glass
cell with coated walls to avoid bad collisions \cite%
{Kash99,Phillips01,Julsgaard01}. Quantum information is stored in the
ground-state manifold of the atoms, such as in Zeeman sublevels with
different atomic spins, or in some hyperfine atomic levels which are stable
or metastable under optical transitions. Long coherence time of the relevant
states has been observed in both kinds of the experimental systems mentioned
above \cite{Hald99,Roch97,Hau99,Liu01,Kash99,Phillips01,Julsgaard01}. The
motivation of using atomic ensembles instead of single-particles for quantum
information processing is three-folds: first, laser manipulation of atomic
ensembles without separate addressing of individual atoms is much easier
than the laser manipulation of single particles; Secondly, the quantum
information encoded in atomic ensembles is robust against some practical
noise. For instance, the lost of few atoms in a large atomic ensemble has
negligible influence on its carried quantum information; Finally and perhaps
most importantly, the use of the atomic ensembles provides a novel way for
enhancing the signal-to-noise ratio with some collective effects existing in
this system. We know that in physical implementations of quantum information
protocols, the central problem is to enhance the signal-to-noise ratio, that
is, to enhance the ratio of the magnitudes between the coherent information
processing and the decoherence process caused by the noisy coupling to the
environment. For instance, in cavity-QED schemes, one needs to build a
high-finesse cavity around the atoms to achieve strong light-atom coupling,
and needs to enter the challenging strong coupling regime for a high
signal-to-noise ratio. However, we will see that due to the collective
enhancement induced by the many-atom coherence, the signal-to-noise ratio in
atomic ensembles can be greatly increased by encoding quantum information
into some collective excitations of the ensembles. As a result of this
effect, quantum information processing is made possible with much simplified
experimental systems, such as atomic ensemble in weak-coupling cavities or
even in free space.

This chapter is arranged as follows: In Sec. 4.2, we describe the
interaction of light with atomic ensembles with various level
configurations. Our attention is focused on collective enhancement of the
signal-to-noise ratio in these systems. For simple demonstration of the
collective enhancement, in the first two level configurations we assume that
there is a weak-coupling cavity around the atomic ensemble, following the
approaches in Refs. \cite{Kuzmich98,Lukin001,Duan001,Duan01b}; and in the
last level configuration, we directly describe the interaction of light with
a free-space atomic ensemble using a one-dimensional light propagation
model, following the approaches in Refs. \cite%
{Raymer81,Fleischhauer00,Duan002}. Some discussions on the three-dimensional
light propagation effects can be found in Ref. \cite{Raymer85}. In these two
different approaches, one can find basically the same kind of collective
enhancement of the signal-to-noise ratio. The systems discussed in Sec. 4.2
provide the basic elements for physical implementation of many quantum
information protocols. In Sec. 4.3, we show how to use atomic ensembles
combined with linear optical elements to realize scalable long-distance
quantum communication, such as quantum key distribution, quantum
teleportation, and Bell-inequality detections. Long-distance quantum
communication is necessarily based on the use of photonic channels. However,
due to losses and decoherence in the channel, the communication fidelity
decreases exponentially with the channel length. To overcome the problem
associated with the exponential fidelity decay, one needs to use quantum
repeaters \cite{Briegel98}, which provide the only known way for robust
long-distance quantum communication. In Sec. 4.3, we review the recent
scheme proposed in Ref. \cite{Duan01b} for physical implementation of
quantum repeaters and robust quantum communication over long lossy channels.
The scheme involves laser manipulation of atomic ensembles, beam splitters,
and single-photon detectors with moderate efficiencies, and therefore well
fits the status of the current experimental technology. The communication
efficiency in this scheme scales polynomially with the channel length
thereby facilitating scalability to very long distances. In Sec. 4.4, we
investigate other applications of atomic ensembles in quantum information
processing. In particular, we show that atomic ensembles can be used to
realize quantum light memory \cite%
{Kozhekin99,Lukin001,Duan001,Fleischhauer00} and single-photon sources with
controllable emission time, direction, and pulse shape. Quantum light memory
and a controllable single-photon source provide the basic elements for
realization of a recent quantum computation scheme \cite{Knill01}. Another
quantum computation scheme using atomic ensembles was proposed in Ref. \cite%
{Lukin002}, which combines laser manipulation of atomic ensembles with the
idea of dipole blockades from Rydberg atoms. In that scheme, one needs to
exploit direct dipole-dipole interactions between atoms which are induced by
exciting atoms to high Rydberg levels. This scheme will not be investigated
in this review since the concept of direct atom-atom interaction is outside
the scope of the current chapter. In Sec. 4.5, we review the applications of
atomic ensembles in continuous variable quantum information processing.
Laser manipulation of atomic ensembles provides a simple way for realizing
continuous variable quantum teleportation between distant atomic ensembles
\cite{Duan002}. Probabilities of using atomic ensembles for realizing
continuous variable quantum computation are also briefly remarked.

\subsection{Interaction of light with atomic ensembles and collective
enhancement of the signal-to-noise ratio}

We consider in the following the interaction of propagating light with
atomic ensembles. In particular, we demonstrate that when a sample of
identical atoms with suitable level configurations are shined by a
propagating optical signal, the signal will only interact with the
(symmetric) collective atomic mode, whereas atomic spontaneous emissions
caused by the coupling to vacuum noise fields are distributed over all the
modes. As a result of this property, the signal-to-noise ratio for the
collective atomic mode can be greatly enhanced. We consider three types of
level configurations for the atoms: two $\Lambda $-level configurations with
different kinds of coupling to the signal light and one four-level
configuration. All these level schemes are useful in subsequent sections for
different purposes of quantum information processing. In all these schemes,
besides the propagating quantum signal, there is some driving field provided
by a classical laser. We always assume that the classical driving field and
the quantum signal are co-propagating to assure a collective coupling
condition. For a simple demonstration of the collective enhancement of the
signal-to-noise ratio, first in the two $\Lambda $-level configurations we
assume that there is a low-Q ring cavity around the ensemble, and a ring
cavity mode, driven by the input and output quantum light signal, is then
coupled to a desired atomic transition. Due to the collective enhancement,
we can get a large signal-to-noise ratio even without a cavity to enhance
the light-atom coupling. As an example to explicitly show this, in the
four-level configuration we assume interaction of quantum light with
free-space atomic ensembles under a one-dimensional light propagation model.
The result confirms that we have the same kind of collective enhancement in
the signal-to-noise ratio.

\subsubsection{$\Lambda $I-level configuration}

\begin{figure}[tbp]
\label{d63reviewp1}
\par
\begin{center}
\includegraphics[width=8.0cm]{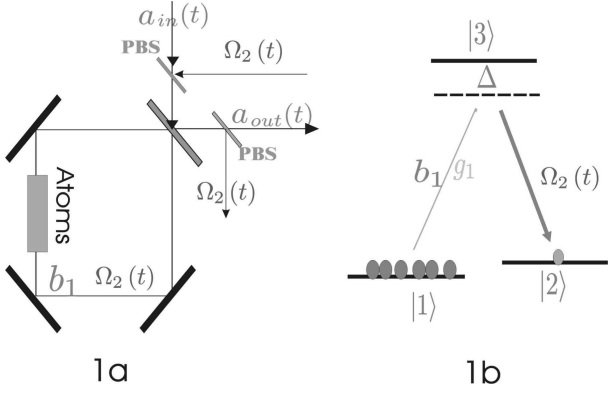}
\end{center}
\caption{(1a) An atomic ensemble in a weak coupling cavity. (1b) The $%
\Lambda $I-level configuration.}
\end{figure}

In the first level configuration, we assume that the atoms have two ground
states $\left| 1\right\rangle ,$ $\left| 2\right\rangle $ and one excited
state $\left| 3\right\rangle $ (see Fig.~ \ref{d63reviewp1}). All the atoms
initially occupy the ground level $\left| 1\right\rangle $. The transition $%
\left| 1\right\rangle \rightarrow \left| 3\right\rangle $ is coupled with a
coupling coefficient $g_{1}$ to a ring cavity mode $b_{1}$, which is then
driven by the input and output quantum signals $a_{in}\left( t\right) $ and $%
a_{out}\left( t\right) $. The transition $\left| 3\right\rangle \rightarrow
\left| 2\right\rangle $ is driven by a classical laser field with a Rabi
frequency $\Omega _{2}$\ ($\Omega _{2}$ can be time-dependent). The driving
laser and the ring cavity mode (quantum signal) are copropagating to satisfy
the collective coupling condition $\left( k_{l}-k_{s}\right) L_{a}\ll \pi $,
where $k_{l}$ and $k_{s}$ are respectively the wave vectors of the laser and
the quantum signal, and $L_{a}$ is the length of the atomic ensemble. We
assume off resonant coupling with a large detuning $\Delta $ as shown in
Fig.~ \ref{d63reviewp1}. This level scheme has been considered in Refs. \cite%
{Kozhekin99,Lukin001,Duan001,Fleischhauer00} for quantum light memory (Refs.
\cite{Lukin001,Fleischhauer00} considered this level scheme with resonant
coupling). Here we follow Ref. \cite{Duan001} for a simple theoretical
description. A description of the resonant coupling based on dark states can
be found in Refs. \cite{Lukin001,Fleischhauer01}, and a free-space
description of this level scheme with the one-dimensional light propagation
model can be found in Refs. \cite{Fleischhauer00}.

In the case of a large detuning $\Delta $, we can adiabatically eliminate
the excited level $\left| 3\right\rangle $, and under the collective
coupling condition, the interaction shown in Fig.~ \ref{d63reviewp1} is
described by the following Hamiltonian in the rotating frame
\begin{equation}
H=\hbar \left( \Omega _{2}g_{1}/\Delta \right) b_{1}^{\dagger
}\sum_{i=1}^{N_{a}}\sigma _{12}^{i}+\text{h.c.,}  \label{1}
\end{equation}
where $N_{a}$ is the total atom number, and $\sigma _{12}^{i}=\left|
1\right\rangle _{i}\left\langle 2\right| $ is the atomic lowering operator
for the $i$th atom. We have neglected the light shift terms such as $\left(
\Omega _{2}/\Delta \right) ^{2}\sum_{i=1}^{N_{a}}\sigma _{22}^{i}$ in Eq. (%
\ref{1}), which can be trivially cancelled by refining the laser frequency.
We assume that the quantum signal is weak so that nearly all the atoms still
remain in the level $\left| 1\right\rangle $\ with $\left\langle \left|
1\right\rangle _{i}\left\langle 1\right| \right\rangle \simeq 1$. Under this
weak excitation condition, we introduce an effective bosonic operator $%
s=\sum_{i=1}^{N_{a}}\sigma _{12}^{i}/\sqrt{N_{a}}$ with $\left[ s,s^{\dagger
}\right] \simeq 1$ for the (symmetric) collective atomic mode. With the
collective atomic operator, the Heisenberg-Langevin equations corresponding
to the Hamiltonian (1) has the form \cite{Gardiner99}
\begin{eqnarray}
\overset{.}{s} &=&-i\left( \sqrt{N_{a}}\Omega _{2}^{\ast }g_{1}^{\ast
}/\Delta \right) b_{1},  \label{2} \\
\overset{.}{b}_{1} &=&-i\left( \sqrt{N_{a}}\Omega _{2}g_{1}/\Delta \right)
s-\left( \kappa /2\right) b_{1}-\sqrt{\kappa }a_{in}\left( t\right) ,
\label{3}
\end{eqnarray}
where $\kappa $ is the cavity decay rate, and $a_{in}\left( t\right) $ is
the input quantum signal with the properties $\left[ a_{in}\left( t\right)
,a_{in}^{\dagger }\left( t^{\prime }\right) \right] =\delta \left(
t-t^{\prime }\right) $. The output quantum signal $a_{out}\left( t\right) $
from the cavity is connected with the input by the input-output relation $%
a_{out}\left( t\right) =a_{in}\left( t\right) +\sqrt{\kappa }b_{1}$. In the
bad-cavity limit with the cavity decay rate $\kappa $ much larger than the
coupling rate $\sqrt{N_{a}}\Omega _{2}g_{1}/\Delta $, we can adiabatically
eliminate the cavity mode $b_{1}$, and obtain a direct Langevin equation for
the collective atomic mode $s$
\begin{equation}
\overset{.}{s}=-\frac{\kappa ^{\prime }}{2}s-\sqrt{\kappa ^{\prime }}%
a_{in}\left( t\right) ,  \label{4}
\end{equation}
where the effective coupling rate $\kappa ^{\prime }=4N_{a}\left| \Omega
_{2}g_{1}\right| ^{2}/\left( \Delta ^{2}\kappa \right) $, and without loss
of generality we have assumed that the phase of the laser is chosen in the
way to make $i\Omega _{2}g_{1}=\left| \Omega g_{c}\right| $. The output
quantum signal, expressed by the atomic operator $s$, has the form $%
a_{out}\left( t\right) =-a_{in}\left( t\right) -\sqrt{\kappa ^{\prime }}s$.
Eq. (\ref{4}) serves as the basic equation in dealing with quantum light
memory in Sec. IV.

To show that we have a collective enhancement of the signal-to-noise ratio,
let us consider the atomic spontaneous emissions. There are two spontaneous
emission processes: first, the (about $N_{a}$) atoms in the level $\left|
1\right\rangle $ can absorb photons from the quantum signal to go up to the
level $\left| 3\right\rangle $, and then go down through spontaneous
emissions; and second, the very few atoms in the level $\left|
2\right\rangle $ can absorb photons from the classical driving laser to go
up and then go down through spontaneous emissions. We assume that the atomic
ensemble is dilute with $k_{s}/\sqrt[3]{\rho _{n}}\geq 1$ (where $\rho _{n}$
is the atomic number density) so that there are no superradiant effects for
spontaneous emissions which go to all the possible directions. In this case,
the total spontaneous emission rate is given by $\gamma _{t1}=N_{a}\left|
g_{1}\right| ^{2}\gamma _{s}/\Delta ^{2}$\ for the first process, and by $%
\gamma _{t2}=\left| \Omega _{2}\right| ^{2}\gamma _{s}/\Delta ^{2}$\ for the
second process, where $\gamma _{s}$ is the resonant spontaneous emission
rate (the natural bandwidth of the level $\left| 3\right\rangle $). When the
classical driving laser is strong with $\left| \Omega _{2}\right|
^{2}\succeq N_{a}\left| g_{1}\right| ^{2}$, the signal-to-noise ratio $%
R_{sn} $ in this configuration is estimated by $R_{sn}\sim \kappa ^{\prime
}/\gamma _{t2}\sim 4N_{a}\left| g_{1}\right| ^{2}/\left( \kappa \gamma
_{s}\right) $. This result should be compared with the corresponding one in
the case with single atoms trapped in a high-Q cavity, where the
signal-to-noise ratio is given by $\left| g_{1}\right| ^{2}/\left( \kappa
\gamma _{s}\right) $ \cite{Ye99}. Therefore, for atomic ensembles with the $%
\Lambda $I-level configuration, the signal-to-noise ratio is greatly
enhanced by the large factor of the atom number $N_{a}$ due to the many-atom
collective effects in coupling to the co-propagating quantum signal.

\subsubsection{$\Lambda $II-level configuration}

\begin{figure}[tbp]
\label{d63reviewp2}
\par
\begin{center}
\includegraphics[width=4.0cm]{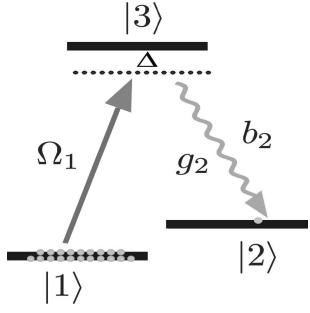}
\end{center}
\caption{The $\Lambda $II-level configuration.}
\end{figure}

In the second level configuration, each atom still has three levels $\left|
1\right\rangle ,$ $\left| 2\right\rangle $ and $\left| 3\right\rangle $. The
difference is now that the classical driving laser is coupling to the
transition $\left| 1\right\rangle \rightarrow \left| 3\right\rangle $, and
the quantum signal to the transition $\left| 3\right\rangle \rightarrow
\left| 2\right\rangle $\ (see Fig.~ \ref{d63reviewp2}). This level
configuration was considered before for a quantum description of the Raman
stimulating process \cite{Raymer81}, and recently it has been shown in \cite%
{Duan01b} to be useful for physical implementation of long-distance quantum
communication. Here, we follow Ref. \cite{Duan01b} for a simple description
and demonstration of the collective enhancement of the signal-to-noise
ratio. In this level configuration, the effective Hamiltonian after
adiabatic elimination of the upper level $\left| 3\right\rangle $ has the
following form
\begin{equation}
H=\hbar \left( \sqrt{N_{a}}\Omega _{1}g_{2}/\Delta \right) s^{\dagger
}b_{2}^{\dagger }+\text{h.c.,}  \label{5}
\end{equation}
where $s$ is the collective atomic annihilation operator defined as before,
and we have neglected the trivial light shift terms. Similar to the
configuration I, we can write the Heisenberg-Langevin equations
corresponding to the Hamiltonian (\ref{5}), and then adiabatically eliminate
the mode $b_{2}$ in the bad-cavity limit to obtain a direct Langevin
equation for the collective atomic mode $s$
\begin{equation}
\overset{.}{s^{\dagger }}=\frac{\kappa ^{\prime }}{2}s^{\dagger }-\sqrt{%
\kappa ^{\prime }}a_{in}\left( t\right) ,  \label{6}
\end{equation}
where the effective coupling rate $\kappa ^{\prime }=4N_{a}\left| \Omega
_{1}g_{2}\right| ^{2}/\left( \Delta ^{2}\kappa \right) $, and without loss
of generality we have assumed $i\Omega _{1}^{\ast }g_{2}^{\ast }=\left|
\Omega _{1}g_{2}\right| $. The output quantum signal $a_{out}\left( t\right)
$ is connected with the input $a_{in}\left( t\right) $ by the input-output
relation $a_{out}\left( t\right) =-a_{in}\left( t\right) +\sqrt{\kappa
^{\prime }}s^{\dagger }$.

The atomic ensembles with the $\Lambda $II-level configuration described
before provide us the basic elements for implementing quantum repeaters and
long-distance quantum communication which will be detailed in the next
section. For the applications there, we explicitly solve the basic equation (%
\ref{6}) with a vacuum input quantum signal $a_{in}\left( t\right) $ which
satisfies the properties $\left\langle a_{in}^{\dagger }\left( t\right)
a_{in}\left( t^{\prime }\right) \right\rangle =0$ and $\left\langle
a_{in}\left( t\right) a_{in}^{\dagger }\left( t^{\prime }\right)
\right\rangle =\delta \left( t-t^{\prime }\right) $. Equation (\ref{6}) is
linear and has the simple solution
\begin{equation}
s^{\dagger }\left( t\right) =s^{\dagger }\left( 0\right) e^{\kappa ^{\prime
}t/2}-\sqrt{\kappa ^{\prime }}\int_{0}^{t}e^{\kappa ^{\prime }\left( t-\tau
\right) /2}a_{in}\left( \tau \right) d\tau .  \label{7}
\end{equation}
What we are interested in the quantum repeater scheme is some measurable
quantity constructed from the output signal $a_{out}\left( t\right) $. If we
use photon detectors for measurement, what we detect is the integration of
the output photon current during the detection time interval $t_{\Delta }$,
which is proportional to the intensity integration of the output field $%
a_{out}\left( t\right) $. So in fact we measure the operator $%
Q_{m}=\int_{0}^{t_{\Delta }}a_{out}^{\dagger }\left( \tau \right)
a_{out}\left( \tau \right) d\tau $. One can find an explicit expression for $%
Q_{m}$ by substituting the solution of $a_{out}\left( t\right) $ from the
input-output relation. To simplify the expression of $Q_{m}$, we define an
effective single-mode bosonic operator $a$ from the continuous field $%
a_{in}\left( t\right) $ by the form
\begin{equation}
a\equiv -\frac{\sqrt{\kappa ^{\prime }}}{\sqrt{e^{\kappa ^{\prime }t_{\Delta
}}-1}}\int_{0}^{t_{\Delta }}e^{\kappa ^{\prime }\left( t_{\Delta }-\tau
\right) /2}a_{in}\left( \tau \right) d\tau .  \label{8}
\end{equation}
With this operator, the measured quantity is expressed as
\begin{equation}
Q_{m}=a_{t_{\Delta }}^{\dagger }a_{t_{\Delta }}+\int_{0}^{t_{\Delta
}}a_{in}^{\dagger }\left( \tau \right) a_{in}\left( \tau \right) d\tau ,
\label{9}
\end{equation}
where $a_{t_{\Delta }}=a\cosh r_{c}+s^{\dagger }\left( 0\right) \sinh r_{c},$
a Bogoliubov transformation of $a$ and $s^{\dagger }\left( 0\right) $ with $%
\cosh r_{c}\equiv e^{\kappa ^{\prime }t_{\Delta }/2}$. The last term of Eq. (%
\ref{9}) is a trivial integration of the intensity of the vacuum field,
which has no contribution to the measurement result. So what we measure is
in fact the photon number in the effective single mode $a_{t_{\Delta }}$.
Note that Eq. (\ref{7}) can also be written in the Bogoliubov form $%
s^{\dagger }\left( t_{\Delta }\right) =s^{\dagger }\left( 0\right) \cosh
r_{c}+a\sinh r_{c}.$ Both of $s^{\dagger }\left( 0\right) $ and $a$ are
initially in vacuum states, which will be denoted by $\left|
0_{a}\right\rangle $ and $\left| 0_{p}\right\rangle $ respectively in the
following (the subscripts ``$a$'' for atoms, and ``$p$'' for photons).
Transferring the solution to the Schr\"{o}dinger picture, we conclude that
after time $t_{\Delta }$ the collective atomic mode $s$ and the effective
single mode for the output quantum signal are in a two-mode squeezed state
\begin{equation}
\left| \phi \right\rangle =\sec r_{c}\sum_{n}\left( S^{\dagger }a\tanh
r_{c}\right) ^{n}/n!\left| 0_{a}\right\rangle \left| 0_{p}\right\rangle .
\label{10}
\end{equation}
This is the basic result which will be used in the next section.

Now let us take into account the atomic spontaneous emissions and calculate
the signal-to-noise ratio for this level configuration. The current
situation is quite different from the $\Lambda $I-level configuration in the
sense that if one directly calculates the total spontaneous emission rate in
this system, one would find that the total rate is given by $N_{a}\left(
\Omega _{1}^{2}/\Delta ^{2}\right) \gamma _{s}$ since almost all the atoms
remain in the level $\left| 1\right\rangle $. If one takes this rate as the
noise rate, there will be no enhancement of the signal-to-noise ratio
compared with the single-atom case. However, in our scheme we only concern
about the collective atomic mode $s$ since the output quantum signal is only
entangled with this mode as shown by Eq. (\ref{10}). So instead of the
calculation of the total spontaneous emission rate, we need to calculate the
noise rate for the mode $s$. The spontaneous emissions are independent for
different atoms without superradiance, and they introduce a coherence decay
term to the Langevin equation of each individual atomic operator $%
s_{i}=\sigma _{12}^{i}$%
\begin{equation}
\overset{.}{s^{\dagger }}_{i}=-\left( \gamma _{s}^{\prime }/2\right)
s_{i}^{\dagger }+\text{noise,}  \label{11}
\end{equation}
where the decay rate $\gamma _{s}^{\prime }=\left( \Omega _{1}^{2}/\Delta
^{2}\right) \gamma _{s}$. The last term in Eq. (\ref{11}) represents the
corresponding fluctuation from the noise field which results in heating, and
we have left out the coherent interaction term from the Hamiltonian (\ref{5}%
). By taking summation of Eq. (\ref{11}) over all the atoms, we immediately
see that there is a coherence decay term to the Langevin equation (\ref{6})
of the collective atomic operator $s^{\dagger }$ with the decay rate still
given by $\gamma _{s}^{\prime }$. The signal-to-noise ratio $R_{sn}$ for the
collective atomic mode is thus given by $R_{sn}=\kappa ^{\prime }/\gamma
_{s}^{\prime }=4N_{a}\left| g_{2}\right| ^{2}/\left( \kappa \gamma
_{s}\right) $. So compared with the single-atom case, the signal-to-noise
ratio is still greatly enhanced by the large factor of the atom number $%
N_{a} $. This collective enhancement comes from the fact that the coherent
interaction producing the output quantum signal involves only the collective
atomic mode $s$, whereas the independent spontaneous emissions distribute
over all the atomic modes, and thus only have small influence on the
interesting mode $s$.

To best understand this point, it is helpful to also have a look at the
master equation for the atomic density operator. The whole density operator $%
\rho _{w}$ for the atomic states and the cavity mode obeys the following
master equation \cite{Gardiner99}
\begin{equation}
\overset{.}{\rho }_{w}=i\left[ \rho _{w},H\right] +\kappa \widehat{L}\left[ b%
\right] \rho _{w}+\gamma _{s}^{\prime }\sum_{i}\widehat{L}\left[
s_{i}^{\dagger }\right] \rho _{w},  \label{12}
\end{equation}
where the Liouville superoperators $\widehat{L}\left[ X\right] $ $\left(
X=b,s_{i}^{\dagger }\right) $ are defined as $\widehat{L}\left[ X\right]
\rho _{w}\equiv X\rho _{w}X^{\dagger }-\left( X^{\dagger }X\rho _{w}+\rho
_{w}X^{\dagger }X\right) /2$. In Eq. (\ref{12}), the first term of the right
hand side (r.h.s.) comes from the coherent Hamiltonian interaction, the
second term represents the cavity output coupling, and the last term
describes independent spontaneous emissions for individual atomic operators.
In the bad cavity limit, after adiabatically eliminating the cavity mode, we
get from Eqs. (\ref{12}) and (\ref{5}) the following master equation for the
traced atomic density operator $\rho _{a}$
\begin{equation}
\overset{.}{\rho }_{a}=\kappa ^{\prime }\widehat{L}\left[ s^{\dagger }\right]
\rho _{a}+\gamma _{s}^{\prime }\sum_{i}\widehat{L}\left[ s_{i}^{\dagger }%
\right] \rho _{a},  \label{13}
\end{equation}
where $\widehat{L}\left[ s^{\dagger }\right] $ is the Liouville
superoperator for the collective atomic mode. The above equation can be
further simplified if we introduce the Fourier transformation to the
individual atomic operators $s_{j}$ $\left( j=0,1,\cdots ,N_{a}-1\right) $
with the form $s_{\mu }\equiv \sum_{j}S_{j}e^{ij\mu /N_{a}}/\sqrt{N_{a}}$,
where $s_{\mu =0}$ gives exactly the collective atomic operator $s$. In
terms of the operators $s_{\mu }$, the master equation has the form
\begin{equation}
\overset{.}{\rho }_{a}=\left( \kappa ^{\prime }+\gamma _{s}^{\prime }\right)
\widehat{L}\left[ s^{\dagger }\right] \rho _{a}+\gamma _{s}^{\prime
}\sum_{\mu \neq 0}\widehat{L}\left[ s_{\mu }^{\dagger }\right] \rho _{a},
\label{14}
\end{equation}
Under the weak excitation condition $\left\langle s_{j}^{\dagger
}s_{j}\right\rangle \ll 1$, the operators $s_{\mu }$ $(\mu =0,1,\cdots ,$ $%
N_{a}-1)$ commute with each other, so they represent independent atomic
modes. We are only interested in the collective atomic mode $s$, and the
populations in all the other modes $s_{\mu }$ with $\mu \neq 0$ have no
influence on the state of the mode $s$. So we can trace over the modes $%
s_{\mu }$ $\left( \mu \neq 0\right) $ and eliminate the last term in Eq. (%
\ref{14}). There are two contributions to the population in the collective
atomic mode $s$: the one with a rate $\kappa ^{\prime }$ produces a coherent
output signal, and the one with a rate $\gamma _{s}^{\prime }$ emits photons
to other random directions. The signal-to-noise ratio for the mode $s$ is
thus given by $R_{sn}=\kappa ^{\prime }/\gamma _{s}^{\prime }\sim
4N_{a}\left| g_{c}\right| ^{2}/\left( \kappa \gamma _{s}\right) $, and we
get exactly the same result as before. It is interesting to note from Eq. (%
\ref{14}) that the total spontaneous emission rate of all the modes is $%
N_{a}\gamma _{s}^{\prime }$, which could be much larger than the coherent
interaction rate $\kappa ^{\prime }$; however, the spontaneous emission rate
for the collective atomic mode is $N_{a}$ times smaller than the total rate.
This is why we still have collective enhancement and a large signal-to-noise
ratio for this level configuration.

\subsubsection{Four-level configuration}

In the above two configurations of the light-atom interaction, the
signal-to-noise ratio is greatly enhanced by the many-atom collective
effects. Due to the collective enhancement, we by no means need a good
cavity in these schemes. In fact, we can even assume to continuously
decrease the cavity finesse down to $1,$ which corresponds to the free-space
limit. In the free-space limit, the cavity decay rate $\kappa $ is estimated
by $c/L_{a}$, the inverse of the travelling time of the optical pulse in the
ensemble. With the well known expressions for the coupling coefficient $%
g_{1} $ (or $g_{2}$) and the resonant spontaneous emission rate $\gamma _{s}$
\cite{Gardiner99}, one can estimate the signal-to-noise ratio in the
free-space limit by $R_{sn}\sim 4N_{a}\left| g_{c}\right| ^{2}/\left( \kappa
\gamma _{s}\right) \sim 3\rho _{n}L_{a}/k_{s}^{2}\sim d_{o}$, where $d_{o}$
denotes the on-resonance optical depth of the atomic ensemble which can be
quite large with the current experimental technology \cite%
{Hald99,Roch97,Hau99,Liu01,Kash99,Phillips01,Julsgaard01}. So we can have a
considerably large signal-to-noise ratio even without a cavity. This
demonstrates that collective effects in many-atom ensembles provide us
another way besides high-Q cavities to achieve strong coherent light-atom
coupling. To show this more directly, we consider another light-atom
interaction configuration with four levels, and in this level scheme, we
solve directly the interaction of light with free-space atomic ensembles by
assuming a one-dimensional light propagation model.

\begin{figure}[tbp]
\label{d63reviewp3}
\par
\begin{center}
\includegraphics[width=6.0cm]{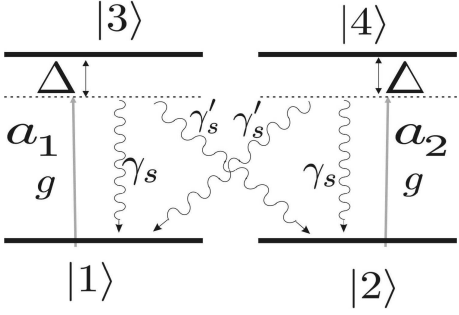}
\end{center}
\caption{The four-level configuration. }
\end{figure}

The relevant atomic level structure is shown by Fig.~ \ref{d63reviewp3}.
Each atom has two degenerate ground states and two degenerate excited
states. The transitions $\left| 1\right\rangle \rightarrow \left|
3\right\rangle $ and $\left| 2\right\rangle \rightarrow \left|
4\right\rangle $ are coupled with a large detuning $\Delta $ to different
circularly polarized propagating light due to the angular-momentum selection
rule. This kind of interaction has been analyzed semiclassically in \cite%
{Happer72,Happer67}, and recently shown to be applicable for quantum
non-demolition measurements \cite{Kuzmich98,Kuzmich001,Takahashi99,Molmer99b}
and for continuous variable quantum teleportation \cite{Duan002,Kuzmich002}.
Here, we follow Ref. \cite{Duan002} for a free-space quantum description of
the light- atomic-ensemble interaction. The atomic spontaneous emissions are
included in the description to demonstrate the collective enhancement of the
signal-to-noise ratio.

We assume a one-dimensional model for the propagating light field. As shown
in Ref.\ \cite{Raymer85}, this is justified if the atomic ensemble is of a
pencil shape with Fresnel number $F=A/\lambda _{0}L_{a}\approx 1$. Here $A$
and $L_{a}$\ are the cross section and the length of the ensemble,
respectively, and $\lambda _{0}$ is the optical wave length. The input laser
pulse is linearly polarized and expressed as $E^{(+)}\left( z,t\right) =%
\sqrt{\frac{\hbar \omega _{0}}{4\pi \epsilon _{0}A}}\underset{i=1,2}{\sum }%
a_{i}\left( z,t\right) e^{i\left( k_{0}z-\omega _{0}t\right) }$, where $%
\omega _{0}=k_{0}c=2\pi c/\lambda _{0}$ is the carrier frequency, and $i$
denotes two orthogonal circular polarizations, with the standard commutation
relations $\left[ a_{i}\left( z,t\right) ,a_{j}\left( z^{\prime },t\right) %
\right] =\delta _{ij}\delta \left( z-z^{\prime }\right) $. The light is
weakly focused with cross section $A$ to match the atomic ensemble. For the
input of a strong coherent light with linear polarization, the initial
condition is expressed as $\left\langle a_{i}\left( 0,t\right) \right\rangle
=\alpha _{t},$ with the total photon number over the pulse duration $T$
satisfies $2N_{p}=2c\int_{0}^{T}\left| \alpha _{t}\right| ^{2}dt\gg 1$. The
Stokes operators are introduced for the free-space input and output light
(light before entering or after leaving the atomic ensemble) by $S_{x}^{p}=%
\frac{c}{2}\int_{0}^{T}\left( a_{1}^{\dagger }a_{2}+a_{2}^{\dagger
}a_{1}\right) d\tau ,$ $S_{y}^{p}=\frac{c}{2i}\int_{0}^{T}\left(
a_{1}^{\dagger }a_{2}-a_{2}^{\dagger }a_{1}\right) d\tau ,$ $S_{z}^{p}=\frac{%
c}{2}\int_{0}^{T}\left( a_{1}^{\dagger }a_{1}-a_{2}^{\dagger }a_{2}\right)
d\tau .$ In free space, $a_{i}\left( z,t\right) $ only depends on $\tau
=t-z/c$, and in this case one can check the Stokes operators satisfy the
spin commutation relations $\left[ S_{y}^{p},S_{z}^{p}\right] =iS_{x}^{p}$.
For our coherent input, we have $\left\langle S_{x}^{p}\right\rangle =N_{p}$
and $\left\langle S_{y}^{p}\right\rangle =$ $\left\langle
S_{z}^{p}\right\rangle =0$. With a very large $N_{p}$, the off-resonant
interaction with atoms is only a small perturbation to $S_{x}^{p}$, and we
can treat $S_{x}^{p}$ classically by replacing it with its mean value $%
\left\langle S_{x}^{p}\right\rangle $. Then, we define two canonical
observables for light by $X^{p}=S_{y}^{p}/\sqrt{\left\langle
S_{x}^{p}\right\rangle },$ $P^{p}=S_{z}^{p}/\sqrt{\left\langle
S_{x}^{p}\right\rangle }$ with a standard commutator $\left[ X^{p},P^{p}%
\right] =i$. These operators are the quantum variables we are interested in.
Similar operators can be introduced for atoms. For an atomic ensemble with
many atoms, it is convenient to define the continuous atomic operators $%
\sigma _{\mu \nu }\left( z,t\right) =\lim_{\delta z\rightarrow 0}\frac{1}{%
\rho A\delta z}\sum_{i}^{z\leq z_{i}<z+\delta z}\left| \mu \right\rangle
_{i}\left\langle \nu \right| $ ($\mu ,\nu =1,2,3,4)$ with the commutation
relations $\left[ \sigma _{\mu \nu }\left( z,t\right) ,\sigma _{\nu ^{\prime
}\mu ^{\prime }}\left( z^{\prime },t\right) \right] =\left( 1/\rho
_{n}A\right) \delta \left( z-z^{\prime }\right) \left( \delta _{\nu \nu
^{\prime }}\sigma _{\mu \mu ^{\prime }}-\delta _{\mu \mu ^{\prime }}\sigma
_{\nu ^{\prime }\nu }\right) $. In the definition, $z_{i}$ is the position
of the $i$ atom, and $\rho _{n}$ is the number density of the atomic
ensemble with the total atom number $2N_{a}=\rho AL_{a}\gg 1$. The
collective spin operators are introduced for the ground states of the atomic
ensemble by $S_{x}^{a}=\frac{\rho A}{2}\int_{0}^{L_{a}}\left( \sigma
_{12}+\sigma _{12}^{\dagger }\right) dz,$ $S_{y}^{a}=\frac{\rho A}{2i}%
\int_{0}^{L_{a}}\left( \sigma _{12}-\sigma _{12}^{\dagger }\right) dz,$ $%
S_{z}^{a}=\frac{\rho A}{2}\int_{0}^{L_{a}}\left( \sigma _{11}-\sigma
_{22}\right) dz$. All the atoms are initially prepared in the equal
superposition of the two ground states $\left( \left| 1\right\rangle +\left|
2\right\rangle \right) /\sqrt{2},$ which is an eigenstate of $S_{x}^{a}$
with a very large eigenvalue $N_{a}$. As before, we treat $S_{x}^{a}$
classically, and define the canonical operators for atoms by $%
X^{a}=S_{y}^{a}/\sqrt{\left\langle S_{x}^{a}\right\rangle },$ $%
P^{a}=S_{z}^{a}/\sqrt{\left\langle S_{x}^{a}\right\rangle }$ with $\left[
X^{a},P^{a}\right] =i$ and an initial vacuum state.

With introduction of the continuous atomic operators, the interaction
between atoms and the propagating light $E^{(+)}\left( z,t\right) $ is
described by the following Hamiltonian in the rotating frame
\begin{eqnarray}
H &=&\hbar \sum_{i=1,2}\int_{0}^{L}\left[ \Delta \sigma _{i+2,i+2}\left(
z,t\right) \right.  \notag \\
&&\left. +\left( ge^{ik_{0}z}a_{i}\left( z,t\right) \sigma _{i+2,i}\left(
z,t\right) +h.c\right) \right] \rho Adz\;,  \label{15}
\end{eqnarray}
where the coupling constant $g=\sqrt{\frac{\omega _{0}}{4\pi \hbar \epsilon
_{0}A}}d$ and $d$ is the dipole moment of the $\left| i\right\rangle
\rightarrow \left| i+2\right\rangle $ transition. Corresponding to this
Hamiltonian, the Maxwell-Bloch equations are written as \cite{Gardiner99}
\begin{eqnarray}
&&\left( \frac{\partial }{\partial t}+c\frac{\partial }{\partial z}\right)
a_{i}\left( z,t\right) =-ig^{\ast }e^{-ik_{0}z}\rho A\sigma _{i,i+2}\left(
z,t\right) ,  \notag \\
&&\frac{\partial }{\partial t}\sigma _{\mu \nu }=-\frac{i}{\hbar }\left[
\sigma _{\mu \nu },H\right] -\frac{\gamma _{\mu \nu }}{2}\sigma _{\mu \nu }
\label{16} \\
&&\text{ \ \ \ \ \ \ \ \ \ \ \ }+\sqrt{\gamma _{\mu \nu }}\left( \sigma
_{\nu \nu }-\sigma _{\mu \mu }\right) F_{\mu \nu }\text{ }\left( \mu <\nu
\right) ,  \notag
\end{eqnarray}
where the spontaneous emission rates (see Fig.~ \ref{d63reviewp3}) are,
respectively, $\gamma _{13}=\gamma _{24}\equiv \gamma _{s}=\frac{\omega
_{0}^{3}\left| d\right| ^{2}}{3\pi \epsilon _{0}\hbar c^{3}},$ $\gamma
_{14}=\gamma _{23}\equiv \gamma _{s^{\prime }},$ and $\gamma _{12}=0$ (the
ground state has a long coherence time). The Doppler broadening caused by
the atomic motion is negligible, since it is eliminated for off-resonant
interactions with the collinear input and output lights. Assuming that the
spontaneous emission is independent for different atoms, the vacuum noise
operators $F_{\mu \nu }$ satisfy the $\delta $-commutation relations $\left[
F_{\mu \nu }\left( z,t\right) ,F_{\mu ^{\prime }\nu ^{\prime }}^{\dagger
}\left( z^{\prime },t^{\prime }\right) \right] =\left( 1/\rho A\right)
\delta _{\mu \mu ^{\prime }}\delta _{\nu \nu ^{\prime }}\delta \left(
z-z^{\prime }\right) \delta \left( t-t^{\prime }\right) $. To simplify Eq.~(%
\ref{16}), first we change the variables by $\tau =t-z/c$, and then
adiabatically eliminate the excited states $\left| 3\right\rangle $ and $%
\left| 4\right\rangle $ of atoms in the case of a large detuning, i.e., $%
\Delta \gg g\left\langle a_{i}\left( z,t\right) \right\rangle \sim g\sqrt{%
N_{p}/\left( cT\right) }$. The resultant equations read
\begin{eqnarray}
&&\frac{\partial }{\partial z}a_{i}\left( z,\tau \right) =\frac{i\left|
g\right| ^{2}\rho A\sigma _{ii}}{\Delta c}a_{i}\left( z,\tau \right) -\frac{%
\left| g\right| ^{2}\rho A\gamma _{s}\sigma _{ii}}{2\Delta ^{2}c}a_{i}\left(
z,\tau \right)  \notag \\
&&\text{ \ \ \ \ \ \ \ \ \ \ \ \ \ \ \ }+\frac{g^{\ast }e^{-ik_{0}z}\rho A%
\sqrt{\gamma _{s}}\sigma _{ii}}{\Delta c}F_{i,i+2}\left( z,\tau \right) ,
\label{17} \\
&&\frac{\partial }{\partial \tau }\sigma _{12}=\frac{i\left| g\right|
^{2}\left( a_{2}^{\dagger }a_{2}-a_{1}^{\dagger }a_{1}\right) }{\Delta }%
\sigma _{12}-\frac{\left| g\right| ^{2}\gamma _{s^{\prime }}\left(
a_{2}^{\dagger }a_{2}+a_{1}^{\dagger }a_{1}\right) }{2\Delta ^{2}}  \notag \\
&&\text{ \ \ \ \ \ }\times \sigma _{12}+\frac{\sqrt{\gamma _{s^{\prime }}}}{%
\Delta }\left( g^{\ast }e^{-ik_{0}z}a_{2}^{\dagger }\sigma
_{11}F_{14}+ge^{ik_{0}z}a_{1}\sigma _{22}F_{23}^{\dagger }\right) .  \notag
\end{eqnarray}
The physical meaning of the above equation is quite clear: The first term at
the right hand side is the phase shift caused by the off-resonant
interaction between light and atoms, and the second and the third terms
represent the damping and the corresponding vacuum noise caused by the
spontaneous emission, respectively. In Eq.~(\ref{17}), the $\sigma _{ii}$
and $a_{i}^{\dagger }a_{i}$ are approximately constant operators, only with
a small damping caused by the spontaneous emission. To consider the
spontaneous emission noise to the first order, it is reasonable to assume
constant $\sigma _{ii}$ and $a_{i}^{\dagger }a_{i}$ for Eq.~(\ref{17}).
Then, this equation can be easily solved by integrating over $z,\tau $ on
both sides. The result, expressed by the canonical atomic and optical
operators $X^{p,a}$ and $P^{p,a}$ introduced before, has the following
simple form
\begin{eqnarray}
X^{p\prime } &=&\sqrt{1-\varepsilon _{p}}\left( X^{p}-\kappa
_{c}P^{a}\right) +\sqrt{\varepsilon _{p}}X_{s}^{p},  \notag \\
P^{p\prime } &=&\sqrt{1-\varepsilon _{p}}P^{p}+\sqrt{\varepsilon _{p}}%
P_{s}^{p},  \notag \\
X^{a\prime } &=&\sqrt{1-\varepsilon _{a}}\left( X^{a}-\kappa
_{c}P^{p}\right) +\sqrt{\varepsilon _{a}}X_{s}^{a},  \label{18} \\
P^{a\prime } &=&\sqrt{1-\varepsilon _{a}}P^{a}+\sqrt{\varepsilon _{a}}%
P_{s}^{a},  \notag
\end{eqnarray}
where the operators with (without) a prime denote the quantities after
(before) the light pulse goes through the atomic ensemble, and $%
X_{s}^{a},P_{s}^{a},X_{s}^{p},P_{s}^{p}$ represent the standard vacuum noise
operators with variance $1/2$, defined from the integration of $F_{\mu \nu
}\left( z,t\right) $, $X_{s}^{p}=\sqrt{\frac{c}{4N_{p}N_{a}\left| g\right|
^{2}}}\int_{0}^{T}\int_{0}^{L}\rho A\left[ ig^{\ast }e^{-ik_{0}z}\left(
a_{2}^{\dagger }\sigma _{11}F_{13}-a_{1}^{\dagger }\sigma _{22}F_{24}\right)
\right. $ $\left. +h.c.\right] dzdt$ for instance. The interaction and
damping coefficients $\kappa _{c},\varepsilon _{p},\varepsilon _{a}$ are
given respectively by $\kappa _{c}=-\frac{2\sqrt{N_{p}N_{a}}\left| g\right|
^{2}}{\Delta c}=\frac{3\sqrt{N_{p}N_{a}}\gamma \lambda _{0}^{2}}{8\pi
^{2}\Delta A},$ $\varepsilon _{p}=\frac{N_{a}\left| g\right| ^{2}\gamma _{s}%
}{\Delta ^{2}c},$ $\varepsilon _{a}=\frac{N_{p}\left| g\right| ^{2}\gamma
_{s^{\prime }}}{\Delta ^{2}c}$. The solution (\ref{18}) is obtained under
the conditions of weak excitation $\kappa _{c}\ll \sqrt{N_{p,a}}$ and small
noise $\varepsilon _{p,a}\ll 1$. For simplicity, we assume $N_{p}\sim N_{a}$
and $\gamma _{s}\sim \gamma _{s^{\prime }}$ so that $\varepsilon _{p}\sim
\varepsilon _{a}\sim \varepsilon $. The interaction parameter $\kappa _{c}$
can be rewritten as $\kappa _{c}=\left( 3\rho _{n}\lambda
_{0}^{2}L_{a}\gamma _{s}\right) /\left( 8\pi ^{2}\Delta \right) $ with $%
N_{p}=N_{a}$. For a atomic sample of density $\rho _{n}\sim 5\times 10^{12}$%
cm$^{-3}$ and of length $L_{a}\sim 2$cm, $\kappa _{c}\sim 5$ is obtainable
with the choice $\Delta \sim 300\gamma $, and at the same time the loss $%
\varepsilon _{p}\sim \varepsilon _{a}\sim \varepsilon <1\%$. The
signal-to-noise ratio for this interaction scheme is quantified by $%
R_{sn}=\kappa _{c}^{2}/\varepsilon \sim 3\rho _{n}L_{a}/k_{0}^{2}\sim d_{o}$%
, which is more than $10^{3}$ for the above example of parameters. We will
see in Sec. V that these numbers are good enough for realizing high-fidelity
continuous variable quantum teleportation. By directly solving the
free-space problem, we get exactly the same signal-to-noise ratio as in the
two previous two level schemes after taking the free-space limit of the
cavity finesses. This clearly shows that collective enhancement of the
signal-to-noise ratio is present for all these types of light-atom
interactions, independent of the presence or absence of the optical
cavities. The collective enhancement of the signal-to-noise ratio is an
important feature of these systems, which facilitate various kinds of
quantum information processing detailed below.

\subsection{Scalable long-distance quantum communication}

Quantum communication is an essential element required for constructing
quantum networks, and it also has the application for absolutely secret
transfer of classical messages by means of quantum cryptography \cite%
{Ekert91} The central problem of quantum communication is to generate nearly
perfect entangled states between distant sites. Such states can be used, for
example, to implement secure quantum cryptography using the Ekert protocol
\cite{Ekert91}, and to faithfully transfer quantum states via quantum
teleportation \cite{Bennett93}. All the known realistic schemes for quantum
communication are based on the use of the photonic channels. However, the
degree of entanglement generated between two distant sites normally
decreases exponentially with the length of the connecting channel due to the
optical absorption and other channel noise. To regain a high degree of
entanglement, purification schemes can be used \cite{Bennett96}. However,
entanglement purification does not fully solve the long-distance quantum
communication problem. Due to the exponential decay of the entanglement in
the channel, one needs an exponentially large number of partially entangled
states to obtain one highly entangled state, which means that for a
sufficiently long distance the task becomes nearly impossible.

To overcome the difficulty associated with the exponential fidelity decay,
the concept of quantum repeaters can be used \cite{Briegel98}. In principle,
it allows to make the overall communication fidelity very close to the
unity, with the communication time growing only polynomially with the
transmission distance. In analogy to a fault-tolerant quantum computing \cite%
{Knill98,Preskill98}, the quantum repeater proposal is a cascaded
entanglement purification protocol for communication systems. The basic idea
is to divide the transmission channel into many segments, with the length of
each segment comparable to the channel attenuation length. First, one
generates entanglement and purifies it for each segment; the purified
entanglement is then extended to a longer length by connecting two adjacent
segments through entanglement swapping \cite{Bennett93,Zukowski93}. After
entanglement swapping, the overall entanglement is decreased, and one has to
purify it again. One can continue the rounds of the entanglement swapping
and purification until a nearly perfect entangled states are created between
two distant sites.

To implement the quantum repeater protocol, one needs to generate
entanglement between distant quantum bits (qubits), store them for
sufficiently long time and perform local collective operations on several of
these qubits. The requirement of quantum memory is essential since all
purification protocols are probabilistic. When entanglement purification is
performed for each segment of the channel, quantum memory can be used to
keep the segment state if the purification succeeds and to repeat the
purification for the segments only where the previous attempt fails. This is
essentially important for polynomial scaling properties of the communication
efficiency since with no available memory we have to require that the
purifications for all the segments succeeds at the same time; the
probability of such event decreases exponentially with the channel length.
The requirement of quantum memory implies that we need to store the local
qubits in the atomic internal states instead of the photonic states since it
is difficult to store photons for a reasonably long time. With atoms as the
local information carriers it seems to be very hard to implement quantum
repeaters since normally one needs to achieve the strong coupling between
atoms and photons with high-finesse cavities for atomic entanglement
generation, purification, and swapping \cite{Cirac97,Enk98}, which, in spite
of the recent significant experimental advances \cite{Ye99,Hood00,Pinkse00},
remains a very challenging technology.

To overcome this difficulty, Ref. \cite{Duan01b} proposes a very different
scheme to realize quantum repeaters based on the use of atomic ensembles
with the $\Lambda $II-level configuration. The laser manipulation of the
atomic ensembles, together with some simple linear optics devices and
moderate single-photon detectors, do the whole work for long-distance
quantum communication. The setup is much simpler compared with the
single-atom and high-Q cavity approach discussed in the previous chapter. To
achieve this, the scheme makes significant advances in each step of
entanglement generation, connection, and applications, with each step having
built-in entanglement purification and resilient to the realistic noise. As
a result, the scheme circumvents the realistic noise and imperfections, and
at the same time keeps the overhead in the communication time increasing
with the distance only polynomially. In this section, we will review the
realization of quantum repeaters and long-distance quantum communication
following the approach in Ref. \cite{Duan01b}.

\subsubsection{Entanglement generation}

To realize long-distance quantum communication, first we need to entangle
two atomic ensembles within the channel attenuation length. The entanglement
generation scheme described here is based on single-photon interference at
photodetectors, and is fault-tolerant to realistic noise. This scheme is an
extension of a proposal first proposed in \cite{Cabrillo99,Bose99} to
entangle single-atoms. The extension was made in \cite{Duan01b} to entangle
atomic ensembles with significant improvements in the communication
efficiency thanks to the collective enhancement of the signal-to-noise ratio
for many-atom ensembles.

The system is a sample of atoms prepared in the ground state $\left|
1\right\rangle $ with the $\Lambda $II-level configuration (see Fig.~ \ref%
{d63reviewp4}). It has been shown in the previous section that one can
define an effective single-mode bosonic annihilation operator $a$ for the
cavity output signal (it is called the forward-scattered Stokes signal in
the free space case). After the light-atom interaction, the signal mode $a$
and the collective atomic mode $s\equiv \left( 1/\sqrt{N_{a}}\right)
\sum_{i}\left| 1\right\rangle _{i}\left\langle 2\right| $ are in a two-mode
squeezed state with the squeezing parameter $r_{c}$ proportional to the
interaction time $t_{\Delta }$\ (see Eq. (\ref{10})). If the interaction
time $t_{\Delta }$ is very small, the whole state of the collective atomic
mode and the signal mode can be written in the perturbative form
\begin{equation}
\left| \phi \right\rangle =\left| 0_{a}\right\rangle \left|
0_{p}\right\rangle +\sqrt{p_{c}}S^{\dagger }a^{\dagger }\left|
0_{a}\right\rangle \left| 0_{p}\right\rangle +o\left( p_{c}\right) ,
\label{19}
\end{equation}
where $p_{c}=\tanh ^{2}r_{c}$ is the small excitation probability and $%
o\left( p_{c}\right) $represents the terms with more excitations whose
probabilities are equal or smaller than $p_{c}^{2}$. The $\left|
0_{a}\right\rangle $ and $\left| 0_{p}\right\rangle $ are respectively the
atomic and optical vacuum states with $\left| 0_{a}\right\rangle \equiv
\bigotimes_{i}\left| 1\right\rangle _{i}$. There is also a fraction of light
from the transition $\left| 3\right\rangle \rightarrow \left| 2\right\rangle
$ emitted in other directions which contributes to spontaneous emissions. We
have shown in the previous section that the contribution to the population
in the collective atomic mode $s$ from the spontaneous emissions is very
small for many-atom ensembles due to the collective enhancement of the
signal-to-noise ratio for this mode.

\begin{figure}[tbp]
\label{d63reviewp4}
\par
\begin{center}
\includegraphics[width=8.0cm]{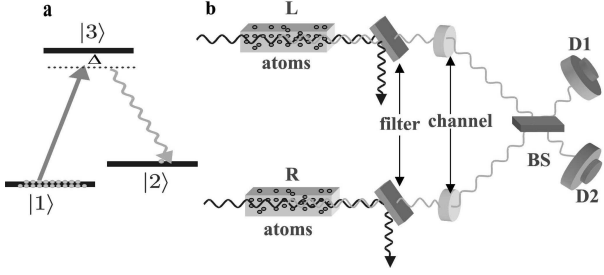}
\end{center}
\caption{(4a) The relevant level structure of the atoms in the ensemble with
$\left| 1\right\rangle $, the ground state, $\left| 2\right\rangle ,$ the
metastable state for storing a qubit, and $\left| 3\right\rangle ,$ the
excited state. The transition $\left| 1\right\rangle \rightarrow \left|
3\right\rangle $ is coupled by the classical laser with the Rabi frequency $%
\Omega $, and the forward scattering Stokes light comes from the transition $%
\left| 3\right\rangle \rightarrow \left| 2\right\rangle $. For convenience,
we assume off-resonant coupling with a large detuning $\Delta $. (4b)
Schematic setup for generating entanglement between the two atomic ensembles
L and R. The two ensembles are pencil shaped and illuminated by the
synchronized classical laser pulses. The forward-scattering Stokes pulses
are collected after the filters (polarization and frequency selective) and
interfered at a 50\%-50\% beam splitter BS after the transmission channels,
with the outputs detected respectively by two single-photon detectors D1 and
D2. If there is a click in D1 \textit{or} D2, the process is finished and we
successfully generate entanglement between the ensembles L and R. Otherwise,
we first apply a repumping pulse to the transition $\left| 2\right\rangle
\rightarrow \left| 3\right\rangle $ on the ensembles L and R to set the
state of the ensembles back to the ground state $\left| 0\right\rangle
_{a}^{L}\otimes \left| 0\right\rangle _{a}^{R}$, then the same classical
laser pulses as the first round are applied to the transition $\left|
1\right\rangle \rightarrow \left| 3\right\rangle $ and we detect again the
forward-scattering Stokes pulses after the beam splitter. This process is
repeated until finally we have a click in the D1 \textit{or} D2 detector. }
\end{figure}

Now we show how to use this setup to generate entanglement between two
distant ensembles L and R using the configuration shown in Fig.~ \ref%
{d63reviewp4}. Here, two laser pulses excited both ensembles simultaneously,
and the whole system is described by the state $\left| \phi \right\rangle
_{L}\otimes \left| \phi \right\rangle _{R}$, where $\left| \phi
\right\rangle _{L}$ and $\left| \phi \right\rangle _{R}$ are given by Eq. (%
\ref{19}) with all the operators and states distinguished by the subscript L
or R. The forward scattered Stokes signal from both ensembles is combined at
the beam splitter and a photodetector click in either D1 \textit{or} D2
measures the combined radiation from two samples, $a_{+}^{\dagger }a_{+}$ or
$a_{-}^{\dagger }a_{-}$ with $a_{\pm }=\left( a_{L}\pm e^{i\varphi
}a_{R}\right) /\sqrt{2}$. Here, $\varphi $ denotes an unknown difference of
the phase shifts in the two-side channels. We can also assume that $\varphi $
has an imaginary part to account for the possible asymmetry of the setup,
which will also be corrected automatically in our scheme. But the setup
asymmetry can be easily made very small, and for simplicity of expressions
we assume that $\varphi $ is real in the following. Conditional on the
detector click, we should apply $a_{+}$ or $a_{-}$ to the whole state $%
\left| \phi \right\rangle _{L}\otimes \left| \phi \right\rangle _{R}$, and
the projected state of the ensembles L and R is nearly maximally entangled
with the form (neglecting the high-order terms $o\left( p_{c}\right) $)
\begin{equation}
\left| \Psi _{\varphi }\right\rangle _{LR}^{\pm }=\left( S_{L}^{\dagger }\pm
e^{i\varphi }S_{R}^{\dagger }\right) /\sqrt{2}\left| 0_{a}\right\rangle
_{L}\left| 0_{a}\right\rangle _{R}.  \label{20}
\end{equation}
The probability for getting a click is given by $p_{c}$ for each round, so
we need repeat the process about $1/p_{c}$ times for a successful
entanglement preparation, and the average preparation time is given by $%
T_{0}\sim t_{\Delta }/p_{c}$. The states $\left| \Psi _{r}\right\rangle
_{LR}^{+}$ and $\left| \Psi _{r}\right\rangle _{LR}^{-}$ can be easily
transformed to each other by a simple local phase shift. Without loss of
generality, we assume in the following that we generate the entangled state $%
\left| \Psi _{r}\right\rangle _{LR}^{+}$.

As will be shown below, the presence of the noise modifies the projected
state of the ensembles to
\begin{equation}
\rho _{LR}\left( c_{0},\varphi \right) =\frac{1}{c_{0}+1}\left( c_{0}\left|
0_{a}0_{a}\right\rangle _{LR}\left\langle 0_{a}0_{a}\right| +\left| \Psi
_{\varphi }\right\rangle _{LR}^{\text{ }+}\left\langle \Psi _{\varphi
}\right| \right) ,  \label{21}
\end{equation}
where the ``vacuum'' coefficient $c_{0}$ is determined by the dark count
rates of the photon detectors. It will be seen below that any state in the
form of Eq. (\ref{21}) will be purified automatically to a maximally
entangled state in the entanglement-based communication schemes. We
therefore call this state an effective maximally entangled (EME) state with
the vacuum coefficient $c_{0}$ determining the purification efficiency.

\subsubsection{Entanglement connection through swapping}

\begin{figure}[tbp]
\label{d63reviewp5}
\par
\begin{center}
\includegraphics[width=8.0cm]{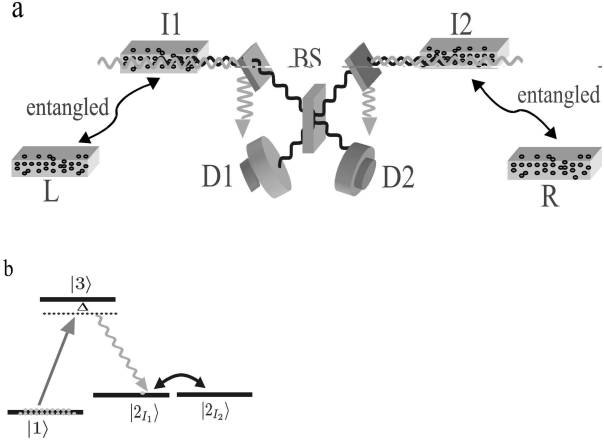}
\end{center}
\caption{(5a) Illustrative setup for the entanglement swapping. We have two
pairs of ensembles L, I$_{1}$ and I$_{2}$, R distributed at three sites L, I
and R. Each of the ensemble-pairs L, I$_{1}$ and I$_{2}$, R is prepared in
an EME state in the form of Eq. (3). The excitations in the collective modes
of the ensembles I$_{1}$ and I$_{2}$ are transferred simultaneously to the
optical excitations by the repumping pulses applied to the atomic transition
$\left| 2\right\rangle \rightarrow \left| 3\right\rangle $, and the
stimulated optical excitations, after a 50\%-50\% beam splitter, are
detected by the single-photon detectors D1 and D2. If either D1 \textit{or}
D2 clicks, the protocol is successful and an EME state in the form of Eq.
(3) is established between the ensembles L and R with a doubled
communication distance. Otherwise, the process fails, and we need to repeat
the previous entanglement generation and swapping until finally we have a
click in D1 or D2, that is, until the protocol finally succeeds. (5b) The
two intermediated ensembles I$_{1}$ and I$_{2}$ can also be replaced by one
ensemble but with two metastable states I$_{1}$ and I$_{2}$ to store the two
different collective modes. The 50\%-50\% beam splitter operation can be
simply realized by a $\protect\pi /2$ pulse on the two metastable states
before the collective atomic excitations are transferred to the optical
excitations. }
\end{figure}

After the successful generation of the entanglement within the attenuation
length, we want to extend the quantum communication distance. This is done
through entanglement swapping with the configuration shown in Fig.~ \ref%
{d63reviewp5}. Suppose that we start with two pairs of the entangled
ensembles described by the state $\rho _{LI_{1}}\otimes \rho _{I_{2}R}$,
where $\rho _{LI_{1}}$ and $\rho _{I_{2}R}$ are given by Eq. (\ref{21}). In
the ideal case, the setup shown in Fig.~ \ref{d63reviewp5} measures the
quantities corresponding to operators $S_{\pm }^{\dagger }S_{\pm }$ with $%
S_{\pm }=\left( S_{I_{1}}\pm S_{I_{2}}\right) /\sqrt{2}$. If the measurement
is successful (i.e., one of the detectors registers one photon), we will
prepare the ensembles L and R into another EME state. The new $\varphi $%
-parameter is given by $\varphi _{1}+\varphi _{2}$, where $\varphi _{1}$ and
$\varphi _{2}$ denote the old $\varphi $-parameters for the two segment EME
states. As will be seen below, even in the presence of the realistic noise
and imperfections, an EME state is still created after a detector click. The
noise only influences the success probability to get a click and the new
vacuum coefficient in the EME state. In general we can express the success
probability $p_{1}$ and the new vacuum coefficient $c_{1}$ as $%
p_{1}=f_{1}\left( c_{0}\right) $ and $c_{1}=f_{2}\left( c_{0}\right) $,
where the functions $f_{1}$ and $f_{2}$ depend on the particular noise
properties.

The above method for connecting entanglement can be cascaded to arbitrarily
extend the communication distance. For the $i$th ($i=1,2,\cdots ,n$)
entanglement connection, we first prepare in parallel two pairs of ensembles
in the EME states with the same vacuum coefficient $c_{i-1}$ and the same
communication length $L_{i-1}$, and then perform the entanglement swapping
as shown in Fig.~ \ref{d63reviewp5}, which now succeeds with a probability $%
p_{i}=f_{1}\left( c_{i-1}\right) $. After a successful detector click, the
communication length is extended to $L_{i}=2L_{i-1}$, and the vacuum
coefficient in the connected EME\ state becomes $c_{i}=f_{2}\left(
c_{i-1}\right) $. Since the $i$th entanglement connection need be repeated
in average $1/p_{i}$ times, the total time needed to establish an EME state
over the distance $L_{n}=2^{n}L_{0}$ is given by $T_{n}=T_{0}\prod_{i=1}^{n}%
\left( 1/p_{i}\right) $, where $L_{0}$ denotes the distance of each segment
in the entanglement generation.

\subsubsection{Entanglement-based communication schemes}

After an EME\ state has been established between two distant sites, we would
like to use it in the communication protocols, such as quantum
teleportation, cryptography, and Bell inequality detection. It is not
obvious that the EME state (\ref{21}), which is entangled in the Fock basis,
is useful for these tasks since in the Fock basis it is experimentally hard
to do certain single-bit operations. In the following we will show how the
EME\ states can be used to realize all these protocols with simple
experimental configurations.

\begin{figure}[tbp]
\label{d63reviewp6}
\par
\begin{center}
\includegraphics[width=8.0cm]{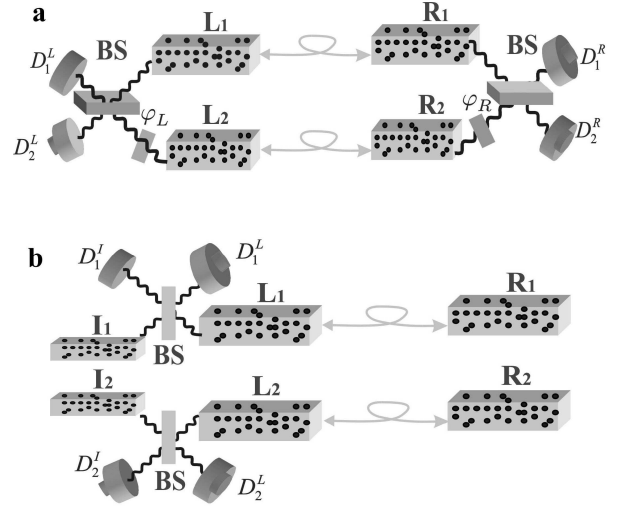}
\end{center}
\caption{(6a) Schematic setup for the realization of quantum cryptography
and Bell inequality detection. Two pairs of ensembles L$_{1}$, R$_{1}$ and L$%
_{2}$, R$_{2}$ (or two pairs of metastable states as shown by Fig.~ (\protect
\ref{d63reviewp2}b)) have been prepared in the EME states. The collective
atomic excitations on each side are transferred to the optical excitations,
which, respectively after a relative phase shift $\protect\varphi _{L}$ or $%
\protect\varphi _{R}$ and a 50\%-50\% beam splitter, are detected by the
single-photon detectors $D_{1}^{L},D_{2}^{L}$ and $D_{1}^{R},D_{2}^{R}$. We
look at the four possible coincidences of $D_{1}^{R},D_{2}^{R}$ with $%
D_{1}^{L},D_{2}^{L}$, which are functions of the phase difference $\protect%
\varphi _{L}-\protect\varphi _{R}$. Depending on the choice of $\protect%
\varphi _{L}$ and $\protect\varphi _{R} $, this setup can realize both the
quantum cryptography and the Bell inequality detection. (6b) Schematic setup
for probabilistic quantum teleportation of the atomic ``polarization''
state. Similarly, two pairs of ensembles L$_{1}$, R$_{1}$ and L$_{2}$, R$%
_{2} $ are prepared in the EME states. We want to teleport an atomic
``polarization'' state $\left( d_{0}S_{I_{1}}^{\dagger
}+d_{1}S_{I_{2}}^{\dagger }\right) \left| 0_{a}0_{a}\right\rangle
_{I_{1}I_{2}}$ with unknown coefficients $d_{0},d_{1} $ from the left to the
right side, where $S_{I_{1}}^{\dagger },S_{I_{2}}^{\dagger }$ denote the
collective atomic operators for the two ensembles I$_{1}$ and I$_{2}$ (or
two metastable states in the same ensemble). The collective atomic
excitations in the ensembles I$_{1}$, L$_{1} $ and I$_{2}$, L$_{2}$ are
transferred to the optical excitations, which, after a 50\%-50\% beam
splitter, are detected by the single-photon detectors $D_{1}^{I},D_{1}^{L}$
and $D_{2}^{I},D_{2}^{L}$. If there are a click in $D_{1}^{I}$ \textit{or }$%
D_{1}^{L}$ and a click in $D_{2}^{I}$ \textit{or }$D_{2}^{I}$, the protocol
is successful. A $\protect\pi $-phase rotation is then performed on the
collective mode of the ensemble R$_{2}$ conditional on that the two clicks
appear in the detectors $D_{1}^{I}$,$D_{2}^{L}$ or $D_{2}^{I}$,$D_{1}^{L}$.
The collective excitation in the ensembles R$_{1}$ and R$_{2}$, if
appearing, would be found in the same ``polarization'' state $\left(
d_{0}S_{R_{1}}^{\dagger }+d_{1}S_{R_{2}}^{\dagger }\right) \left|
0_{a}0_{a}\right\rangle _{R_{1}R_{2}}$. }
\end{figure}

Quantum cryptography and the Bell inequality detection are achieved with the
setup shown by Fig.~ \ref{d63reviewp6}a. The state of the two pairs of
ensembles is expressed as $\rho _{L_{1}R_{1}}\otimes \rho _{L_{2}R_{2}}$,
where $\rho _{L_{i}R_{i}}$ $\left( i=1,2\right) $ denote the same EME state
with the vacuum coefficient $c_{n}$ if we have done $n$ times entanglement
connection. The $\varphi $-parameters in $\rho _{L_{1}R_{1}}$ and $\rho
_{L_{2}R_{2}}$ are the same provided that the two states are established
over the same stationary channels. We register only the coincidences of the
two-side detectors, so the protocol is successful only if there is a click
on each side. Under this condition, the vacuum components in the EME states,
together with the state components $S_{L_{1}}^{\dagger }S_{L_{2}}^{\dagger
}\left| \text{vac}\right\rangle $ and $S_{R_{1}}^{\dagger
}S_{R_{2}}^{\dagger }\left| \text{vac}\right\rangle $, where $\left| \text{%
vac}\right\rangle $ denotes the ensemble state $\left|
0_{a}0_{a}0_{a}0_{a}\right\rangle _{L_{1}R_{1}L_{2}R_{2}}$, have no
contributions to the experimental results. So, for the measurement scheme
shown by Fig.~ \ref{d63reviewp3}, the ensemble state $\rho
_{L_{1}R_{1}}\otimes \rho _{L_{2}R_{2}}$ is effectively equivalent to the
following ``polarization'' maximally entangled (PME) state (the terminology
of ``polarization'' comes from an analogy to the optical case)
\begin{equation}
\left| \Psi \right\rangle _{\text{PME}}=\left( S_{L_{1}}^{\dagger
}S_{R_{2}}^{\dagger }+S_{L_{2}}^{\dagger }S_{R_{1}}^{\dagger }\right) /\sqrt{%
2}\left| \text{vac}\right\rangle .  \label{22}
\end{equation}
The success probability for the projection from $\rho _{L_{1}R_{1}}\otimes
\rho _{L_{2}R_{2}}$ to $\left| \Psi \right\rangle _{\text{PME}}$ (i.e., the
probability to get a click on each side) is given by $p_{a}=1/[2\left(
c_{n}+1\right) ^{2}]$. One can also check that in Fig.~ \ref{d63reviewp6},
the phase shift $\psi _{\Lambda }$ $\left( \Lambda =L\text{ or }R\right) $
together with the corresponding beam splitter operation are equivalent to a
single-bit rotation in the basis $\left\{ \left| 0\right\rangle _{\Lambda
}\equiv S_{\Lambda _{1}}^{\dagger }\left| 0_{a}0_{a}\right\rangle _{\Lambda
_{1}\Lambda _{2}},\text{ }\left| 1\right\rangle _{\Lambda }\equiv S_{\Lambda
_{2}}^{\dagger }\left| 0_{a}0_{a}\right\rangle _{\Lambda _{1}\Lambda
_{2}}\right\} $ with the rotation angle $\theta =\psi _{\Lambda }/2$. Now,
it is clear how to do quantum cryptography and Bell inequality detection
since we have the PME state and we can perform the desired single-bit
rotations in the corresponding basis. For instance, to distribute a quantum
key between the two remote sides, we simply choose $\psi _{\Lambda }$
randomly from the set $\left\{ 0,\pi /2\right\} $ with an equal probability,
and keep the measurement results (to be $0$ if $D_{1}^{\Lambda }$ clicks,
and $1$ if $D_{1}^{\Lambda }$ clicks) on both sides as the shared secret key
if the two sides become aware that they have chosen the same phase shift
after the public declare. This is exactly the Ekert scheme \cite{Ekert91}
and its absolute security follows directly from the proofs in \cite%
{Lo99,Shor00}. For the Bell inequality detection, we infer the correlations $%
E\left( \psi _{L},\psi _{R}\right) \equiv
P_{D_{1}^{L}D_{1}^{R}}+P_{D_{2}^{L}D_{2}^{R}}-P_{D_{1}^{L}D_{2}^{R}}-P_{D_{2}^{L}D_{1}^{R}}=\cos \left( \psi _{L}-\psi _{R}\right)
$ from the measurement of the coincidences $P_{D_{1}^{L}D_{1}^{R}}$ etc. For
the setup shown in Fig.~ \ref{d63reviewp6}a, we would have $\left| E\left(
0,\pi /4\right) +E\left( \pi /2,\pi /4\right) +E\left( \pi /2,3\pi /4\right)
-E\left( 0,3\pi /4\right) \right| =2\sqrt{2}$, whereas for any local hidden
variable theories, the CHSH inequality \cite{Clauser69} implies that this
value should be below $2$.

We can also use the established long-distance EME states for faithful
transfer of unknown quantum states through quantum teleportation, with the
setup shown by Fig.~ \ref{d63reviewp6}b. In this setup, if two detectors
click on the left side, there is a significant probability that there is no
collective excitation on the right side since the product of the EME states $%
\rho _{L_{1}R_{1}}\otimes \rho _{L_{2}R_{2}}$ contains vacuum components.
However, if there is a collective excitation appearing from the right side,
its ``polarization'' state would be exactly the same as the one input from
the left. So, as in the Innsbruck experiment \cite{Boumeester97}, the
teleportation here is probabilistic and needs posterior confirmation; but if
it succeeds, the teleportation fidelity would be nearly perfect since in
this case the entanglement is equivalently described by the PME state (\ref%
{22}). The success probability for the teleportation is also given by $%
p_{a}=1/[2\left( c_{n}+1\right) ^{2}]$, which determines the average number
of repetitions for a successful teleportation.

\subsubsection{Noise and built-in entanglement purification}

We next discuss noise and imperfections in the schemes for entanglement
generation, connection, and applications. In particular we show that each
step contains built-in entanglement purification which makes the whole
scheme resilient to the realistic noise and imperfections.

In the entanglement generation, the dominant noise is the photon loss, which
includes the contributions from the channel attenuation, the spontaneous
emissions in the atomic ensembles (which results in the population of the
collective atomic mode $s$ with the accompanying photon going to other
directions), the coupling inefficiency of the Stokes signal into and out of
the channel, and the inefficiency of the single-photon detectors. The loss
probability is denoted by $1-\eta _{p}$ with the overall efficiency $\eta
_{p}=\eta _{p}^{\prime }e^{-L_{0}/L_{\text{att}}}$, where we have separated
the channel attenuation $e^{-L_{0}/L_{\text{att}}}$ ($L_{\text{att}}$ is the
channel attenuation length) from other noise contributions $\eta
_{p}^{\prime }$ with $\eta _{p}^{\prime }$ independent of the communication
distance $L_{0}$. The photon loss decreases the success probably for getting
a detector click from $p_{c}$ to $\eta _{p}p_{c}$, but it has no influence
on the resulting EME state. Due to this noise, the entanglement preparation
time should be replaced by $T_{0}\sim t_{\Delta }/\left( \eta
_{p}p_{c}\right) $. The second source of noise comes from the dark counts of
the single-photon detectors. The dark count gives a detector click, but
without population of the collective atomic mode, so it contributes to the
vacuum coefficient in the EME state. If the dark count comes up with a
probability $p_{dc}$ for the time interval $t_{\Delta }$, the vacuum
coefficient is given by $c_{0}=p_{dc}/\left( \eta _{p}p_{c}\right) $, which
is typically much smaller than $1$ since the Raman transition rate is much
larger than the dark count rate. The final source of noise, which influences
the fidelity to get the EME state, is caused by the event that more than one
atom are excited to the collective mode $S$ whereas there is only one click
in D1 or D2. The conditional probability for that event is given by $p_{c}$,
so we can estimate the fidelity imperfection $\Delta F_{0}\equiv 1-F_{0}$
for the entanglement generation by
\begin{equation}
\Delta F_{0}\sim p_{c}.  \label{23}
\end{equation}
Note that by decreasing the excitation probability $p_{c}$, one can make the
fidelity imperfection closer and closer to zero with the price of a longer
entanglement preparation time $T_{0}$. This is the basic idea of the
entanglement purification. So, in this scheme, the confirmation of the click
from the single-photon detector generates and purifies entanglement at the
same time.

In the entanglement swapping, the dominant noise is still the losses, which
include the contributions from the detector inefficiency, the inefficiency
of the excitation transfer from the collective atomic mode to the optical
mode \cite{Liu01,Phillips01}, and the small decay of the atomic excitation
during the storage \cite{Liu01,Phillips01}. Note that by introducing the
detector inefficiency, we have automatically taken into account the
imperfection that the detectors cannot distinguish the single and the two
photons. With all these losses, the overall efficiency in the entanglement
swapping is denoted by $\eta _{s}$. The loss in the entanglement swapping
gives contributions to the vacuum coefficient in the connected EME state,
since in the presence of loss a single detector click might result from two
collective excitations in the ensembles I$_{1}$ and I$_{2}$, and in this
case, the collective modes in the ensembles L and R have to be in a vacuum
state. After taking into account the realistic noise, we can specify the
success probability and the new vacuum coefficient for the $i$th
entanglement connection by the recursion relations $p_{i}\equiv f_{1}\left(
c_{i-1}\right) =\eta _{s}\left( 1-\frac{\eta _{s}}{2\left( c_{i-1}+1\right) }%
\right) /\left( c_{i-1}+1\right) $ and $c_{i}\equiv f_{2}\left(
c_{i-1}\right) =2c_{i-1}+1-\eta _{s}$. The coefficient $c_{0}$ for the
entanglement preparation is typically much smaller than $1-\eta _{s}$, then
we have $c_{i}\approx \left( 2^{i}-1\right) \left( 1-\eta _{s}\right)
=(L_{i}/L_{0}-1)\left( 1-\eta _{s}\right) $, where $L_{i}$ denotes the
communication distance after $i$ times entanglement connection. With the
expression for the $c_{i}$, we can easily evaluate the probability $p_{i}$
and the communication time $T_{n}$ for establishing a EME state over the
distance $L_{n}=2^{n}L_{0}$. After the entanglement connection, the fidelity
of the EME state also decreases, and after $n$ times connection, the overall
fidelity imperfection $\Delta F_{n}\sim 2^{n}\Delta F_{0}\sim \left(
L_{n}/L_{0}\right) \Delta F_{0}$. We need fix $\Delta F_{n}$ to be small by
decreasing the excitation probability $p_{c}$ in Eq. (\ref{23}).

It is important to point out that our entanglement connection scheme also
has built-in entanglement purification function. This can be understood as
follows: Each time we connect entanglement, the imperfections of the setup
decrease the entanglement fraction $1/\left( c_{i}+1\right) $ in the EME
state. However, the entanglement fraction decays only linearly with the
distance (the number of segments), which is in contrast to the exponential
decay of the entanglement for the connection schemes without entanglement
purification. The reason for the slow decay is that in each time of the
entanglement connection, we need repeat the protocol until there is a
detector click, and the confirmation of \ a click removes part of the added
vacuum noise since a larger vacuum components in the EME state results in
more times of repetitions. The built-in entanglement purification in the
connection scheme is essential for the polynomial scaling law of the
communication efficiency.

As in the entanglement generation and connection schemes, our entanglement
application schemes also have built-in entanglement purification which makes
them resilient to the realistic noise. Firstly, we have seen that the vacuum
components in the EME states are removed from the confirmation of the
detector clicks and thus have no influence on the fidelity of all the
application schemes. Secondly, if the single-photon detectors and the
atom-to-light excitation transitions in the application schemes are
imperfect with the overall efficiency denoted by $\eta _{a}$, one can easily
check that these imperfections only influence the efficiency to get the
detector clicks with the success probability replaced by $p_{a}=\eta _{a}/%
\left[ 2\left( c_{n}+1\right) ^{2}\right] $, and have no effects on the
communication fidelity. Finally, we have seen that the phase shifts in the
stationary channels and the small asymmetry of the stationary setup are
removed automatically when we project the EME state to the PME state, and
thus have no influence on the communication fidelity.

The noise not correctable by our scheme includes the detector dark count in
the entanglement connection and the non-stationary channel noise and set
asymmetries. The resulting fidelity imperfection from the dark count
increases linearly with the number of segments $L_{n}/L_{0}$, and form the
non-stationary channel noise and set asymmetries increases by the
random-walk law $\sqrt{L_{n}/L_{0}}$. For each time of entanglement
connection, the dark count probability is about $10^{-5}$ if we make a
typical choice that the collective emission rate is about $10$MHz and the
dark count rate is $10^{2}$Hz. So this noise is negligible even if we have
communicated over a long distance ($10^{3}$ the channel attenuation length $%
L_{\text{att}}$ for instance). The non-stationary channel noise and setup
asymmetries can also be safely neglected for such a distance. For instance,
it is relatively easy to control the non-stationary asymmetries in local
laser operations to values below $10^{-4}$ with the use of accurate
polarization techniques \cite{Budker98} for Zeeman sublevels (as in Fig.~ %
\ref{d63reviewp5}b).

\subsubsection{Scaling of the communication efficiency}

We have shown that each of our entanglement generation, connection, and
application schemes has built-in entanglement purification, and as a result
of this property, we can fix the communication fidelity to be nearly
perfect, and at the same time keep the communication time to increase only
polynomially with the distance. Assume that we want to communicate over a
distance $L=L_{n}=2^{n}L_{0}$. By fixing the overall fidelity imperfection
to be a desired small value $\Delta F_{n}$, the entanglement preparation
time becomes $T_{0}\sim t_{\Delta }/\left( \eta _{p}\Delta F_{0}\right) \sim
\left( L_{n}/L_{0}\right) t_{\Delta }/\left( \eta _{p}\Delta F_{n}\right) $.
For an effective generation of the PME state (\ref{22}), the total
communication time $T_{\text{tot}}\sim T_{n}/p_{a}$ $\ $with $T_{n}\sim
T_{0}\prod_{i=1}^{n}\left( 1/p_{i}\right) $. So the total communication time
scales with the distance by the law
\begin{equation}
T_{\text{tot}}\sim 2\left( L/L_{0}\right) ^{2}/\left( \eta _{p}p_{a}\Delta
F_{T}\Pi _{i=1}^{n}p_{i}\right) ,  \label{24}
\end{equation}
where the success probabilities $p_{i},p_{a}$ for the $i$th entanglement
connection and for the entanglement application have been specified before.
The expression (\ref{24}) has confirmed that the communication time $T_{%
\text{tot}}$ increases with the distance $L$ only polynomially. We show this
explicitly by taking two limiting cases. In the first case, the inefficiency
$1-\eta _{s}$ for the entanglement swapping is assumed to be negligibly
small. One can deduce from Eq. (\ref{24}) that in this case the
communication time $T_{\text{tot}}\sim T_{\text{con}}\left( L/L_{0}\right)
^{2}e^{L_{0}/L_{\text{att}}}$, with the constant $T_{\text{con}}\equiv
2t_{\Delta }/\left( \eta _{p}^{\prime }\eta _{a}\Delta F_{T}\right) $ being
independent of the segment and the total distances $L_{0}$ and $L$. The
communication time $T_{\text{tot}}$ increases with $L$ quadratically. In the
second case, we assume that the inefficiency $1-\eta _{s}$ is considerably
large. The communication time in this case is approximated by $T_{\text{tot}%
}\sim T_{\text{con}}(L/L_{0})^{[\log _{2}\left( L/L_{0}\right) +1]/2+\log
_{2}(1/\eta _{s}-1)+2}e^{L_{0}/L_{\text{att}}}$, which increases with $L$
still polynomially (or sub-exponentially in a more accurate language, but
this makes no difference in practice since the factor $\log _{2}\left(
L/L_{0}\right) $ is well bounded from above for any reasonably long
distance). If $T_{\text{tot}}$ increases with $L/L_{0}$ by the $m$th power
law $\left( L/L_{0}\right) ^{m}$, there is an optimal choice of the segment
length to be $L_{0}=mL_{\text{att}}$ to minimize the time $T_{\text{tot}}$.
As a simple estimation of the improvement in the communication efficiency,
we assume that the total distance $L$ is about $100L_{\text{att}}$, for a
choice of the parameter $\eta _{s}\approx 2/3$, the communication time $T_{%
\text{tot}}/T_{\text{con}}\sim 10^{6}$ with the optimal segment length $%
L_{0}\sim 5.7L_{\text{att}}$. This result is a dramatic improvement compared
with the direct communication case, where the communication time $T_{\text{%
tot}}$ for getting a PME state increases with the distance $L$ by the
exponential law $T_{\text{tot}}\sim T_{\text{con}}e^{L/L_{\text{att}}}$. For
the same distance $L\sim 100L_{\text{att}}$, one needs $T_{\text{tot}}/T_{%
\text{con}}\sim 10^{43}$ for direct communication, which means that for this
example the present scheme is $10^{37}$ times more efficient .

In summary, in this section we explained the recent atomic ensemble scheme
for implementation of quantum repeaters and long-distance quantum
communication. The proposed technique allows to generate and connect the
entanglement and use it in quantum teleportation, cryptography, and tests of
Bell inequalities. All of the elements of the scheme are within the reach of
current experimental technology, and have the important property of built-in
entanglement purification which makes them resilient to the realistic noise.
As a result, the overhead required to implement the scheme, such as the
communication time, scales polynomially with the channel length. This is in
remarkable contrast to direct communication where the exponential overhead
is required. Such an efficient scaling, combined with a relative simplicity
of the proposed experimental setup, opens up realistic prospective for
quantum communication over long distances.

\subsection{Other Applications: quantum light memory and single-photon source%
}

In this section, we investigate two other significant applications of atomic
ensembles in quantum information processing: laser manipulation of atomic
ensembles provides a simple experimentally feasible way to realize quantum
light memory and single-photon source with controllable emission time,
direction, and pulse shape. The realization of quantum light memory and
controllable single-photon source constitutes two important steps towards
implementation of a recently proposed quantum computation scheme \cite%
{Knill01}. In \cite{Knill01}, a potentially scalable fault-tolerant quantum
computation scheme is proposed based on the use of single-photon source,
linear optics devices, and single-photon detectors. To realize that scheme,
one is required (i) to have the ability of storing qubits (that is, quantum
light memory is required to store optical qubits involved in that scheme),
(ii) to have desired single-photon source with controllable emission time
and direction, (iii) and to maintain noise and imperfections in the involved
physical setups below the percent level. Except the last requirement, which
is still very challenging with the current experimental technology, atomic
ensembles provide an ideal system for realization of the first two elements.
Besides this potential application in quantum computation, quantum light
memory and single-photon source are also important by themselves. For
instance, single-photon source is important in the BB84 scheme for quantum
key distribution to assure the absolute security and to increase the
distribution efficiency \cite{Brassard00}; and quantum light memory provides
a powerful tool for eavesdropping in quantum cryptography.

\subsubsection{Quantum light memory}

It is very hard to directly store photons for a reasonably long time.
However, we know that coherence of atomic internal states can be maintained
for a quite long time with the current technology. The basic idea of quantum
light memory is to transfer the photonic excitation to the excitation in
atomic internal states so that it can be saved, and afterwards, it should be
possible to restore the excitation to photons without change of its quantum
state. Quantum light memory has been investigated theoretically in Refs.
\cite{Kozhekin99,Lukin001,Duan001,Fleischhauer00}, and its experimental
realization has been recently reported with either an ultracold
Bose-Einstein condensate \cite{Liu01} or a hot atomic ensemble \cite%
{Phillips01} as the storing medium. Ref. \cite{Kozhekin99} described a
method for irreversible mapping of the light state to the atomic state, and
Refs. \cite{Lukin001,Duan001} proposed the first quantum light memory
schemes with revisable mapping between photonic and atomic states. Both
schemes use the $\Lambda $I-level configuration with a weak coupling cavity
around the ensemble as described in Sec. II. The difference is that Ref.
\cite{Lukin001} is based on resonant coupling through adiabatic passages of
dark states, and Ref. \cite{Duan001} is based on off-resonant coherent Raman
absorption. Ref. \cite{Fleischhauer00} described the first scheme for
storing light in a free-space atomic ensemble, which has been realized in
the recent experiments \cite{Liu01,Phillips01}. Here, to be consistent with
other sections in this chapter, we will follow the off-resonant approach in
Ref. \cite{Duan001} to review the principle for implementing quantum light
memory. The readers interested in the schemes based on adiabatic passages
are referred to Refs. \cite{Lukin001,Fleischhauer00}.

We consider an atomic ensemble with the $\Lambda $I-level configuration as
shown in Sec. 4.2 (see Fig.~ \ref{d63reviewp1}). The input quantum optical
signal is described by a continuous operator $a_{in}\left( t\right) $, with $%
\left[ a_{in}\left( t\right) ,a_{in}^{+}\left( t^{^{\prime }}\right) \right]
=\delta \left( t-t^{^{\prime }}\right) $. We assume that input signal has a
definite pulse shape $f_{in}\left( t\right) $. This is the case in most of
the applications in quantum information processing. For instance, if we know
that the signal comes from the output of a free cavity mode, the signal
pulse has the shape $f_{in}\left( t\right) =\frac{\sqrt{\eta }}{\sqrt{%
1-e^{-\kappa T}}}e^{-\kappa t/2},$ $\left( 0\leq t\leq T\right) $, where $%
\kappa $ is the cavity decay rate and $T$ denotes the pulse duration. For
the input pulse with a definite shape, we can define an effective
single-mode bosonic operator $c_{in}=\int_{0}^{T}f_{1}\left( t\right)
a_{in}\left( t\right) dt$ with $\left[ c_{in},c_{in}^{\dagger }\right] =1$
and a normalized shape function $f_{in}\left( t\right) $. Similarly for the
output optical signal $a_{out}\left( t\right) $ with a definite shape $%
f_{out}\left( t\right) $, we can also define a single-mode operator $%
c_{out}=\int_{0}^{T}f_{out}\left( t\right) a_{out}\left( t\right) dt$. The
purpose of quantum light memory is to faithfully transfer the quantum state
of the input optical mode $c_{in}$ to the state of the collective atomic
mode $s\equiv \left( 1/\sqrt{N_{a}}\right) \sum_{i=1}^{N_{a}}\left|
1\right\rangle _{i}\left\langle 2\right| $. The state can be stored in the
atomic mode $s$, and afterwards we need to have ability to read out this
state again to the output optical mode $c_{out}$ with an arbitrary
intentionally chosen pulse shape $f_{out}\left( t\right) $.

The light-atom interaction is described by the basic Langevin equation (\ref%
{4}), with the effective coupling rate $\kappa ^{\prime }\left( t\right)
=4N_{a}\left| \Omega _{2}\left( t\right) g_{1}\right| ^{2}/\left( \Delta
^{2}\kappa \right) $ adjustable by controlling the time dependence of the
Rabi frequency $\Omega _{2}\left( t\right) $ through change of the classical
laser intensity. Equation (\ref{4}) is linear and has the following simple
solution
\begin{eqnarray}
s\left( T\right) &=&s\left( 0\right) e^{-\int_{0}^{T}\kappa ^{\prime }\left(
t\right) dt/2}  \notag \\
&&-\int_{0}^{T}e^{-\int_{t}^{T}\kappa ^{\prime }\left( \tau \right) d\tau
/2}a_{in}\left( t\right) \sqrt{\kappa ^{\prime }\left( t\right) }dt.
\label{25}
\end{eqnarray}
To map the state of the input optical mode $c_{in}$ to the atomic mode $s$,
we choose the form of $\Omega _{2}\left( t\right) $\ so that the coupling
rate $\kappa ^{\prime }\left( t\right) $ satisfies the differential equation
\begin{equation}
\overset{.}{\kappa ^{\prime }}=\frac{2\overset{.}{f}_{in}}{f_{in}}\kappa
^{\prime }-\kappa ^{\prime 2},  \label{26}
\end{equation}
where $f_{in}\left( t\right) $ is the shape of the input pulse. This
differential equation is sometimes called the impedance matching condition
\cite{Lukin001,Fleischhauer01}. Under this condition, it can be seen from
Eq. (\ref{25}) that the collective atomic operator $s_{T}$ at time $T$ is
given by
\begin{equation}
s_{T}=s\left( T\right) =\sqrt{M}s\left( 0\right) -\sqrt{1-M}c_{in}
\label{27}
\end{equation}
with the mapping inefficiency $M=e^{-\int_{0}^{T}\kappa ^{\prime }\left(
t\right) dt}$. Since $\kappa ^{\prime }\left( t\right) $ is positive, the
mapping inefficiency $M$ quickly tends to zero after a sufficiently long
time $T$. With a zero $M$, the photonic state is faithfully mapped to the
atomic state with $s_{T}=-c_{in}$. For a non-zero $M$, there will be a small
inherent loss to the state mapping caused by the vacuum noise $s\left(
0\right) $.

After mapping of the photonic state to the atomic state, the classical laser
is turned off and the information can be stored in the atomic internal mode $%
s_{T}$. Then, after some time we want to read out this state to an output
optical pulse with an intentionally chosen pulse shape $f_{out}\left(
t\right) ,$ that is, we would like to map the state of the atomic mode $%
s_{T} $ to the output optical mode $c_{out}=\int_{0}^{T}f_{out}\left(
t\right) a_{out}\left( t\right) dt.$ To attain this goal, we turn on the
classical laser to the transition $\left| 2\right\rangle \rightarrow \left|
3\right\rangle $ (see Fig.~ \ref{d63reviewp1}), and control its intensity so
that the effective coupling rate $\kappa ^{\prime }\left( t\right) $
satisfies the equation $\overset{.}{\kappa ^{\prime }}=\frac{2\overset{.}{f}%
_{out}}{f_{out}}\kappa ^{\prime }+\kappa ^{\prime 2}$, which is the time
reverse of the impedance matching condition (\ref{26}) for write-in of the
photonic state. With this condition, one can deduce from Eq. (\ref{25}) and
the input-output relation $a_{out}\left( t\right) =-a_{in}\left( t\right) -%
\sqrt{\kappa ^{\prime }}s$ that after time $T$, the outgoing optical mode is
given by $c_{out}=-\sqrt{M}c_{T}-\sqrt{1-M}s_{T}$, where $c_{T}$ is a
single-mode vacuum noise operator. The matching inefficiency $M$ has the
same form as before, and tends to zero for a sufficiently long time $T$. In
this case, the atomic state is faithfully read out with $c_{out}=-s_{T}$.

\begin{figure}[tbp]
\label{d63reviewp7}
\par
\begin{center}
\includegraphics[width=8.0cm]{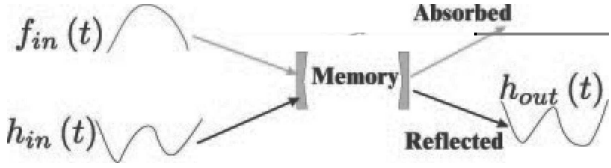}
\end{center}
\caption{Schematic setup to illustrate pulse shape splitter.}
\end{figure}

The physical setup of quantum light memory discussed above could have other
applications. We note that the output pulse shape $f_{out}\left( t\right) $
need not be the same as the input pulse shape $f_{in}\left( t\right) $. This
means that the setup can work as a pulse shape modulator to change the shape
of an optical pulse without alternation of its quantum state. The pulse
shape modulator could be useful in quantum communication between two
cavities \cite{Cirac97}. For instance, if one wants to input a pulse from
the free decay of a cavity to another cavity with the same decay rate, the
pulse will be nearly completely reflected at the mirror of the second cavity
\cite{Cirac97}. However, if between the two cavities we insert a pulse shape
modulator to change the pulse shape to be its time reversal, the pulse will
be nearly completely absorbed by the second cavity since the input process
is exactly the time reversal of the output process. Besides the application
as a pulse shape modulator, the quantum light memory setup can also be used
as a pulse shape splitter illustrated by Fig.~ \ref{d63reviewp7}. Consider
that we have two independent pulse modes superposed in the same time window $%
\left[ 0,T\right] $, with the shapes denoted by $f_{in}\left( t\right) $ and
$h_{in}\left( t\right) $ respectively. The pulse shape functions are
orthogonal to each other with $\int_{0}^{T}f_{in}^{\ast }\left( t\right)
h_{in}\left( t\right) dt=0$ for independent modes. Now we want to split
these two modes by selectively transferring one of the optical modes $%
c_{in}=\int_{0}^{T}f_{in}\left( t\right) a_{in}\left( t\right) dt$ to the
atomic mode $s$ with the impedance matching condition. In this case, one can
show from Eq. (\ref{25}) and the input-output relation that the other mode $%
d_{in}=\int_{0}^{T}h_{in}\left( t\right) a_{in}\left( t\right) dt$ is
completely reflected. In fact, one has $d_{out}\equiv
\int_{0}^{T}h_{out}\left( t\right) a_{out}\left( t\right) dt=d_{in}$, where $%
d_{out}$ is the reflected optical mode with its pulse shape changed to $%
h_{out}\left( t\right) =h_{in}\left( t\right) -\left( e^{R\left( T\right)
}-1\right) e^{-R(t)}f_{in}\left( t\right) \int_{0}^{t}f_{in}^{\ast }\left(
\tau \right) h_{in}\left( \tau \right) d\tau $, where $R(t)\equiv
\int_{0}^{t}\kappa ^{\prime }\left( \tau \right) d\tau $. In this way, we
obtain a pulse shape splitter to separate different shapes, just like a
polarization beam splitter to separate different polarizations.

In the above we have shown how to store an effectively one-mode optical
field in an atomic ensemble. It is also possible to store many optical modes
in the same atomic ensemble with a step-by-step method to increase its
memory capacity. For this we consider a one-dimensional atomic ensemble with
a length $L_{a}$. The coordinate of the $j$ atom is denoted by $x_{j}$. We
introduce a Fourier transformation to the individual atomic operator $\sigma
_{12}^{j}=\left| 1\right\rangle _{j}\left\langle 2\right| $ with the form $%
s_{\mu }\equiv \sum_{j}\sigma _{12}^{j}e^{i\mu x_{j}/L_{a}}/\sqrt{N_{a}}$ $%
\left( \mu =0,1,\cdots \right) $. Under the weak excitation condition $%
\left\langle \left| 2\right\rangle _{j}\left\langle 2\right| \right\rangle
\ll 1$, the new modes $s_{\mu }$ are independent and satisfy bosonic
commutation relations $\left[ s_{\mu },s_{\mu ^{\prime }}^{\dagger }\right]
=\delta _{\mu \mu ^{\prime }}$. We can induce a transition $s_{\mu
}\rightarrow s_{\mu +1}$ $\left( \mu =0,1,\cdots \right) $\ between the new
modes by applying an electric field with spatial gradient for a suitable
time to give a coordinate-dependent phase kick $\sigma _{12}^{j}\rightarrow
\sigma _{12}^{j}e^{ix_{j}/L_{a}}$. To store in one atomic ensemble many
optical pulse modes which come one after another, we transfer the first
optical mode to the collective atomic mode $s=s_{0}$ with the method
described above, and then induce a transition $s_{\mu }\rightarrow s_{\mu
+1} $ by applying a phase kick. After the phase kick, the mode $s$ is free
to be used for storing the second optical mode. This process can be
continued until many optical modes are stored in the atomic modes $%
s,s_{1},s_{2},\cdots $. For releasing these atomic modes, we can simply
reverse the above process.

In the real experiments, one cannot realize ideal quantum light memory with
perfect state mapping between photons and atoms. As has been shown above,
there is some inherent loss if the interaction time $T$ is not much longer
than the average effective coupling rate $\overline{\kappa ^{\prime }}$.
Additional to this, atomic spontaneous emissions will also contribute to
loss with a signal-to-noise ratio given by $R_{sn}\sim 4N_{a}\left|
g_{1}\right| ^{2}/\left( \kappa \gamma _{s}\right) $ as described in Sec.
4.2. Assume that the overall inefficiency for all the loss effects in
quantum light memory is denoted by $\eta $. In this case, if one inputs an
optical mode in a coherent state $\left| \alpha \right\rangle $ (which is
commonly taken in experimental demonstrations), the read-out state from the
memory is given by $\left| \sqrt{1-\eta }\alpha \right\rangle $ with a state
fidelity $F=\left( \left\langle \alpha \right| \left| \sqrt{1-\eta }\alpha
\right\rangle \right) ^{2}=e^{-x_{\eta }\left| \alpha \right| ^{2}}$, where $%
x_{\eta }=\left( 1-\sqrt{1-\eta }\right) ^{2}$. With this imperfection, one
would like to ask how to experimentally verify that one indeed realizes a
quantum light memory, which should be better than any classical memory
protocol. For instance, in a classical protocol, one can measure the input
optical state to get some classical information, and then according to this
information prepare a similar state after some time. The problem here is
very similar to that in continuous variable quantum teleportation \cite%
{Braunstein981,Furusawa98}, and we can use the result there to provide an
experimentally testable criterion for quantum light memory. This criterion
is given by measuring the input-output state fidelity $F$. Assume that the
input state to the memory is randomly chosen from the set of coherent states
$\left\{ \left| \alpha \right\rangle \right\} $ with a probability
distribution $p\left( \alpha \right) =\left( \lambda /\pi \right)
e^{-\lambda \left| \alpha \right| ^{2}}$, where $\lambda $ is a positive
parameter, quantum light memory is verified with the confirmation to be
better than any classical memory protocol if the measured average state
fidelity $F>F_{\text{cri}}=\left( 1+\lambda \right) /\left( 2+\lambda
\right) $ \cite{Braunstein00}. With an overall loss inefficiency $\eta $ for
quantum light memory, the calculated average state fidelity in this case is
given by $F_{\text{cal}}=\lambda /\left( \lambda +x_{\eta }\right) $. One
can see that it is always possible to make $F_{\text{cal}}>F_{\text{cri}}$
by choosing a large parameter $\lambda $, that is, by choosing the input
coherent states close enough to the vacuum state. In a real experiment,
there might be some technique noise which induces an additional fidelity
decrease $F_{\text{tec}}$ to the calculated value $F_{\text{cal}}$. In this
case, one can verify that to meet the criterion $F=F_{\text{cal}}-F_{\text{%
tec}}>F_{\text{cri}}$, one has an optimal choice of the parameter $\lambda $
to be $\lambda =\left( 1-x_{\eta }\right) /\left( 2F_{\text{tec}}\right)
-1-x_{\eta }/2$, and the technique fidelity decrease $F_{\text{tec}}$ needs
to be approximately below $F_{\text{tec}}<\left( 1-x_{\eta }\right) ^{2}/4$
for a successful demonstration of quantum light memory. In the experimental
demonstrations \cite{Liu01,Phillips01}, the above criterion has not been
checked, but it seems possible to meet this criterion with the current
experimental technology.

\subsubsection{Single-photon source}

As has been mentioned before, single-photon source has many applications in
quantum information processing. It is desirable to have a single-photon
source with controllable emission time, direction, and pulse shape. This is
required in some quantum information processing schemes since one needs to
interfere two single-photon pulses, and this is generally available only
when we can control the emission time, direction, and shape of the pulses.
It is possible to produce single-photons with the setup of optical
spontaneous parametric down conversion, which has been commonly used now in
quantum communication experiments \cite{Zeilinger99}. In this setup, photons
are always generated in pairs due to the nonlinearity in the optical
crystal. If we measure one output beam with a single-photon detector,
conditional on a detector click the other output beam will be projected to a
single-photon state. However, in this setup single photons will be produced
at random times which are not controllable due to the randomness of
spontaneous emissions. There are also proposals and experiments of using
blockade mechanism in semiconductors or other solid-state materials to
produce a source of single-photons \cite{Jim99,Michler00,Beveratos01}, with
the emission time fully controllable. To require the emitted single-photon
pulses to be also directional, it seems that one needs to build a good
cavity around the material. A fully controllable single-photon source is
achievable if one could trap single-atoms in high-Q cavities for a
sufficiently long time \cite{Gheri98}. However, as we have mentioned before,
this is possible but a experimentally challenging work. We show here that
atomic ensembles can provide another way to achieve a fully controllable
single-photon source, which seems to be much easier for experimental
demonstrations.

The principle of using an atomic ensemble to produce a single-photon source
can be easily understood with the ideas explained in this chapter. The
atomic ensemble working in the $\Lambda $II-level configuration generates a
correlated state in the form of Eq. (\ref{19}) in the perturbative limit,
which is an exact analogy of the optical spontaneous parametric down
conversion process. We can measure the forward scattered signal with a
single-photon detector, and conditional on a detector click, the collective
atomic mode is projected to a single-excitation state. Since excitations can
be stored for a reasonably long time in the ground-state manifold of the
atoms, we can transfer the single-atomic excitation to the single-photonic
excitation at any desired time with the method described in the previous
subsection. The emission time is controllable by when the repumping pulse is
applied, and the emitted single-photon pulse is directed to the forward
direction. The pulse shape is controllable by changing the time dependence
of the Rabi frequency of the repumping pulse as in quantum light memory.
This shows that the relatively simple system of an atomic ensemble can be
used to produce single-photon pulses with fully controllable properties.

\subsection{Applications in continuous variable quantum information
processing}

In quantum information protocols, quantum information is normally carried by
qubits, that is, by two-dimensional quantum systems. Similar to the
classical case, quantum information can also be carried by some observables
with continuous values, the canonical observables $X$ and $P$ for instance.
The bosonic field (such as the optical field) provides a natural physical
system to carry the continuous variable quantum information. Because of
this, continuous variable quantum information processing has recently
aroused a lot of interests. There have been proposals for continuous
variable quantum teleportation \cite{Braunstein981,Furusawa98}, cryptography
\cite{Gottesman01}, computation \cite{Braunstein99}, error correction \cite%
{Braunstein982}, entanglement purification \cite{Duan003}, cloning \cite%
{Cerf00}, and etc. , and continuous variable teleportation has been
experimentally demonstrated by using single-mode optical fields \cite%
{Furusawa98}. We have seen from Sec. 4.2 that one can define a pair of
canonical continuous-valued observables for atomic ensembles with suitable
level schemes (the four-level scheme for instance). This property, combined
with the ability of storing (qubit or continuous variable) quantum
information in the ground-state manifold of atomic ensembles and the
collectively enhanced coupling of the atomic mode in the ensemble to the
optical mode, provides many possibilities for using atomic ensembles in
continuous variable quantum information processing.

A good existing example to show these possibilities is given by the recent
scheme for realizing continuous variable atomic quantum teleportation with
laser manipulation of atomic ensembles \cite{Duan002}. Quantum teleportation
of atomic states has not been realized yet due to the difficulty of
achieving strong light-atom coupling. As we have seen, collective
enhancement of the signal-to-noise ratio in atomic ensembles provides a
possible way to go around this problem. There are two proposals to realize
continuous variable teleportation with free-space atomic ensembles. Ref.
\cite{Kuzmich002} is based on the use of an external source of entanglement
(non-classical light). Ref. \cite{Duan002} eliminates this requirement, and
proposes a quantum teleportation scheme with the use of only coherent light.
Based on the method in Ref. \cite{Duan002}, a recent experiment has
successfully generated entanglement between two distant macroscopic atomic
ensembles \cite{Julsgaard01}, which is an important first step towards the
final realization of atomic quantum teleportation. The following of this
section is mainly devoted to a review of the scheme proposed in Ref. \cite%
{Duan002}, and we will also briefly remark at the end of this section the
possibilities of using atomic ensembles for realization of other continuous
variable quantum information protocols.

\begin{figure}[tbp]
\label{d63reviewp8}
\par
\begin{center}
\includegraphics[width=8.0cm]{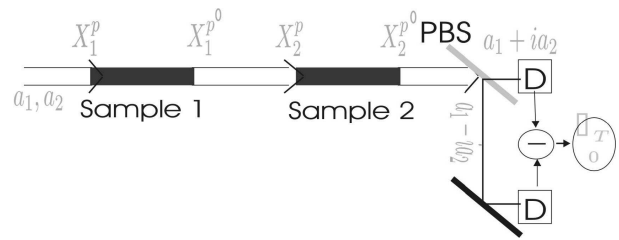}
\end{center}
\caption{Schematic setup for Bell measurements. A linearly polarized strong
laser pulse (decomposed into two circular polarization modes $a_{1},a_{2}$)
propagates successively through the two atomic samples. The two polarization
modes $\left( a_{1}+ia_{2}\right) /\protect\sqrt{2}$ and $\left(
a_{1}-ia_{2}\right) /\protect\sqrt{2}$ are then split by a polarizing beam
splitter (PBS), and finally the difference of the two photon currents
(integrated over the pulse duration $T$) is measured.}
\end{figure}

The scheme in Ref. \cite{Duan002} is based on the four-level scheme
described and analyzed in details in Sec. 4.2. The transformations (\ref{18}%
) of the canonical atomic and optical observables induced by the light-atom
interaction serve as the basic equations for understanding this scheme. To
teleport continuous variable states from one atomic ensemble to the other,
first we need to generate entanglement between the continuous observables $%
X_{1}^{a},P_{1}^{a}$ and $X_{2}^{a},P_{2}^{a}$ of two distant ensembles 1
and 2. This is done through a nonlocal Bell measurement of the EPR operators
$X_{1}^{a}-X_{2}^{a}$ and $P_{1}^{a}+P_{2}^{a}$ with the setup depicted by
Fig.~ \ref{d63reviewp8}. This setup measures the Stokes operator $%
X_{2}^{p\prime }$ of the output light. Using Eq.~(\ref{18}) and neglecting
the small loss terms, we have $X_{2}^{p\prime }=X_{1}^{p}+\kappa _{c}\left(
P_{1}^{a}+P_{2}^{a}\right) $, so we get a collective measurement of $%
P_{1}^{a}+P_{2}^{a}$ with some inherent vacuum noise $X_{1}^{p}$. The
efficiency $1-\eta $ of this measurement is determined by the parameter $%
\kappa _{c}$ with $\eta =1/\left( 1+2\kappa _{c}^{2}\right) $. After this
round of measurements, we rotate the collective atomic spins around the $x$
axis to get the transformations $X_{1}^{a}\rightarrow -P_{1}^{a},$ $%
P_{1}^{a}\rightarrow X_{1}^{a}$ and $X_{2}^{a}\rightarrow P_{2}^{a},$ $%
P_{2}^{a}\rightarrow -X_{2}^{a}$. The rotation of the atomic spin can be
easily obtained with negligible noise by applying classical laser pulses
with detuning $\Delta \gg \gamma $. After the rotation, the measured
observable of the first round of measurement is changed to $%
X_{1}^{a}-X_{2}^{a}$ in the new variables. We then make another round of
collective measurement of the new variable $P_{1}^{a}+P_{2}^{a}$. In this
way, both the EPR operators $X_{1}^{a}-X_{2}^{a}$ and $P_{1}^{a}+P_{2}^{a}$
are measured, and the final state of the two atomic ensembles is collapsed
into a two-mode squeezed state with variance $\delta \left(
X_{1}^{a}-X_{2}^{a}\right) ^{2}=\delta \left( P_{1}^{a}+P_{2}^{a}\right)
^{2}=e^{-2r}$, where the squeezing parameter $r$ is given by
\begin{equation}
r=\frac{1}{2}\ln \left( 1+2\kappa _{c}^{2}\right) .  \label{28}
\end{equation}
Thus, using only coherent light, we generate continuous variable
entanglement \cite{Duan004} between two nonlocal atomic ensembles. With the
interaction parameter $\kappa _{c}\approx 5$, a high squeezing (and thus a
large entanglement) $r\approx 2.0$ is obtainable. Note that entanglement
generation is the key step for many quantum information protocols, and as an
example we show in the following how to use it to achieve quantum
teleportation.

To achieve quantum teleportation, first the ensembles 1 and 2 are prepared
in a continuous variable entangled state using the nonlocal Bell measurement
described above. Then, a Bell measurement with the same setup as shown by
Fig.~\ref{d63reviewp8} on the two local ensembles 1 and 3, together with a
straightforward displacement of $X_{3}^{a},$ $P_{3}^{a}$ on the sample 3,
will teleport an unknown collective spin state from the atomic ensemble 3 to
2. The teleported state on the ensemble 2 has the same form as that in the
original proposal of continuous variable teleportation using squeezing light
\cite{Braunstein981}, with the squeezing parameter $r$ replaced by Eq.~(\ref%
{28}) and with an inherent Bell detection inefficiency $\eta =1/\left(
1+2\kappa _{c}^{2}\right) $. The teleportation quality is best described by
the fidelity, which, for a pure input state, is defined as the overlap of
the teleported state and the input state. For any coherent input state of
the sample 3, the teleportation fidelity is given by
\begin{equation}
F=1/\left( 1+\frac{1}{1+2\kappa _{c}^{2}}+\frac{1}{2\kappa _{c}^{2}}\right) .
\label{29}
\end{equation}
Equation (\ref{29}) shows, if there is no extra noise, a high fidelity $%
F\approx 96\%$ would be possible for the teleportation of the collective
atomic spin state with the interaction parameter $\kappa _{c}\approx 5$.

Finally, we need to incorporate several sources of noise in this scheme and
analyze their influence on the teleportation fidelity. The noise includes
the spontaneous emission noise described by Eq. (\ref{18}), the detector
inefficiency, and the transmission loss of the light from the first ensemble
to the second one. The spontaneous emission noise can be included partly in
the transmission loss and partly in the detector efficiency, so we do not
analyze it separately. The effect of the detector inefficiency $\eta _{d}$
is to replace $\kappa _{c}^{2}$ in Eqs.~(\ref{28}) and (\ref{29}) with $%
\kappa _{c}^{2}\left( 1-\eta _{d}\right) $, and the teleportation fidelity
is decreased by a term $\eta _{d}/\kappa _{c}^{2}$, which is very small and
can be safely ignored. The most important noise comes from the transmission
loss. The transmission loss is described by $X_{2}^{p}=\sqrt{1-\eta _{t}}%
X_{1}^{p^{\prime }}+\sqrt{\eta _{t}}X_{s}^{t}$ (see Fig.~ \ref{d63reviewp8}%
), where $\eta _{t} $ is the loss rate and $X_{s}^{t}$ is the standard
vacuum noise. The transmission loss changes the measured observables to be $%
\sqrt{1-\eta _{t}}X_{1}^{a}-X_{2}^{a}$ and $\sqrt{1-\eta _{t}}%
P_{1}^{a}+P_{2}^{a}$. These two observables do not commute, and the two
rounds of measurements influence each other. To minimize the influence on
the teleportation fidelity, we choose the following configuration (for
simplicity, we assume we have the same loss rate $\eta _{t}$ from the sample
1 to 2 and from 1 to 3): In the nonlocal Bell measurements on the samples 1
and 2 (the entanglement generation process), we choose a suitable
interaction coefficient $\kappa _{c2}$ (where its optimal value will be
determined below) for the second round measurement, whereas $\kappa _{c1}$
for the first round of measurement is large with $\kappa _{c1}^{2}\gg \kappa
_{c2}^{2}$\ (the interaction coefficient can be easily adjusted, for
instance, by changing the detuning). In the local Bell measurement, we
choose the same $\kappa _{c2}$ for the first round of measurement and the
large $\kappa _{c1}$ for the second round of measurement. For a coherent
input state of the ensemble 3, the teleported state on the ensemble 2 is
still a Gaussian state, and the teleportation fidelity $F$ is found to be
\begin{equation}
F\approx 2/\left( 2+\frac{1}{\kappa _{c2}^{2}}+\kappa _{c2}^{2}\eta
_{t}\right) \leq 1/\left( 1+\sqrt{\eta _{t}}\right) ,  \label{30}
\end{equation}
which is independent of the coherent input state with suitable gain for the
displacements \cite{Braunstein981}. The optimal value for $\kappa _{c2}$ is
thus given by $\kappa _{c2}=1/\sqrt[4]{\eta _{t}}$. Even with a significant
transmission loss rate $\eta _{t}\sim 0.2$, quantum teleportation with a
remarkable high fidelity $F\sim 0.7$ is still achievable, which well exceeds
the fidelity criterion $1/2$ for continuous variable quantum teleportation
with arbitrary coherent input states \cite{Braunstein00}.

We have described the scheme in Ref. \cite{Duan002} for achieving continuous
variable quantum teleportation of atomic spin state by laser manipulation of
several atomic ensembles. The proposed scheme is within the reach of the
current experimental technology, and has been partially demonstrated by the
first-step experiment \cite{Julsgaard01}. In general, one concerns about the
possibilities of using laser manipulation of atomic ensembles to realize
other continuous variable quantum information schemes. The four-level
configuration used here and the $\Lambda $II-level configuration used in
Sec. 4.3 have the ability to produce any squeezing operation. The $\Lambda $%
1-level configuration can be used to realize any beam-splitter like
operation. One can also use the phase-kick technique discussed in quantum
light memory subsection to manipulate many bosonic modes in one atomic
ensemble, and the number of controllable modes can be further extended and
is well scalable by connecting many atomic ensembles through optical pulses
making use of the collective enhancement of the desired light-atom coupling
in the ensembles. With these abilities, in principle one can use laser
manipulation of atomic ensembles to realize any scheme which is based on the
following physical requirements, that is, a series of well controllable
bosonic modes, and the ability of preforming any desired squeezing or
beam-splitter like operation on these modes. The schemes belong to this
class include some continuous variable quantum cryptography scheme \cite%
{Gottesman01}, the continuous variable quantum error correction scheme in
Ref. \cite{Braunstein982}, and the scheme in Ref. \cite{Cerf00} for quantum
cloning of Gaussian continuous variable states. To realize the continuous
variable quantum computation scheme in Ref. \cite{Braunstein99} and the
continuous variable entanglement purification scheme in Ref. \cite{Duan003},
one needs another kind of operation, i.e., the Kerr operation, which is
described by the single-mode Hamiltonian $H_{k}=\chi _{k}\left( a^{\dagger
}a\right) ^{2}$. There have been proposals of some level configurations to
realize the Kerr operation in atomic ensembles using the phenomenon of
electromagnetic field induced transparency (EIT) \cite{Imamoglu97,Lukin003}.
Unfortunately, unlike the three configurations discussed in Sec. 4.2, the
signal-to-noise ratio for the Kerr operation is not collectively enhanced
\cite{Gheri99}, and as a result one basically still needs to build a high-Q
cavity around the ensemble and has to enter the strong coupling regime for
getting a good signal-to-noise ratio, which is experimentally very
challenging. Therefore, it seems to be hard to realize continuous variable
quantum computation and entanglement purification solely based on laser
manipulation of atomic ensembles, but studies are going on in this
direction, and possibly one can realizes it with a combination of some other
technique, using direct interactions between atoms as in Refs. \cite%
{Lukin002} for instance.

\subsection{Summary}

This chapter has reviewed the recent advances of using atomic ensembles for
quantum information processing. We put emphasis on the collective
enhancement of the signal-to-noise ratio for the coupling between light and
atomic ensembles with suitable level configurations. Due to the collectively
enhanced coupling, we can do various kinds of interesting quantum
information processing simply by laser manipulation of atomic ensembles in
weak coupling cavities or even in free space, which greatly simplifies their
experimental demonstration. All the theoretical schemes for quantum
information processing reviewed in this chapter are within the scope of the
near-future experiments. We hope that these example schemes have shown the
promising future of using atomic ensembles for quantum information
processing, with a combination of the advantages of a long coherence time
and a collectively-enhanced coupling to light. The progress in this area is
fast, and the important open questions include a full understanding of the
interaction of light with hot and cold atomic ensembles in three-dimensional
free space with various level configurations, and applications of these
interaction configurations in physical realization of more quantum
information protocols.


\begin{thebibliography}{999}
\bibitem{Ni01} A very good introduction to quantum information and its
applications can be found in: PERES, A. \textit{Quantum Theory: Concepts and
Methods}, (Kluwer Academic) 1993; NIELSEN M. and CHUANG I., \textit{Quantum
computation and quantum information}, (Cambridge University Press), 2000;
\textit{The Physics of Quantum Information: Quantum Cryptography, Quantum
Teleportation, Quantum Computation}, Edited by BOUWMEESTER D., EKERT A., and
ZEILINGER A. (Springer--Verlag), 2000. On the other hand, most of the
articles on the field can be found in http: //xxx.lanl.gov/ archive /
quant-ph.

\bibitem{Sh94} SHOR P. W., \textit{Proc. of the 3rd Annual Symposium on the
Foundations f Computer Science} (IEEE Computer Society Press, Los Alamitos,
Ca) 1994, pp. 124.

\bibitem{Ek96} EKERT A. and JOSZA R., \textit{Rev. Mod. Phys.}, \textbf{68}
(1995) 733.


\bibitem{Briegel98} BRIEGEL H. J., DUR W., CIRAC J. I., ZOLLER P., \textit{%
Phys. Rev. Lett.}, \textbf{81} (1998) 5932; DUR W., BRIEGEL H. J., CIRAC J.
I., ZOLLER P., \textit{Phys. Rev. A}, \textbf{59} (1999) 169.

\bibitem{Be64} BELL J. S., \textit{Physics} \textbf{1} (1964) 195.

\bibitem{Be96} BENNETT C. H., BERNSTEIN H. J., POPESCU S., and SCHUMACHER
B., \textit{Phys. Rev. A}, \textbf{53} (1996) 2046.

\bibitem{Po97} POPESCU S. and ROHRLICH D., \textit{Phys. Rev. A}, \textbf{56}
(1997) 3319.

\bibitem{Li98} LINDEN N. and POPESCU S., \textit{Fortsch. Phys.}, \textbf{46}
(1998) 567.

\bibitem{Gr89} GREENBERGER D. M., HORNE M., and ZEILINGER A., in \textit{%
Bell's theorem, Quantum Theory and Conceptions of the Universe}, edited by
M. Kafatos (Kluwer Academic, Dordrecht, The Netherlands) 1989, pp 69.

\bibitem{Du00W} DUR W., VIDAL G., and CIRAC J. I., \textit{Phys. Rev. A},
\textbf{62} (2000) 62314.

\bibitem{We89} WERNER R. F., \textit{Phys. Rev. A}, \textbf{40} (1989) 4277.

\bibitem{Pe96} PERES A., \textit{Phys. Rev. Lett}, \textbf{77} (1996) 1413.

\bibitem{Ho97} HORODECKI P., \textit{Phys. Lett. A}, \textbf{232} (1997) 333.

\bibitem{Ho96} HORODECKI R., HORODECKI P., HORODECKI M., \textit{Phys. Lett.
A}, \textbf{210} (1996) 377.

\bibitem{Le00} LEWENSTEIN W. \textit{et al.}, \textit{J. Mod. Opt.}, \textbf{%
77} (2000) 2481; TERHAL B. M., quant-ph/0101032.


\bibitem{Bennett96} BENNETT C. H., BRASSARD G., POPESCU S., SCHUMACHER B.,
SMOLIN J. A., and WOOTTERS W. K. \textit{Phys. Rev. Lett.}, \textbf{76}
(1996) 722; BENNETT C. H., DIVINCENZO D. P., SMOLIN J. A., and WOOTTERS W.
K., \textit{Phys. Rev. A}, \textbf{54} (1996) 3824.

\bibitem{Gi96} GISIN N., \textit{Phys. Lett. A}, \textbf{210} (1996) 151.

\bibitem{Gi98} GIEDKE G., BRIEGEL H. J., CIRAC J. I., and ZOLLER P., \textit{%
Phys. Rev. A}, \textbf{59} (1999) 2641.

\bibitem{Di00} See, for example, DIVINCENZO D. P., \textit{Fortschr. Phys.},
\textbf{48} (2000) 771.

\bibitem{fort} Special issue of \textit{Fortschr. Phys.} (edited by
BRAUNSTEIN S. and LO H.-K.), \textbf{48} (2000) 769.

\bibitem{Sh95} SHOR P. W., \textit{Phys. Rev. A}, \textbf{52} (1995) 2493.

\bibitem{St96} STEANE A. M., \textit{Phys. Rev. Lett.}, \textbf{77} (1996)
793.

\bibitem{La96} LAFLAMME R. \textit{et al.}, \textit{Phys. Rev. Lett.},
\textbf{77} (1996) 198.

\bibitem{Go96} GOTTESMAN D., \textit{Phys. Rev. A}, \textbf{54} (1996) 1862.

\bibitem{Sh96} SHOR P. W., in \textit{37th Symposium on Foundations of
Computing}, (IEEE Computer Society Press) 1996, pp. 56-65.

\bibitem{Bennett93} BENNETT C. H., BRASSARD G., CREPEAU C., JOZSA R., PERES
A., and WOOTTERS W. K. \textit{Phys. Rev. Lett.}, \textbf{70} (1993) 1895.

\bibitem{Boumeester97} BOUWMEESTER D., PAN J.-W., MATTLE K., EIBL M.,
WEINFURTER H., ZEILINGER A., \textit{Nature}, \textbf{390} (1997) 575.

\bibitem{Boschi98} BOSCHI D., BRANCA S., DE MARTINI F., HARDY L., and
POPESCU S., \textit{Phys. Rev. Lett.}, \textbf{80} (1998) 1121.

\bibitem{Furusawa98} FURUSAWA A., SORENSEN J. L., BRAUNSTEIN S. L., FUCHS C.
A., KIMBLE H. J., and POLZIK E. S., \textit{Science}, \textbf{282} (1998)
706.

\bibitem{Cirac95} CIRAC J. I. and ZOLLER P., \textit{Phys. Rev. Lett.},
\textbf{74} (1995) 4091.

\bibitem{Pellizzari95} PELLIZZARI T., GARDINER S.A., CIRAC J.I., and ZOLLER
P., \textit{Phys. Rev. Lett.}, \textbf{75} (1995) 3788.


\bibitem{Cirac97} CIRAC J. I., ZOLLER P., KIMBLE H. J., MABUCHI H., \textit{%
Phys. Rev. Lett.} \textbf{78} (1997) 3221.

\bibitem{Molmer99} SORENSEN A. and MOLMER K., \textit{Phys. Rev. Lett.},
\textbf{82} (1999) 1971.

\bibitem{Poyatos98} POYATOS J.F., CIRAC J.I., and ZOLLER P.
\textit{Phys. Rev. Lett.}, \textbf{81} (1998) 1322.


\bibitem{Jaksch99} JAKSCH D., BRIEGEL H.J., CIRAC J.I., GARDINER C.W., and
ZOLLER P., \textit{Phys. Rev. Lett.}, \textbf{82} (1999) 1975.


\bibitem{Brennen99} BRENNEN G., CAVES C., JESSEN P., and DEUTSCH I., \textit{%
Phys. Rev. Lett.}, \textbf{82} (1999) 1060.


\bibitem{Jaksch00} JAKSCH D., CIRAC J. I., and ZOLLER P., ROLSTON S.L., COTE
R. and LUKIN M. D., \textit{Phys. Rev. Lett.}, \textbf{85} (2000) 2208.

\bibitem{Duan01} DUAN L.-M., CIRAC J. I., and ZOLLER P., \textit{Science},
\textbf{292} (2001) 1695.

\bibitem{Zanardi00} PACHOS J., ZANARDI P., RASETTI M., \textit{Phys. Rev. A}%
, \textbf{61} (2000) 10305(R).


\bibitem{Cirac96} CIRAC J.I., PARKINS A. S., BLATT R., and ZOLLER P.,
\textit{Adv. Atom. Mol. Opt. Phys}, \textbf{37} (1996) 237.


\bibitem{Steane97} STEANE A., 
\textit{Appl. Phys. B}, \textbf{64} (1997) 623.


\bibitem{thebible} WINELAND D. J., MONROE C., ITANO W. M., LEIBFRIED D.,
KING B. E., and MEEKHOF D. M., \textit{RES J. NIST}, \textbf{103} (1998) 259.



\bibitem{Cirac00} CIRAC J. I., ZOLLER P., \textit{Nature}, \textbf{404}
(2000) 579.


\bibitem{Calarco01} CALARCO T., CIRAC J. I., and ZOLLER P., \textit{Phys.
Rev. A}, \textbf{63} (2001) 62304.


\bibitem{Law96} LAW C.K. and EBERLY J.H., \textit{Phys. Rev. Lett.}, \textbf{%
76} (1996) 1055.


\bibitem{Gardiner97} GARDINER S. A., CIRAC J. I., and ZOLLER P., \textit{%
Phys. Rev. A}, \textbf{55} (1997) 1683.

\bibitem{Gardiner99} GARDINER C. W., ZOLLER P., \textit{Quantum Noise}
(Springer) 1999.


\bibitem{Stenholm86} STENHOLM S., \textit{Rev. Mod. Phys.}, \textbf{58}
(1986) 699.

\bibitem{Cirac92} CIRAC J. I., BLATT R., ZOLLER P., and PHILLIPS W. D.,
\textit{Phys. Rev. A}, \textbf{46} (1992) 2668.


\bibitem{Marzoli94} MARZOLI I., CIRAC J. I., BLATT R., and ZOLLER P.,
\textit{Phys. Rev. A}, \textbf{49} (1994) 2771.


\bibitem{Monroe95} MONROE C., MEEKHOF D. M., KING B. E., ITANO W. M., and
WINELAND D. J., \textit{Phys. Rev. Lett.}, \textbf{75} (1995) 4714.


\bibitem{King98} KING B. E., WOOD C. S., MYATT C. J., TURCHETTE Q.A.,
LEIBFRIED D., ITANO W. M., MONROE C., and WINELAND D. J., {it Phys. Rev.
Lett.}, \textbf{7} (1998) 1525.


\bibitem{Roos00} ROOS C.F., LEIBFRIED D., MUNDT A., SCHMIDT-KALER F.,
ESCHNER J., BLATT R., \textit{Phys. Rev. Lett}, \textbf{85} (2000) 5547.


\bibitem{Nagerl99} NAGERL H. C., LEIBFRIED D., ROHDE H., THALHAMMER G.,
ESCHNER J., SCHMIDT-KALER F., BLATT R., \textit{Phys. Rev. A}, \textbf{60}
(1999) 145.


\bibitem{Turchette98} TURCHETTE Q. A., WOOD C. S., KING B. E., MYATT C. J.,
LEIBFRIED D., ITANO W. M., MONROE C., and WINELAND D. J., \textit{Phys. Rev.
Lett.}, \textbf{81} (1998) 3631.


\bibitem{Sackett00} SACKETT C.A. KIELPINSKI D., KING B.E., LANGER C., MEYER
V., MYATT C.J., ROWE M., TURCHETTE Q.A., ITANO W.M., WINELAND D.J., \textit{%
Nature}, \textbf{404} (2000) 256.


\bibitem{Kielpinski01} KIELPINSKI D., MEYER V., ROWE M. A., SACKETT C. A.,
ITANO W. M., MONROE C., and WINELAND D. J., \textit{Science}, \textbf{291}
(2001) 1013.


\bibitem{Rowe01} ROWE M. A., KIELPINSKI D., MEYER V., SACKETT C. A. ITANO,
W. M., MONROE C., WINELAND D. J., \textit{Nature}, \textbf{409} (2001) 791.


\bibitem{Knight00} JONATHAN D., PLENIO M. B., and KNIGHT P. L., \textit{%
Phys. Rev. A}, \textbf{62} (2000) 42307.


\bibitem{Zanardi99} ZANARDI P. and RASETTI M., \textit{Phys. Lett. A},
\textbf{264} (1999) 94.

\bibitem{Pachos01} PACHOS J., ZANARDI P., LANL preprint available at
http://xxx.lanl.gov/abs/quant-ph/0007110.


\bibitem{Cirac96b} CIRAC J.I., PELLIZZARI T., ZOLLER P., \textit{Science},
\textbf{273} (1996) 1207.



\bibitem{Pellizzari97} PELLIZZARI T., \textit{Phys. Rev. Lett.}, \textbf{79}
(1997) 5242.


\bibitem{Turchette95} TURCHETTE Q.A., HOOD C.J., LANGE W., MABUCHI H., and
KIMBLE H. J., \textit{Phys. Rev. Lett.}, \textbf{75} (1995) 4710.


\bibitem{Maitre97} MAITRE X., HAGLEY E., NOGUES G., WUNDERLICH C., GOY P.,
BRUNE M., RAIMOND J. M., and HAROCHE S., \textit{Phys. Rev. Lett.} \textbf{79%
} (1997) 769.


\bibitem{Rauschenbeutel99} RAUSCHENBEUTEL A., NOGUES G., OSNAGHI S., BERTET
P., BRUNE M., RAIMOND J. M., and HAROCHE S., \textit{Phys. Rev. Lett.},
\textbf{83} (1999) 5166.


\bibitem{Rauschenbeutel00} RAUSCHENBEUTEL A., NOGUES G., OSNAGHI S., BERTET
P., BRUNE M., RAIMOND J. M., and HAROCHE S., \textit{Science}, \textbf{288}
(2000) 2024.


\bibitem{Osnaghi01} OSNAGHI S., BERTET P., AUFFEVES A., MAIOLI P., BRUNE M.,
RAIMOND J.M., and HAROCHE S., \textit{Phys. Rev. Lett.}, \textbf{87} (2001)
37902.


\bibitem{Walther01} BRATTKE S., VARCOE B.T.H., and WALTHER H., \textit{Phys.
Rev. Lett.}, \textbf{86} (2001) 3534.

\bibitem{Zheng00} S.-B. ZHENG, and G.-C. GUO, \textit{Phys. Rev. Lett.},
\textbf{85} (2000) 2392.


\bibitem{Gardiner93} GARDINER C.W., \textit{Phys. Rev. Lett.}, \textbf{70}
(1993) 2269.

\bibitem{Carmichael93} CARMICHAEL H.J., \textit{Phys. Rev. Lett.}, \textbf{70%
} (1993) 2273.


\bibitem{Bose99} BOSE S., KNIGHT P.L., PLENIO M.B., and VEDRAL V., {Phys.
Rev. Lett.}, \textbf{83} (1999) 5158.




\bibitem{Enk97} VAN ENK S.J., CIRAC J.I., and ZOLLER P., \textit{Phys., Rev.
Lett.}, \textbf{78} (1997) 4293.


\bibitem{Enk98} VAN ENK S.J., CIRAC J.I., and ZOLLER P., \textit{Science},
\textbf{279} (1998) 205.




\bibitem{Calarco00} CALARCO T., HINDS E.A., JAKSCH D., SCHMIEDMAYER J.,
CIRAC J.I., and ZOLLER P., \textit{Phys. Rev. A}, \textbf{61} (2000) 22304.


\bibitem{Brennen00} BRENNEN G.K., DEUTSCH I.H., JESSEN P.S., \textit{Phys.
Rev. A}, \textbf{61} (2000) 62309.


\bibitem{Brennen01} BRENNEN G.K., DEUTSCH I.H., WILLIAMS C.J.,
quant-ph/0107136 (Phys. Rev. A, in press)


\bibitem{Weiner99} WEINER J., BAGNATO V.S. and ZILIO S., JULIENNE P.S.,
\textit{Rev. Mod. Phys.}, \textbf{71} (1999) 1.


\bibitem{Jaksch98} JAKSCH D., BRUDER C., CIRAC J.I., GARDINER C.W., and
ZOLLER P., \textit{Phys. Rev. Lett.}, \textbf{81} (1998) 3108.


\bibitem{Grangier01} SCHLOSSER N., REYMOND G., PROTSENKO I., GRANGIER P.,
\textit{Nature}, \textbf{411} (2001) 1024.

\bibitem{Birkl01} BUCHKREMER F.B.J., DUMKE R., VOLK M. , MUETHER T., BIRKL
G., and ERTMER W., quant-ph/0110119


\bibitem{Tiesinga00} TIESINGA E., WILLIAMS C.J., MIES F.H., and JULIENNE
P.S., \textit{Phys. Rev. A}, \textbf{61} (2000) 63416.

\bibitem{Gallagher94} GALLAGHER T.F., Rydberg Atoms (Cambridge University
Press, New York, 1994).

\bibitem{Hald99} Hald J., SORENSEN J. L., SCHORI C., and POLZIK E.S.,
\textit{Phys. Rev. Lett.}, \textbf{83} (1999) 1319.

\bibitem{Roch97} ROCH J.-F., VIGNERON K., GRELU Ph., SINATRA A., POIZAT J.
-Ph., and GRANGIER Ph. \textit{Phys. Rev. Lett.}, \textbf{78} (1997) 634.

\bibitem{Hau99} HAU L. V., HARRIS S. E., DUTTON Z., and BEHROOZI C. H.,
\textit{Nature}, \textbf{397} (1999) 594.

\bibitem{Liu01} LIU C., DUTTON Z., BEHROOZI C. H., and HAU L. V., \textit{%
Nature}, \textbf{409} (2001) 490.

\bibitem{Kash99} KASH M. M., SAUTENKOV V. A., ZIBROV A. S., HOLLBERG L.,
WELCH G. R., LUKIN M. D.,  ROSTOVTSEV Y., FRY E. S., and SCULLY M. O.,
\textit{Phys. Rev. Lett.}, \textbf{82} (1999) 5229.

\bibitem{Phillips01} PHILLIPS D. F., FLEISCHHAUER A., MAIR A., and WALSWORTH
R. L., and LUKIN M. D.,  \textit{Phys. Rev. Lett.}, \textbf{86} (2001) 783.

\bibitem{Julsgaard01} JULSGAARD B., KOZHEKIN A., POLZIK E. S.,
quant-ph/0106057, to appear in \textit{Nature}.

\bibitem{Kuzmich98} KUZMICH A., BIGELOW N. P., and MANDEL L.,  \textit{%
Europhys. Lett. A}, \textbf{42} (1998) 481.

\bibitem{Lukin001} LUKIN M. D., YELIN S. F., and FLEISCHHAUER M., \textit{%
Phys. Rev. Lett.}, \textbf{84} (2000) 4232.

\bibitem{Duan001} DUAN L.-M., CIRAC J. I., and ZOLLER P., talk at the
IQEC-CLEO/Europe, (Nice, France) 2000.

\bibitem{Duan01b} DUAN L.-M., LUKIN M. D., CIRAC J. I., and ZOLLER P.,
quant-ph/0105105, to appear in \textit{Nature}.

\bibitem{Raymer81} RAYMER M. A. and MOSTOWSKI J., \textit{Phys. Rev. A},
\textbf{24} (1981) 1980.

\bibitem{Fleischhauer00} FLEISCHHAUER M. and LUKIN M. D., \textit{Phys. Rev.
Lett.}, \textbf{84} (2000) 5094.

\bibitem{Duan002} DUAN L.-M., CIRAC J. I., ZOLLER P., and POLZIK E. S.,
\textit{Phys. Rev. Lett.}, \textbf{85} (2000) 5643.

\bibitem{Raymer85} RAYMER M. G., WALMSLEY I. A., MOSTOWSKI J., and
SOBOLEWSKA B., \textit{Phys. Rev. A}, \textbf{32} (1985) 332.

\bibitem{Kozhekin99} KOZHEKIN A. E., MOLMER K., POLZIK E. S.,
quant-ph/9912014.

\bibitem{Knill01} KNILL E., LAFLAMME R., and MILBURN G. J.,  \textit{Nature}%
, \textbf{409} (2001) 46.

\bibitem{Lukin002} LUKIN M. D., FLEISCHHUAER M., COTE R., DUAN L.-M., JAKSCH
D., CIRAC J. I., and ZOLLER P., \textit{Phys. Rev. Lett.}, \textbf{87}
(2001) 037901.

\bibitem{Fleischhauer01} FLEISCHHAUER M. and LUKIN M. D., quant-ph/0106066.

\bibitem{Ye99} YE J., VERNOOY D. W., and KIMBLE H. J., \textit{Phys. Rev.
Lett.}, \textbf{83} (1999) 4987.

\bibitem{Happer72} HAPPER W., \textit{Rev. Mod. Phys.}, \textbf{44} (1972)
169.

\bibitem{Happer67} HAPPER W. and ATHUR B. S., \textit{Phys. Rev. Lett.},
\textbf{18} (1967) 577.

\bibitem{Kuzmich001} KUZMICH A., MANDEL L., and BIGELOW N. P., \textit{Phys.
Rev. Lett.}, \textbf{85} (2000) 1594.

\bibitem{Takahashi99} TAKAHASHI Y., HONDA K., TANAKA N., TOYODA K., ISHIKAWA
K., and YABUZAKI T., \textit{Phys. Rev. A}, \textbf{60} (1999) 4974.

\bibitem{Molmer99b} MOLMER K., \textit{Eur. Phys. J. D}, \textbf{5} (1999)
301.

\bibitem{Kuzmich002} KUZMICH A. and POLZIK E. S., \textit{Phys. Rev. Lett.},
\textbf{85} (2000) 5639.

\bibitem{Ekert91} EKERT A., \textit{Phys. Rev. Lett.}, \textbf{67} (1991)
661.

\bibitem{Knill98} KNILL E., R. LAFLAMME R., and ZUREK W. H., \textit{Science}%
, \textbf{279} (1998) 342.

\bibitem{Preskill98} PRESKILL J., \textit{Proc. R. Soc. Lond. A}, \textbf{454%
} (1998) 385.

\bibitem{Zukowski93} ZUKOWSKI M., ZEILINGER A., HORNE M. A., and EKERT A.,
\textit{Phys. Rev. Lett.}, \textbf{71} (1993) 4287.

\bibitem{Hood00} HOOD C. J., LYNN T. W., DOHERTY A. C., PARKINS A. S., and
KIMBLE H. J., \textit{Science} \textbf{287} (2000) 1447.

\bibitem{Pinkse00} PINKSE P. W. H., FISCHER T., MAUNZ T. P., and REMPE G.,
\textit{Nature}, \textbf{404} (2000) 365.

\bibitem{Cabrillo99} CABRILLO C., CIRAC J. I., G-FERNANDEZ P., and ZOLLER
P., \textit{Phys. Rev. A}, \textbf{59} (1999) 1025.

\bibitem{Lo99} LO H. K. and CHAU H. F., \textit{Science}, \textbf{283}
(1999) 2050.

\bibitem{Shor00} SHOR P. W. and PRESKILL J., \textit{Phys. Rev. Lett.},
\textbf{85} (2000) 441.

\bibitem{Clauser69} CLAUSER J. F., HORNE M. A., SHIMONY A., and HOLT R. A.,
\textit{Phys. Rev. Lett.}, \textbf{23} (1969) 880.

\bibitem{Budker98} BUDKER D., YASHUK V., and ZOLOTOREV M., \textit{Phys.
Rev. Lett.}, \textbf{81} (1998) 5788.

\bibitem{Brassard00} BRASSARD G., L\"{U}TKENHAUS N., MOR T., and SANDERS B.
C., \textit{Phys. Rev. Lett.}, \textbf{85} (2000) 1330.

\bibitem{Zeilinger99} ZEILINGER A., \textit{Rev. Mod. Phys.}, \textbf{71}
(1999) S288.

\bibitem{Braunstein981} BRAUNSTEIN S. L. and KIMBLE H. J., \textit{Phys.
Rev. Lett.} \textbf{80} (1998) 869.


\bibitem{Braunstein00} BRAUNSTEIN S. L., FUCHS C. A., and KIMBLE H. J.,
quant-ph/9910030.

\bibitem{Jim99} KIM J., BENSON O., KAN H., and YAMAMOTO Y., Nature, \textbf{%
397} (1999) 500.

\bibitem{Michler00} MICHLER P., KIRAZ A., BECHER C., SCHOENFELD W. V.,
PETROFF P. M., ZHANG L., HU E., and  IMAMOGLU A., \textit{Science}, \textbf{%
290} (2000) 2282.

\bibitem{Beveratos01} BEVERATOS A., BROURI R., GACOIN T., POIZAT J.-P., and
GRANGIER P., quant-ph/0104028.

\bibitem{Gheri98} GHERI K. M., SAAVEDRA C., TORMA P., CIRAC J. I., and
ZOLLER P., \textit{Phys. Rev. A}, \textbf{58} (1998) R2627.

\bibitem{Braunstein99} BRAUNSTEIN S. L. and LLOYD S., \textit{Phys. Rev.
Lett.} \textbf{82} (1999) 1789.

\bibitem{Gottesman01} GOTTESMAN D. and PRESKILL J., \textit{Phys. Rev. A},
\textbf{63} (2001) 022309 and refs. therein.

\bibitem{Duan003} DUAN L.-M., GIEDKE G., CIRAC J. I., and ZOLLER P., \textit{%
Phys. Rev. Lett.}, \textbf{84} (2000) 4002; \textit{Phys. Rev. A} \textbf{62}
(2000) 032304.

\bibitem{Cerf00} CERF N. J. and ROTTENBERG A. X., \textit{Phys. Rev. Lett.},
\textbf{85} (2000) 1754.

\bibitem{Braunstein982} BRAUNSTEIN S. L., \textit{Nature}, \textbf{394}
(1998) 47.

\bibitem{Duan004} DUAN L.-M., GIEDKE G., CIRAC J. I., and ZOLLER P., \textit{%
Phys. Rev. Lett.} \textbf{84} (2000) 2722.

\bibitem{Imamoglu97} IMAMOGLU A., SCHMIDT H., WOODS G., and DEUTSCH M.,
\textit{Phys. Rev. Lett.} \textbf{79} (1997) 1467.

\bibitem{Lukin003} LUKIN M. D., IMAMOGLU A., \textit{Phys. Rev. Lett.},
\textbf{84} (2000) 1419.

\bibitem{Gheri99} GHERI K. M., ALGE W., and GRANGIER P., \textit{Phys. Rev. A%
}, \textbf{60} (1999) R2673.
\end{thebibliography}
\end{document}